\documentclass[twocolumn]{aastex631}

\usepackage{amsmath}
\usepackage{appendix}
\usepackage{bbding} 
\usepackage{pifont}
\usepackage{bm}
\usepackage{xcolor}
\usepackage{cuted}
\shorttitle{X-Ray Observations of Old Nearby Supernovae}

\shortauthors{Ahlvind et al.}

\graphicspath{{./}{figures/}}

\begin{document}

\newcommand{\KTHOKC}{Department of Physics, KTH Royal Institute of Technology, The Oskar Klein Centre, AlbaNova, SE-106 91 Stockholm, Sweden}

\newcommand{\lr}[1]{\left(#1\right)}
\newcommand{\vdag}{(v)^\dagger}
\newcommand\aastex{AAS\TeX}
\newcommand\latex{La\TeX}
\newcommand{\jj}[1]{\textcolor{red}{X: #1}}

\title{X-Ray Observations of Old Nearby Supernovae -- Constraints on 
Compact Object Populations and Late Interaction} 

\correspondingauthor{J. Ahlvind}
\email{ahlvind@kth.se}

\author[0009-0002-2740-9570]{Julia Ahlvind}
\affiliation{\KTHOKC}

\author[0000-0003-0065-2933]{Josefin Larsson}
\affiliation{\KTHOKC}

\author[0000-0002-0427-5592]{Dennis Alp}
\affiliation{\KTHOKC}

\received{May 09, 2025} \revised{ October 23, 2025} \accepted{November 4, 2025}
\submitjournal{ApJ}

\begin{abstract}
The properties of the population of compact objects created in core-collapse supernovae (SNe) are uncertain. X-ray observations years to decades after the explosions offer a way to gain insight into this, as hard X-ray emission from the central regions will emerge as the ejecta absorption decreases. Here we analyze and place upper limits on late-time X-ray emission in 242 nearby SNe, using 607 observations from Chandra, XMM-Newton, Swift, and NuSTAR. We use absorption models based on 3D simulations of neutrino-driven explosions to account for absorption of emission from the compact objects by the asymmetric ejecta. We detect X-ray emission from 12 SNe, including four for the first time (SN~1982R, SN~1984J, SN~1992bu, and SN~2003gk), and several of the others at later epochs than before. The X-ray spectra of these SNe are consistent with interaction with the circumstellar medium (CSM), with the possible exception of SN~1979C, which shows an additional hard component, as also noted in previous studies at earlier epochs. This emission may be due to a pulsar wind nebula. Using the upper limits in the full sample, we also perform a population synthesis to constrain the fraction of SNe that produce pulsars and the properties of the pulsars themselves. We find that pulsar populations with mean initial spin periods $\gtrsim100\rm~ ms$ are favored. Finally, we note that the high luminosities of several of the SNe with CSM interaction imply interactions with dense shells. 

\end{abstract}


\section{Introduction} \label{sec:intro}
\setcounter{footnote}{1}
Many massive stars end their lives as core-collapse supernovae (SNe), leaving behind compact objects (COs) in the form of neutron stars (NSs) or black holes (BHs), while others may collapse directly into BHs. The type of CO formed depends on several factors, including progenitor mass, mass-loss history, and metallicity (e.g., \citealt{2020Vink}). While progenitor stars with initial masses above 20--25$~\rm M_\Sol$ are more likely to form BHs, most successful core-collapse events are expected to form NSs \citep{1999Fryer,2003Heger,2011OConnor,2012Ugliano,2015Kochanek,2016Sukhbold}, often with strong magnetic fields and rapid spin, leading to pulsar formation. However, there are major uncertainties in the relationship between the SN properties and the nature and properties of the newly created COs. 

These uncertainties include the properties of the newly created pulsars, with population synthesis of Galactic radio pulsars giving different results, particularly regarding the distribution of initial spin periods \citep{2006Kaspi,2008Perna,2020Cieslar,2024Graber,2024Pardo_Araujo}.
Adding observations of younger COs would be valuable, but this is challenging due to the greater distances and the heavy obscuration by the SN ejecta. Aside from the recently discovered infrared signal associated with the CO in SN~1987A \citep{2024Fransson}, there are only a small number of extragalactic SNe showing evidence of an emerging pulsar wind nebula (PWN) or accreting BH \citep{2011Patnaude,2017Bietenholz,2018Milisavljevic}. 

One way to address this question is to study X-ray emission in SNe several years to decades after the explosions. In the months to years following a SN, the X-ray emission primarily originates from the shock interaction with the circumstellar medium (CSM) \citep{2017Chevalier}. At these early times, the ejecta are optically thick, preventing direct observations of the newly formed CO. However, as the ejecta expand and become optically thin, X-ray emission from the CO may become detectable, providing an opportunity to infer its properties. Systematic X-ray studies of SNe at these late times could therefore offer valuable insights into the connection between SNe and their compact remnants. Importantly, a young PWN is expected to emerge as a hard X-ray component due to the non-thermal spectrum and higher ejecta absorption at low energies, making it stand out from any soft thermal emission from late CSM interaction. 

A previous search for such emission by \cite{2008Perna} did not reveal any convincing candidates, but the available sample has grown substantially since that study was carried out. The four X-ray telescopes–Chandra X-ray Observatory (Chandra) \citep{2000Weisskopf}, XMM-Newton (XMM) \citep{2001Jansen}, Neil Gehrels Swift Observatory (Swift) \citep{2004Gehrels} and the Nuclear Spectroscopic Telescope Array (NuSTAR) \citep{2013Harrison}–-have serendipitously observed the locations of many SNe in nearby galaxies for decades, creating a rich data set for such studies. These observations often span multiple epochs, including some up to $\sim100$~yrs post-explosion, and cover a broad X-ray range, extended to $\sim 70$~keV with NuSTAR, making them well suited for investigating late-time X-ray emission from COs.

In this work, we analyze 607 X-ray observations of 242 nearby SNe (within 60~Mpc), from the four aforementioned X-ray telescopes, obtained before 2021 (or 2022 for Swift). This sample includes previously studied events, as well as many newly detected SNe. We use the absorption models from \cite{2018Alp}, which are based on 3D simulations of neutrino-driven SN explosions \citep{2015Wongwathanarat}, to account for the effects of X-ray absorption within the asymmetric SN ejecta. Based on the CO luminosity limits and detections, we perform a population synthesis and compare the results to previous radio population studies to place constrains on the pulsar birth properties.

This paper is organized as follows. Section~\ref{sec:sample_selection} introduces the SN sample, and Section~\ref{sec:data_reduction} describes the data reduction. In Section~\ref{sec:method}, we outline the methodology, including descriptions of the spectral models. The main results are presented in Section~\ref{sec:results}. In Section~\ref{sec:discussion}, we discuss the implications of our findings, including a population synthesis of pulsar birth properties and constraints on progenitor mass-loss rates in SNe showing evidence for CSM interaction. Finally, Section~\ref{sec:summary_conclusion} summarizes our conclusions.

\section{Sample Selection} \label{sec:sample_selection}
We generate a sample of SNe within 60~Mpc from the Open supernova Catalog (OSNC) \citep{2017Guillochon} from July 2022. The sample includes all SNe types except type IIn, which are known to produce strong X-ray emission due to CSM interaction. We also exclude SN~2014C and the similar events listed in \cite{2022Brethauer}, which were initially classified as type II SNe but later developed into interacting SNe. The sample also excludes SN~1987A, for which the possible contribution from a compact object to the X-ray spectrum has previously been investigated in detail \citep{2018Alp,2021Alp,2021Greco, 2022Greco}, as well as SN~1993J, which is a type IIb that shows clear signs of interaction at late times \citep{1996Fransson,2019Kundu,2009Chandra,2009Nymark,2004Zimmermann}. The resulting SN sample consists of 940 unique entries, 271 stripped type (IIb, Ib and Ic) and 669 type {II} ({II}, {II}P, {II}L, {II}Pec) after the first cut and is summarized in the first row of Table~\ref{tab:sample_cuts}.

\begin{deluxetable}{l l l l}[htbp!] 
\tablenum{1}
\tablecaption{Sample Selection Cuts \label{tab:sample_cuts}}
\tablewidth{0pt} 
\tablehead{ 
\colhead{Cut} & \colhead{Total} & \colhead{Stripped} &\colhead{Type II} }
\startdata
\colhead{Distance} & 940 & 271 & 669 \\
\colhead{X-ray observations} & 620 & 197 & 423\\
\colhead{Epoch} & 276 & 139 & 137\\
\colhead{Data quality} & 242 & 134 & 108
\enddata
\tablecomments{\footnotesize{Each line shows the number of SNe left in the sample after the corresponding selection cut. The last row shows the final number of analyzed SNe.}}
\end{deluxetable}

We cross-match the SN sample with X-ray observations from the Chandra data archive (in March 2023), the XMM-Newton 4XMM-DR11 catalog \citep{2020Webb}, complete up to December 2020, and the Swift Master Catalog from September 2022 \citep{2018Nasa}. The Chandra observations used in this study are taken by the Advanced CCD Imaging Spectrometer (ACIS–S:1-4 and ACIS-I:0, 2 \& 3, where the majority of the sources are detected with ACIS-S3; \citealt{2003Garmire}). For XMM, we use data from the European Photon Imaging Camera (EPIC--pn). Observations from these three telescopes cover approximately similar energy ranges between 0.5 and 10~keV. While observations from the Swift X-ray telescope (XRT) are typically less constraining due to short exposure times and lower effective area, we include Swift observations for completeness and to ensure that we do not miss any interesting candidates. To also cover the hard X-ray range (3--78~keV), the SN sample is further cross-matched with the NuSTAR Master Catalog \citep{2013Harrison,2015Madsen} downloaded in September 2021. For all instruments, we exclude data obtained in timing modes that remove one of the spatial dimensions.

SNe are typically not the targets of these X-ray observations. A successful cross-match, therefore, denotes that the optical SN coordinate is within the field of view (FOV) of the respective X-ray observation. These FOVs are approximately:  $17\arcmin\times17\arcmin$ and $50\arcmin\times8\arcmin$ for Chandra's ACIS-I and ACIS-S respectively \footnote{\url{https://cxc.harvard.edu/cal/Acis/}}, $27\arcmin\times27\arcmin$~ for EPIC of XMM \citep{2001AFreyberg}, $24\arcmin\times24\arcmin$ for XRT on board Swift \citep{2005Burrows} and $13\arcmin\times13\arcmin$ for NuSTAR (FPMA and FPMB) \citep{2015Madsen}. We note that the majority of the Chandra observations are done using the ACIS-S3 chip with a smaller FOV of $8.3\arcmin \times 8.3\arcmin$. Ultimately, we find X-ray cross matches for 620 SNe as summarized in Table~\ref{tab:sample_cuts}.

\begin{figure*}[ht]
\includegraphics[width=\textwidth]{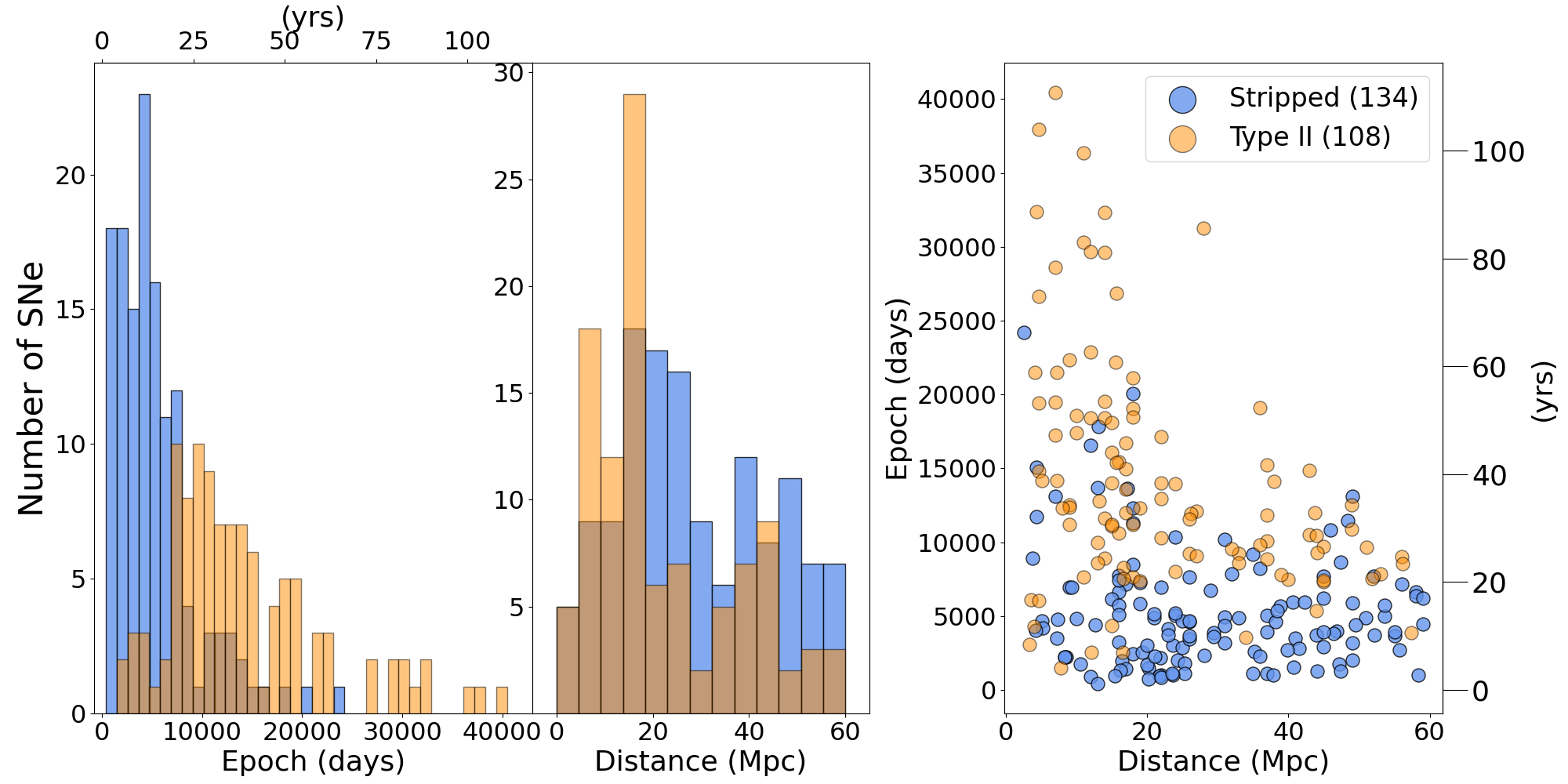}
\caption{\label{fig:hist_1}\footnotesize{Distribution of epochs and distances in the SN sample. The epochs correspond to the latest available observation for each SN. All distances are taken from the OSNC with the exception of SN~2017eaw, for which we use the updated distance from \cite{2019VanDyk}.}}
\end{figure*}

\begin{deluxetable*}{l l l l l l l }[ht] 
\tablenum{2}
\tablecaption{Summary of X-Ray Observations in the SN Sample\label{tab:sample_selection}}
\tablewidth{0pt} 
\tablehead{ 
\colhead{Type} & \colhead{Nr. of SNe} & \colhead{Tot. X-ray obs.} &\colhead{Chandra } &\colhead{XMM} &\colhead{Swift} &\colhead{NuSTAR } }
\startdata
\colhead{Stripped} & 134 & 229 & 113 & 17 & 90 & 9\\
\colhead{Type II} & 108 & 378 & 183 & 91 & 63 & 41
\enddata
\tablecomments{\footnotesize{The number of stripped and type II SNe in our final sample, as well as the corresponding number of X-ray observations in total and per telescope.}}
\end{deluxetable*}

As our main objective is to constrain the X-ray luminosities emanating from newly formed COs, we also make cuts on the observing epochs to remove observations close in time to the SNe explosions. At early epochs, the X-ray emission from CSM interaction is expected to dominate over potential emission from the CO, which is likely heavily absorbed by the ejecta. The optical depth ($\tau$) of the absorbing material in the ejecta decreases quadratically with time, $\tau \propto M_{\rm ej}/(E_{\rm ej}t^2)$, where $M_{\rm ej}$ and $E_{\rm ej}$ are the ejecta mass and energy, and $t$ is the time since explosion. This relation comes from $\tau \propto \rho\kappa\ell_r \propto 1/r^2$, where $ \rho(\propto r^{-3})$ is the mass density, $\kappa$ is the opacity, which is the absorption cross-section per unit mass and independent of $r$, $\ell_r(\propto r)$ is the photon travel distance through the ejecta, and $r$ is the radius of the ejecta. The radius follows $r=v\times t$ for homologous expansion with constant velocity $v$, where $v\propto \sqrt{E_{\rm ej}/M_{\rm ej}}$. We also assume the expanding non-interacting ejecta to be neutral (see Section~\ref{sec:ejecta_models} and \citealt{2018Alp}). Furthermore, the optical depth is typically at least one order of magnitude lower for stripped SNe due to their higher expansion velocities as a result of having lost their envelopes, with the details depending on the progenitor model, line of sight, and energy interval \citep{2018Alp}. 

We estimate approximate times after the explosion when emission from the COs is likely to first become detectable by using our models for ejecta absorption (Section~\ref{sec:ejecta_models}) and considering the effective areas and energy ranges for the different telescopes. The resulting epoch cuts for Chandra, XMM and Swift, which operate in similar soft energy ranges, are $\sim2$ and $\sim$20~yrs for stripped and type {II} SNe respectively. For NuSTAR, which covers harder X-ray energies, we instead attain $\sim~$200~days for stripped SNe and 4~yrs for type~{II}. The earlier epoch cuts for NuSTAR are connected to the fact that the optical depth decreases with photon energy, approximately following $\tau \propto E^{-2}$, but with deviations due to absorption edges, a contribution from Compton scattering, and the ejecta composition in a given model (see Figure~6 in \citealt{2018Alp}). As discussed in \citealt{2018Alp}, the high metallically of the SN ejecta results in a shallower dependence on the photon energy compared to the standard interstellar medium (ISM) case, for which the scaling is closer to $\tau \propto E^{-3}$ (e.g., \citealt{Morrison1983}). We note that we have made these epoch cuts conservative in order to not miss any COs. This implies that detections at the earliest epochs would only be expected for luminous COs in the most nearby SNe with low levels of absorption, corresponding to favorable viewing angles through the asymmetric ejecta.

Finally, we impose data quality cuts and exclude observations where the majority of the observing time is dominated by high background levels, or where there are obvious instrumental artifacts. Furthermore, we exclude observations where the SN position is near a bright X-ray source such as an active galactic nucleus. This is done by visual inspection of the generated images (Section~\ref{sec:data_reduction}). This last cut to our SN sample leads to a final SN sample of 242 unique SN (Table~\ref{tab:sample_cuts}).

A majority of the SNe in our sample are observed by more than one telescope and often multiple times by the same telescope. This provides an opportunity to study the time evolution and to cross-check results from different telescopes. In particular, we make use of the superior angular resolution of Chandra to check for contamination by nearby sources. If the source regions in XMM, Swift or NuSTAR contain an unrelated source in the Chandra image, the observation is excluded from the sample. The resulting total number of X-ray observations from the different telescopes is summarized in Table~\ref{tab:sample_selection}.

Chandra and XMM were both launched in 1999, but XMM has a larger FOV and is therefore expected to serendipitously observe more SNe. However, as seen in Table~\ref{tab:sample_selection}, Chandra has more observations in our sample. This is partly due to our removal of XMM observations when there are nearby contaminating sources seen in Chandra, but it is also likely that the better angular resolution of Chandra has resulted in more targeted observations of SNe and nearby galaxies.
For Swift and NuSTAR, we have much fewer observations. This is because the previously mentioned reasons, but also that they have been employed for a shorter time (launched in 2004 and 2012, respectively).

The final sample is illustrated in Figure~\ref{fig:hist_1}, where we only plot the latest observation for each SN. The latest epochs reach approximately 100~yrs and correspond to the most nearby SNe. The aforementioned epoch cuts are also evident in Figure~\ref{fig:hist_1}, where more stripped type SNe are seen at earlier epochs and more type II at later epochs. However, the total number of SNe of each type are similar owing to the different rates of occurrence in combination with the epoch cuts. We also stress that the older SNe in the sample, with epochs $\gtrsim60$~yrs, have more uncertain classification \citep{1941Minkowski}.

\section{Data Reduction and Source Detection} \label{sec:data_reduction}

We perform data reduction of all observations following mostly standard procedures. Below we first discuss some common steps before describing the detailed procedure for each telescope. 

Different image detection algorithms are used to identify sources detected at 3$\sigma$ significance in Chandra, XMM, and Swift. No equivalent tool is available for NuSTAR, so in case we use spectral analysis as described in Section~\ref{sec:Analysis_of_detected_sources,nustar}.

Since most of the SNe in the sample are undetected or faint in X-rays, we limit the size of the source regions to obtain constraining upper limits and to maximize the signal-to-noise ratio in the case of detections. At the same time, we must account for astrometric uncertainties from the X-ray telescopes and potential positional errors in the optical coordinates, particularly for older SNe from the OSNC. To balance these considerations, we define circular source regions (or squares for Swift) with radii ranging from $\sim5\arcsec$ to 30$\arcsec$, depending on the telescope resolution and nearby emission. We use smaller source regions when the SN is positioned on prominent galaxies or in proximity to strong nearby sources. Background regions are selected as the largest-allowed region without other point sources or diffuse emission, situated as close to the SN source region as possible, always on the same CCD chip. The background regions are circular, typically with radii $\gtrsim$ 2--3 times the source regions.

Spectra of detected sources are subsequently generated from these source and background regions. The regions are instead used to obtain upper limits from the images when no point source is detected (Section~\ref{sec:Upper_limits_for_undetected_sources}). 
For a small number of SNe that are detected in galaxies with strong diffuse X-ray emission, we also extract a spectrum of the galaxy itself, using a similar region size as for the SN. We analyze this spectrum to more precisely account for the host galaxy contribution to the SN spectrum (Section~\ref{sec:Analysis_of_detected_sources}). 
All spectra are grouped with one count per spectral bin, allowing for use of Poisson statistics with the C-stat statistic in XSPEC. The grouping is performed during spectral extraction using \texttt{specgroup} for XMM data, \texttt{specextract} for Chandra, and \texttt{grpmincounts} for NuSTAR, which are part of the software packages for each telescope described below.

\subsection{Chandra}\label{sec:chandra}
Chandra observations taken by the ACIS-S and ACIS-I instruments are retrieved from the Chandra Source Catalog (ver. 1.3.10, CSC 2.0) using the CSCview application \citep{2011VanStone}. The data reduction uses CIAO (ver. 4.13) \citep{2006SCIAO} with CALDB 4.9.4. Event files and light curves are created using the CIAO tools \texttt{chandra\_repro} and \texttt{dmextract} and corrected for flares using the \texttt{deflare} CIAO tool. Simultaneously, high-resolution images (bin=1) are produced across five energy ranges: 0.5--8 (full band), 2--8 (hard band), 0.5--2, 2--6, and 6--8~keV. The full-band image serves to demarcate the source, background, and galaxy regions. The images in the different energy intervals are useful for placing upper limits on SN and CO luminosities (Section~\ref{sec:Upper_limits_for_undetected_sources}).

We performed source detection on the images in the 0.5--8~keV energy range using the CIAO tool \texttt{wavdetect}. This tool correlates the image with wavelets of varying scales in search for significant correlations. Since we are looking for point sources at specific locations, we reduce the runtime by limiting the search to an image 10 times the size of the source region (described below). Similarly, we adjust the wavelet scales to 1.0, 2.0 and 3.0, more suitable for point sources and our image sizes. A significance threshold of \texttt{sigthresh}=$10^{-5}$ is used, otherwise we follow standard procedures. The output includes a source list with positions and detection significances. We require a $3\sigma$ significance and that the angular separation between the detected source and the SN position is sufficiently small to indicate a correlation. This separation is typically set to 0.6\arcsec. Finally, a visual inspection is also done to confirm that the identified source is indeed a point source.

Spectra of detected sources were extracted using the \texttt{specextract} tool following standard procedures with the exception of the two parameters \texttt{weight} and \texttt{correctpsf}, which were changed from default settings. The auxiliary response file (arf) is not weighted, as is appropriate for point source analysis, and a point-source aperture correction is applied to the unweighted arf. 

Typically, the source regions are circular with a radius of $\sim$5\arcsec. However, sources positioned further off-axis require larger regions due to the significant dependence of the Point Spread Function (PSF) on the off-axis angle. We adopt the CIAO tool \texttt{psfsize\_srcs} to compute the region size that encompasses a PSF fraction of 90\% for the broad band. To account for uncertainties in the optical position of the SNe, we refrain from using excessively small source regions. Therefore, we adopt the source radius derived by the \texttt{psfsize\_srcs} tool only if the radius is $\ge5$\arcsec, which corresponds to off-axis angles $\ge6$\arcmin. The \texttt{psfsize\_srcs} tool does not account for the fact that the PSF becomes increasingly elliptical with the off-axis angle, but this likely does not have a major impact on the final outcome due to the large source sizes and the faint nature of our sources.

\subsection{XMM}\label{sec:XMM}
The XMM EPIC-pn observational data files are downloaded from the XMM Science Archive and processed with the calibration files (CCF, from 2021). The data are reduced following the recommended data reduction guidelines for Science Analysis Software (SAS, ver. 19.1.0) and HEASOFT (ver. 6.30.1) \citep{2014Nasa}. Similarly as for Chandra, high-resolution images (bin=1) are produced across five energy ranges: 0.5--10 (full band), 2--10 (hard band), 0.5--2, 2--7, and 7--10~keV. Analogous to Chandra, the full-band image serves to demarcate the source, background, and galaxy regions.

We used the SAS tool \texttt{ewavelet} to perform source detection on images in the 0.5--10~keV energy range. Although \texttt{edetect\_chain} is commonly used for source detection in XMM, we opt for \texttt{ewavelet} to maintain methodological consistency between telescopes, as \texttt{ewavelet} uses a similar method as \texttt{wavdetect}. This tool also employs wavelet correlations with images, utilizing wavelet scales ranging from one to ten. Unlike \texttt{wavdetect}, \texttt{ewavelet} does not provide a detection significance in its output. Therefore, we set a $3\sigma$ detection threshold as an input parameter. As for the Chandra analysis, a small separation is required to identify a detected source with a SN, though a slightly larger offset is accepted for XMM due to the lower angular resolution and larger astrometric uncertainties. This separation is typically set to 3\arcsec.

The source and background spectra of detected sources are created from filtered event files, where time intervals affected by background flaring activity have been removed with a \texttt{mode} value around 0.4 counts/s, but adjusted for each individual observation. This adjustment is based on examination of the light curve generated within 10--12~keV. Aperture correction and weighting redistribution matrix (rmf) settings are configured for point sources by default, differing from Chandra data reduction.

The source region radii are determined using the SAS task \texttt{eregionanalyse}, which computes the optimum radius of a circular region to maximize the signal-to-noise ratio based on source counts, the PSF fraction and background region. The majority of optimal radii are set at 10\arcsec, with few sources requiring larger radii. Unlike Chandra, the optimum radius for XMM is not correlated with the off-axis angle but is based on signal-to-noise considerations. For SNe located on X-ray bright galaxies, we manually select regions, which are sometimes smaller than the derived optimal radius.

\subsection{Swift}\label{sec:swift}

\textit{Swift}-XRT data products are generated from Swift archival data using the \texttt{xrt\_prods} module (v1.10) from the \texttt{swifttools} (v3.0, 2022 August 31) Python package \citep{2009Evans}. Images for each SN are generated from the most recent observation and grouped with all previous observations from within one year. This grouping is useful as most XRT observations have short exposure times and low count rates. Swift-XRT images are created in the same five energy intervals as for XMM as well as a full image in 0.3--10~keV used in the source detection. 

Source detection in Swift images is done using the \texttt{xrt\_products} module \citep{2020Evansswift}. This provides a list of sources found in the image of the full 0.3-10~keV energy band. To identify a detected source with a SN, we require a maximum separation of 5\arcsec\ between the source and SN position and a detection flag "GOOD", corresponding to a spurious detection of $\le0.3~\%$,  equivalent to $\ge3\sigma$ significance. We find no significant Swift detections and, therefore, do not create any spectra. All Swift results are therefore derived from image limits as described in Section~\ref{sec:Upper_limits_for_undetected_sources}.

\subsection{NuSTAR}\label{sec:nustar}
NuSTAR FPMA and FPMB data are taken from the NuSTAR Master Catalog \citep{2013Harrison} and reduced following the recommended pipeline using NuSTARDAS (2.1.2) and CALDB 4.9.4. The pipeline process includes applying livetime and vignetting corrections. Images from both FPMA and FPMB are simultaneously created over the full 3--78~keV energy range. These are used to select the source- and background regions. The source spectra from both detectors are generated from regions of typical size 30\arcsec\ as recommended for faint sources by the NuSTAR observatory guide\footnote{\url{https://heasarc.gsfc.nasa.gov/docs/nustar/NuSTAR_observatory_guide-v1.0.pdf}}, while the background regions have at least twice this radius.

\section{Methods}\label{sec:method} 
The goal of this study is to determine X-ray luminosities of the SNe and COs in our sample, or establish upper limits in cases of non-detections. The CO luminosities are provided under various degrees of ejecta absorption, accounting for differences expected due to asymmetries in the explosions (see Section~\ref{sec:ejecta_models}). We base the luminosity calculations on distances from the OSNC, with the exception of SN~2017eaw, for which we found a significant deviation from a recent distance estimate \citep{2019VanDyk}. While the OSNC distances do carry uncertainties, we find no systematic biases among them and do not expect a substantial impact on the results for the sample as a whole. We also report all corresponding flux values for reference.

The data analysis is conducted using XSPEC version 12.12.1 \citep{1996Arnaud}, including spectral fitting as well as model scaling based on image limits. Due to typically low-count spectra, we employ simple models and assume Poisson statistics to utilize cstat in XSPEC. All luminosities and fit parameters are reported with 90\% confidence intervals, corresponding to $\Delta$cstat 2.706, while upper limits  are given for 3$\sigma$ (or 99.7\%) uncertainty, corresponding to $\Delta$cstat 7.74. 

Below, we first describe the models used for the analysis of detected sources and NuSTAR data in Section~\ref{sec:models}. We then describe the spectral analysis and the methods for placing upper limits on undetected sources in Sections~\ref{sec:Analysis_of_detected_sources} and \ref{sec:Upper_limits_for_undetected_sources}, respectively.

\subsection{Models}\label{sec:models}

\subsubsection{Galactic Absorption (\texttt{tbabs})}
\label{sec:nH}
The absorption from the Milky Way is modeled by the XSPEC component \texttt{tbabs} \citep{2000Wilms}. The abundances are set according to \cite{2000Wilms} and the column density of hydrogen for each SN is taken from the NHtot tool\footnote{\url{https://www.Swift.ac.uk/analysis/nhtot/}} \citep{2013Willingale}. We do not model absorption by the SN host galaxy in this study as it is usually unknown and would add further uncertainty by introducing an additional free parameter. All fluxes and luminosities presented in this work have been corrected for Galactic absorption.

\subsubsection{CSM Interaction (\texttt{mekal})} \label{sec:CSM_interaction}
The X-ray emission of a SN partly arises from interaction between the ejecta and the CSM formed by the previous stellar wind \citep{2017Chevalier, 2022Margalit}. As the expanding ejecta run into the near-stationary CSM, a forward and a reverse shock are created. The shocks heat the gas as they propagate, resulting in thermal X-ray emission. Different progenitor stars have different wind properties and may also have experienced episodes of eruptive mass loss, resulting in a wide range of CSM properties. Consequently, the duration of the interaction and the properties of the resulting X-ray emission can vary significantly, though the basic picture is that the temperature and X-ray luminosity decrease over time (e.g., \citealt{2012Dwarkadas}). 

For the SN types and epochs considered in our analysis, any remaining detectable X-ray emission from CSM interaction is likely to peak in the soft X-ray regime (e.g., \citealt{2020Ramakrishnan}). To approximate this emission we use the XSPEC \texttt{mekal} model \citep{1985Mewe,1986Mewe,kaastra1992,1995Liedahl}, which models the emission from a thermal plasma in ionization equilibrium. A more realistic model would consider non-equilibrium ionization and a range of different temperatures, but such a model cannot be constrained with our low count-rate spectra. The free parameters of the \texttt{mekal} component are the temperature ($kT$) and the normalization.

\subsubsection{Host Galaxy (\texttt{mekal})}
\label{sec:mekal_galax_comp}
Some of the SNe in the sample are located in X-ray- bright galaxies, implying a more complex background. For detected SNe with such background, we introduce an additional plasma component (\texttt{mekal}) to characterize the X-ray emission from the host galaxy. Diffuse X-ray emission in all types of galaxies is usually well described by a thermal plasma with $kT$ in the range 0.2--1~keV \citep{2005Grimes,2009Owen,2011Boroson}, motivating the use of \texttt{mekal}. 

\subsubsection{Pulsar and Pulsar Wind Nebulae (\texttt{pow})} \label{sec:pow_comp}
Pulsars and PWNe produce X-ray radiation due to synchrotron emission. The NS surface also emits thermal blackbody emission, but this is weak and not expected to be detected for the SNe in our sample.  We therefore focus on the non-thermal synchrotron emission. The X-ray spectra of both pulsars and PWNe are well described by power laws with power-law index ($\Gamma$) in the range $\sim$ 0.9--2.5, with more luminous systems displaying softer spectra \citep{2008Li}. Several studies have shown that PWNe are typically brighter than the pulsars \citep{2008Li,2021Hsiang,2008Kargaltsev}, though with large variations in the PWN/pulsar luminosity ratios (e.g., \citealt{2021Hsiang} find values in the range 0.06--73). It is thus likely that the PWNe would dominate over the pulsars also in our sample, with the significant caveat that the studies above are based on much older and less luminous systems. We also assume that bright young PWNe will have $\Gamma \sim 2$, similar to the Crab \citep{1981Pravdo,2000Zhang}.

\subsubsection{Ejecta Absorption (\texttt{tbvarabs})}\label{sec:ejecta_models}

We model the X-ray absorption from the ejecta using the results of \cite{2018Alp}, which are based on 3D simulations of neutrino-driven explosions \citep{2015Wongwathanarat,2017Wongwathanarat} for various progenitor star models. We adopt the blue supergiant model (B15) for type II Pec and the red supergiant model (W15) for type II, while the W15-IIb model is used as a starting point to create appropriate models for different types of stripped SNe. The models are summarized in Table~\ref{tab:abs_models}. We directly use the relative elemental abundances and asymmetries as provided by these models, but scale the total $\tau$ to better match typical parameters of different SN types. Specifically, we use the large samples studied in \cite{2021Martinez} and \cite{2018Taddia} to obtain typical ejecta masses and explosion energies for type II and stripped types, respectively.  The scale factor for $\tau$, $f_{\rm scale}$, is given by

\begin{equation}
\label{eqn:abs_scale}
f_{\rm scale} = 
\frac{M_{\rm ej, SN} }{M_{ \rm ej,mod}}
\left( \frac{v_{\rm ej,SN}}{v_{\rm ej, mod}} \right)^{-2},
\end{equation}

where $v_{\rm ej,mod/SN}$ is the ejecta velocity and $M_{\rm ej,mod/SN}$ is the ejecta mass, with subscripts “mod" and “SN" denoting values from the original models and average values inferred from observations, respectively. The ejecta velocity is given by

\begin{equation} \label{eqn:v_ej}
v_{\rm ej, mod/SN}=\sqrt{2E_{\rm ej, mod/SN}/ M_{ \rm ej, mod/SN}},
\end{equation}

where $E_{\rm ej, mod/SN}$ are the explosion energies of the models and those inferred from observations.

For stripped SNe, we remove the hydrogen and helium masses from the total ejecta mass before scaling according to eq.~\ref{eqn:abs_scale}, and the values shown for $M_{\rm ej, mod}$ in Table~\ref{tab:abs_models} reflect this adjustment. However, for ejecta velocity calculations (eq.~\ref{eqn:v_ej}), we use the original model mass of 3.7~$M_\Sun$ from \cite{2017Wongwathanarat}, which better reflects the explosion dynamics. The final scale factors and parameters for each SN type are listed in Table~\ref{tab:abs_models}. 

For type II SNe, the scale factor is close to unity ($f_{\rm scale}\approx1$) as the differences between the model ejecta mass and energy compared to typical observed values balance out. The type II Pec model is not rescaled ($f_{\rm scale} = 1$) since this model is already tuned to SN~1987A and its progenitor \citep{2018Alp,2019Alp}. The stripped models have higher $f_{\rm scale}$ in the range $\sim$1.3–1.5, primarily due to higher $M_{\rm ej, SN}$, compared to  $M_{\rm ej, mod}$. For stripped SNe with uncertain classification (e.g. Ib/c), we apply the scaling for type Ib. We note, however, that the difference in the absorption between Ib and Ic has little effect on the spectral fitting, as indicated by the $f_{\rm scale}$ values in Table~\ref{tab:abs_models}. This is because the opacity contribution of helium is small, especially above 2~keV. Finally, we scale the $\tau$ to each observation in the sample with the time since the explosion in units of days $\tau \propto (t/10,000)^{-2}$. The ejecta absorption is implemented in XSPEC using the model \texttt{tbvarabs}, where we set specific elemental abundances in the ejecta composition, such as zero hydrogen and/or helium for stripped SNe (see \citealt{2018Alp} for details).

A major benefit of using 3D models for the absorption is that we can investigate how different viewing angles through the asymmetric ejecta affect the level of absorption. We therefore consider the distributions of $\tau$ over all directions for these models, as presented in \cite{2018Alp}. To probe a range of scenarios, we perform the scaling described above and all subsequent analysis for $\tau$ corresponding to the 10th, 50th and 90th percentiles of the full distributions, which represent lines of sight with low, typical, and high absorption, respectively. 

These absorption models assume a neutral medium, which is reasonable for ejecta at these stages and supported by detailed modeling of SN 1987A, which shows that the ejecta are mainly neutral, with some contribution from low-ionization ions in singly ionized states \citep{Jerkstrand2011}. The most important exception to this is the scenario where a highly energetic PWN ionizes the ejecta, making it transparent to X-rays \citep{2014Metzger}.  We account for this possibility by also performing the analysis without any ejecta absorption. This unabsorbed case can alternatively be interpreted to represent the scenario of a highly asymmetric explosion with a an almost clean line of sight to the observer.

\begin{deluxetable*}{l l l l l l l}[ht] 
\tablenum{3}
\tablecaption{Parameters for Ejecta Absorption\label{tab:abs_models}}
\tablewidth{0pt}
\tablehead{ 
\colhead{} & \colhead{} & \colhead{Type {II}} & \colhead{Type {II} Pec} & \colhead{Type {II}b} & \colhead{Type {I}b} & \colhead{Type {I}c}\\
 \colhead{Parameter} & \colhead{Unit} & \colhead{Model values} & \colhead{Model values} & \colhead{Model values} & \colhead{Model values} & \colhead{Model values}
 }
\startdata
 Progenitor model & & W15& B15 &W15-IIb &W15-IIb &W15-IIb\\
    M$_{\rm ej,mod}$\tablenotemark{a} & $M_\Sun$ & 14.0 & 14.2 & 3.7 & 3.2 & 1.1\\
    E$_{\rm ej,mod}$\tablenotemark{a} & $\times 10^{51}~\rm erg$& 1.45 & 1.43 & 1.52 & 1.52 & 1.52\\
    v$_{\rm ej,mod}$ &  km~s$^{-1}$& 2281 & 2250 & 4545 & 4545 & 4545 \\
    \hline
    M$_{\rm ej, SN}$ & $M_\Sun$ & 9.2\tablenotemark{b} &14.2& 4.3\tablenotemark{c} & 3.8\tablenotemark{c} & 2.1\tablenotemark{c}\\
    E$_{\rm ej, SN}$ & $\times 10^{51}~\rm erg$ & 0.63\tablenotemark{b} & 1.43 & 1.3\tablenotemark{c}& 1.4\tablenotemark{c} & 1.2\tablenotemark{c}\\
    v$_{\rm ej, SN}$ &$\rm km~s^{-1}$& 1855 & 2250 & 3899 & 4304 & 5360\\
    $f_{\rm scale}$ & & 0.99 & 1 & 1.58 & 1.32 & 1.37    
\enddata
\tablecomments{\footnotesize{The upper half of the table lists the parameters used for the absorption models in \cite{2018Alp}. They are: the name of the progenitor model, the ejecta mass (M$_{\rm ej,mod}$), the explosion energy (E$_{\rm ej,mod}$) and the ejecta velocity (v$_{\rm ej,mod}$). The parameters in the last two columns were obtained by modifying the W15-IIb model as described in the text. The typical observed values adopted for different classes of SNe are given in the lower part of the table below the horizontal line. They are: the ejecta mass (M$_{\rm ej, SN}$), ejecta energy (E$_{\rm ej, SN}$) and ejecta velocity (v$_{\rm ej, SN}$). Finally, the scale factor ($f_{\rm scale}$) used to scale the optical depth from the original models is presented.
\tablenotetext{a}{\cite{2018Alp,2019Alp} and references therein.}
\tablenotetext{b}{\cite{2021Martinez}}
\tablenotetext{c}{\cite{2018Taddia}} }}
\end{deluxetable*}

\subsection{Spectral Analysis}\label{sec:Analysis_of_detected_sources}

\subsubsection{Chandra and XMM}
\label{sec:Analysis_of_detected_sources,lowe}
We fit the  Chandra and XMM spectra of all sources identified by the image detection algorithms described in Sections~\ref{sec:chandra}~\&~\ref{sec:XMM}. All spectral fitting is conducted over the 0.5--8~keV energy range for Chandra data and over the 0.5--10~keV range for XMM data. To facilitate comparison between Chandra and XMM results, we extrapolate the model fits of the Chandra observations to obtain fluxes and luminosities up to 10~keV. Fluxes are retrieved using \texttt{cflux}, and all models include a \texttt{tbabs} component with absorption fixed at the Galactic value (Section~\ref{sec:nH}). As a null hypothesis we assume that the X-ray emission is dominated by CSM interaction. We therefore start by fitting the spectra with a \texttt{tbabs*mekal} model, where $kT$ is free to vary between 0.1--10~keV. These fits confirm that all the sources are detected at $>3\sigma$ significance. 

In a small number of cases (5 observations of 4 SNe), we include an additional \texttt{mekal} component in the fits to account for bright X-ray emission from the host galaxy (Section~\ref{sec:mekal_galax_comp}). The temperature of this component is first found in a separate \texttt{tbabs*mekal} fit to a spectrum of the galaxy itself, where $kT$ is allowed to vary between 0.1--2~keV \citep{2005Grimes,2009Owen,2011Boroson}. In these fits we use the same background spectrum extracted away from the host galaxy as used in the analysis of the SN spectrum.  When subsequently fitting the SN spectrum, the normalization of the galaxy \texttt{mekal} is free, while the temperature is fixed to the value found in the fit to the host galaxy spectrum. This method allows us to more accurately account for the host galaxy contribution than simply subtracting it as a background. We find no systematic differences between the small number of SNe analyzed in this way and the rest of the sample. 

For SNe with multiple observations obtained within one year by the same telescope, we improve statistics by fitting the spectra simultaneously with $kT$ of the CSM \texttt{mekal} component tied, while allowing the individual normalizations to vary. The temperature is not expected to vary significantly on this time scale, and we verified that we obtained consistent results when fitting the spectra individually. In these fits we also include observations that are below the $3\sigma$ detection criterion.

The simple \texttt{mekal} model provides a good fit to the majority of the spectra and we use the best-fit models to derive the SN luminosities ($L_{\rm SN}$). To account for possible emission from a CO, we then add a power-law component to the model, modified by different degrees of ejecta absorption. We let $\Gamma$ of the power law vary between 0.5--3, but fix it at 2 when it could not be constrained. The resulting models are \texttt{tbabs(mekal + pow)} for the scenario without significant ejecta absorption and \texttt{tbabs(mekal + tbvarabs*pow)} for ejecta absorption at the 10th, 50th and 90th percentiles of the asymmetric explosion models (Section~\ref{sec:ejecta_models}). The $\Gamma$ of the power law was tied for observations obtained close in time, as for $kT$ of the \texttt{mekal} component. The resulting luminosities or upper limits on the compact objects are denoted by $L_{\rm CO, 0,abs}$, $L_{\rm CO,10,abs}$, $L_{\rm CO,50,abs}$ and $L_{\rm CO,90,abs}$. 

We performed Monte Carlo simulations to assess the statistical significance of the added power law in all cases where it improved the fit statistic and where the parameters ($\Gamma$ and  normalization) could be constrained. The simulations were carried out using XSPEC's \texttt{fakeit} tool. We created 10,000 fake spectra per observation based on the best-fit \texttt{tbabs*mekal} model. These 10,000 spectra were then fitted with both \texttt{tbabs*mekal} and \texttt{tbabs(mekal+pow)} and the difference in fit statistic ($\Delta$cstat) was recorded. This difference represents the improvement in fit statistic when adding an additional component to a spectrum generated by the null hypothesis spectrum. The resulting $\Delta$cstat distribution was used to determine the value of $\Delta$cstat for which the null hypothesis (\texttt{tbabs*mekal}) could be rejected at 3$\sigma$ confidence, implying that the power law significantly improves the fit. These tests were done on the individual observations where $kT$ and $\Gamma$ were not tied between observations.

In the majority of cases where the power law was not significant, we instead obtain upper limits on its luminosity from the fits. In these fits we fix $\Gamma$ at 2 when it could not be constrained, which is a typical spectral index expected for PWNe \citep{2021Hsiang,2008Li,2000Chevalier}. This was the case for $\sim 85$\% of spectra in the scenario without ejecta absorption and for all spectra when ejecta absorption was included. 

\subsubsection{NuSTAR}
\label{sec:Analysis_of_detected_sources,nustar}

NuSTAR spectra were fitted with a power law over the energy range 3--78~keV. We used a power law instead of a \texttt{mekal} component as our baseline model in this hard energy band because these models are similar when $kT$ is high. We also verified that the fits did not improve by replacing the power law with \texttt{mekal}. Observations obtained within one year were fitted simultaneously with $\Gamma$ tied, as for the fits in Section~\ref{sec:Analysis_of_detected_sources,lowe}. The photon index was fixed at  $\Gamma=2$ in  cases when it could not be constrained. This analysis indicated a 3$\sigma$ detection for some sources, but inspection of the images did not reveal a point source in any of these cases. We attribute this to statistical fluctuations in the spectra of the source and background regions, which will sometimes cause spurious detections, and treat all luminosities derived from such NuSTAR detections as upper limits. The limits on the power law without ejecta absorption from these fits correspond to  $L_{\rm SN} = L_{\rm CO, 0, abs}$. We also obtained luminosity limits on compact objects modified by different levels of ejecta absorption by introducing a \texttt{tbvarabs} model as in Section~\ref{sec:Analysis_of_detected_sources,lowe}.

\subsection{Upper Limits for Undetected Sources}
\label{sec:Upper_limits_for_undetected_sources}

For the undetected SNe, we derive upper limits on the SN and CO luminosities based on count rates in images in the five different energy bands mentioned in Section~\ref{sec:data_reduction}. The full-and hard bands are used to estimate SN luminosities, while the narrow ones are used to estimate the CO luminosities. For Chandra observations, the former ones are 0.5/2--8~keV and the narrow ones are 0.5--2, 2--6, and 6--8~keV. Similarly we have 0.5/2--10~keV, and 0.5--2, 2--7, 7--10~keV for XMM and Swift.

Count rates from Chandra data are found using the CIAO tool \texttt{srcflux}, which utilizes Bayesian statistics to derive count rates within a specified region. We provide the SN source region and background regions as identified in Section~\ref{sec:data_reduction}. A credible interval of \texttt{conf=0.997} is set to obtain the upper bound of the interval corresponding to $3\sigma$. 

For XMM analysis, we use the same tool as used for finding the optimal source region sizes, \texttt{eregionanalyse}. Similar to the Chandra analysis, \texttt{eregionanalyse} derives count rates given a source and background region. The regions were selected as described in Section ~\ref{sec:data_reduction}. The significance level of the upper limit is set to \texttt{ulsig=0.997} and the output upper count rate is the Bayesian upper limit if the source region has $<80$ counts \citep{1991Kraft} or derived according to the equation of the statistical upper limit provided by HEADAS \footnote{\url{https://xmm-tools.cosmos.esa.int/external/sas/current/doc/eregionanalyse/node21.html}} for more counts. The majority of our sources have less than 80 counts.

Swift count rates are manually derived based on the number of counts and the corresponding exposure time. The source region is defined as a 20\arcsec\ × 20\arcsec\ box centered on the SN coordinates. The background is estimated from a larger, square box (60\arcsec\ × 60\arcsec) surrounding the source region, excluding the source box itself. The 3$\sigma$ upper limits on the net count rates are calculated using Poisson statistics, following a Bayesian approach implemented via the \texttt{astropy.statistics} Python module, consistent with the method applied to the XMM data.

The upper limits on the luminosity are derived using a \texttt{tbabs*pow} model with $\Gamma=2.0$. This is done using XSPEC, where we use the response and exposure time of each observation, and scale the power-law normalization so that the count rate from the model matches the count rate limit derived from the images. A power law is used since we assume that the emission at these late times would be dominated by a CO.

To derive limits on COs affected by ejecta absorption, we use a \texttt{tbabs*tbvarabs*pow} model for three different levels of ejecta absorption, as described in Sections ~\ref{sec:ejecta_models} and \ref{sec:Analysis_of_detected_sources}. We scale the power-law normalization in XSPEC to find the limits, but now use the narrower energy intervals. This method provides more insight into the spectral distribution and becomes particularly important for more absorbed models, where the count rate in the highest energy interval (7--10~keV for XMM and SWIFT, and 6--8~keV for Chandra) becomes the most constraining one for a large fraction of the sample due to the hard predicted spectra. 

We cross-checked the limits derived in this way by also directly fitting spectra of undetected sources that had a sufficient number of counts. This showed a generally good agreement (typically within a factor $\sim 2$) between spectroscopic limits on $L_{\rm SN}$ and limits derived from the image count rates in the full and hard bands. Similarly for the CO luminosity limits, we find good agreement between the spectroscopically derived limits and those estimated from the image count rates in narrow energy intervals. However, the spectroscopically derived limits from these very low count-rate spectra show larger variations, including extreme outliers. We therefore base all our upper limits of undetected sources on the image count rates.

\section{Results}\label{sec:results}
We present SN luminosities and upper limits for 242 SNe from a total of 607 observations, alongside CO luminosities and limits assuming different degrees of ejecta absorption. We detected 12 SNe, one of which (SN~1979C) showed a significant improvement in fit statistic when a power-law component associated with the CO was included, provided negligible ejecta absorption. In the following section, we present all results regarding SN and CO luminosities and limits in more detail.

\subsection{Supernova Detections and Upper Limits}
\label{sec:result_obsflux}

\begin{figure*}[ht]
\hspace{0cm}\includegraphics[width=0.99\textwidth]{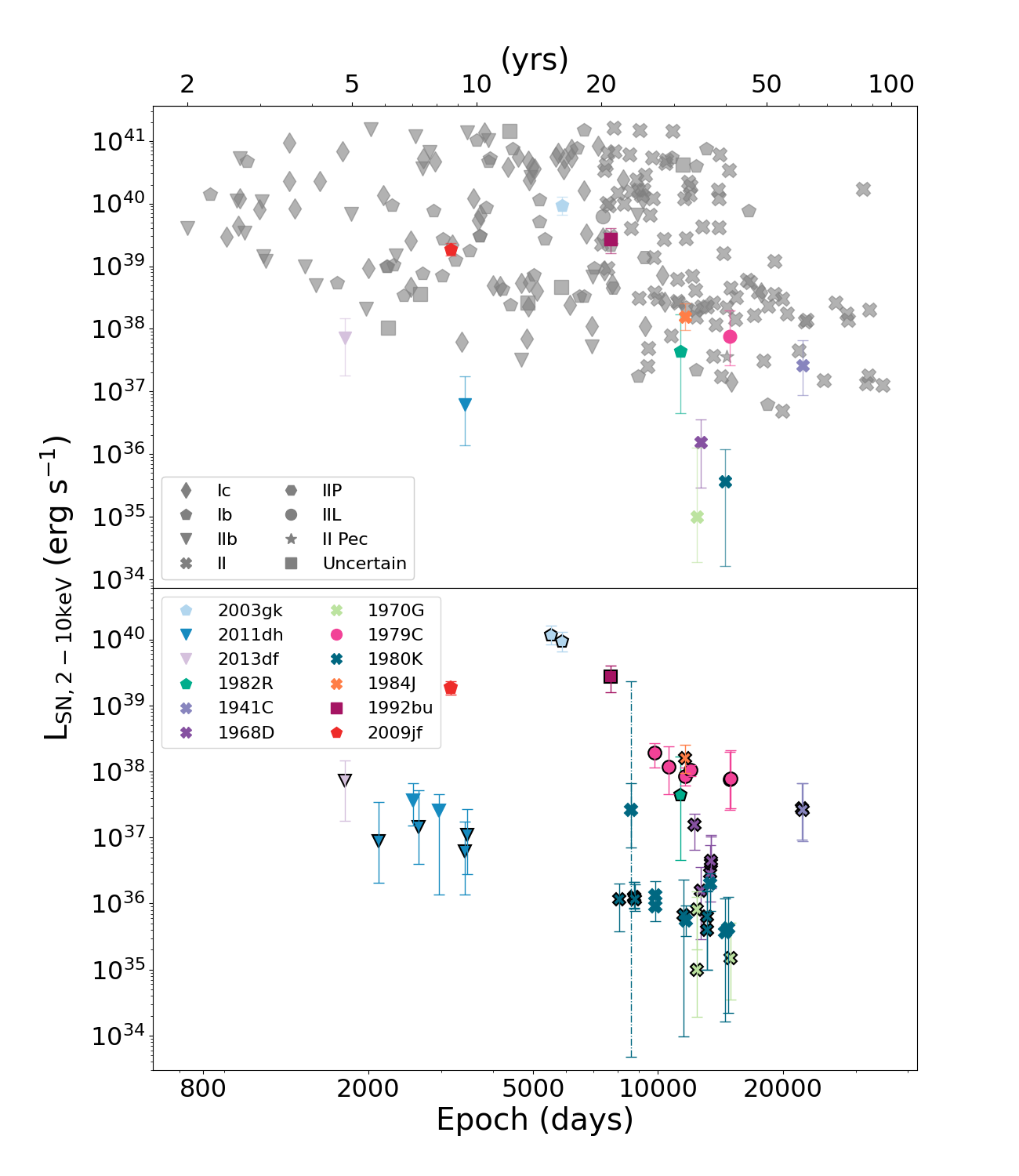} 
\caption{\footnotesize{SN luminosities and upper limits in the energy range 2--10~keV. The top panel displays the most constraining limit (grey) or the lowest detected luminosity (color) for each SN. The bottom panel presents all observations that resulted in detected SNe, with Chandra observations outlined in black. All detections are shown with $90\%$ error bars and the grey markers correspond to 3$\sigma$ upper limits. The marker symbols represent different types of SNe as described in the legend in the upper panel. For the outlying observation of SN~1980K we also show the $99.7~\%$ error bars with dashed lines.}} \label{fig:obsflux}
\end{figure*}

\begin{figure}[ht]
\includegraphics[width=\columnwidth]{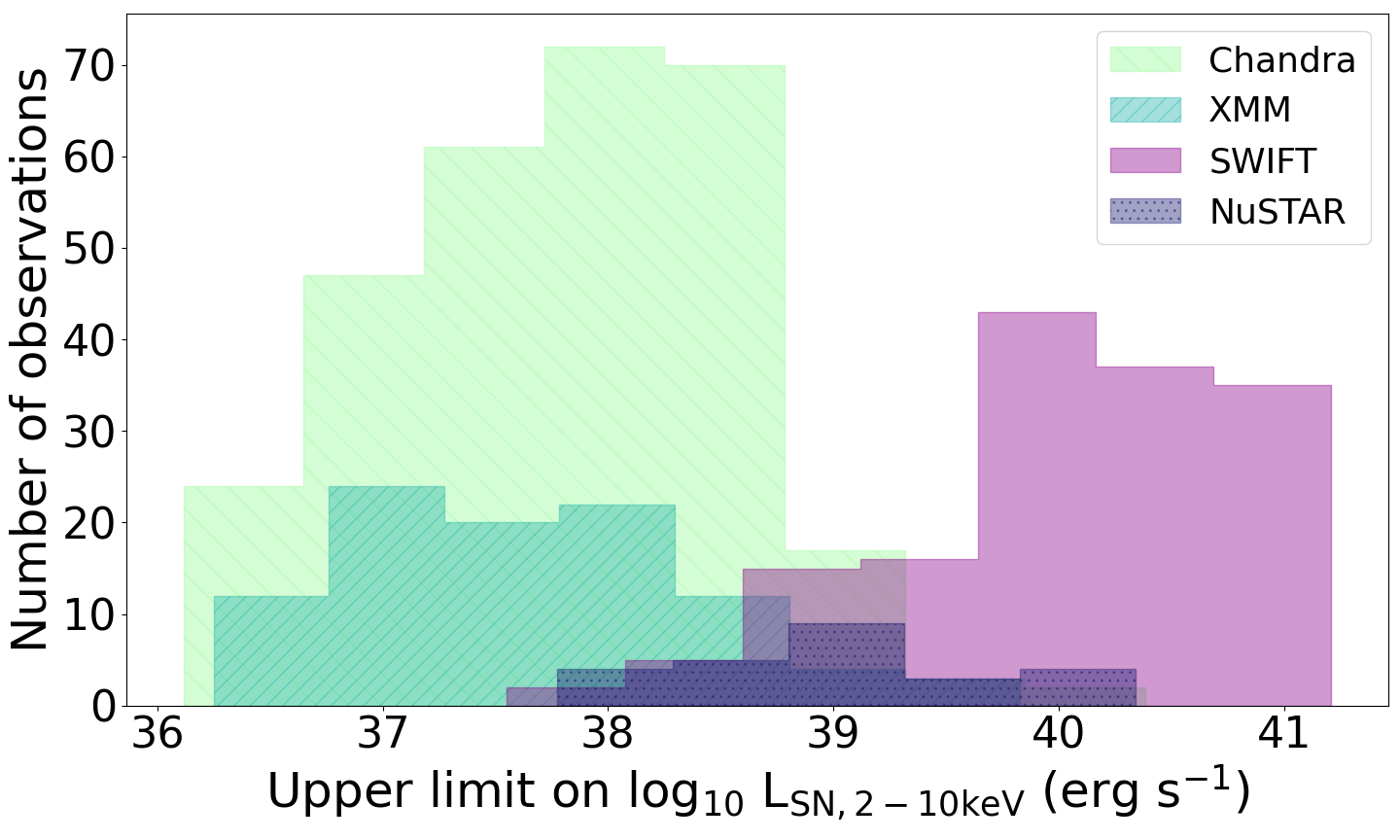} 
\caption{\footnotesize{A histogram of 3$\sigma$ upper limits of SN luminosities categorized into four groups based on the telescope used for the observations.} \label{fig:telescope_hist}}
\end{figure}

Figure~\ref{fig:obsflux} shows the resulting SNe luminosities and upper limits in the energy interval 2--10~keV as a function of time since explosion. Only the most constraining limit or lowest detected luminosity is shown for SNe with multiple observations in the top panel, while all detections are shown in the lower panel. The least constraining luminosity limits, typically $\gtrsim10^{39}\rm~erg~s^{-1}$, primarily originate from Swift observations, as shown in Figure~\ref{fig:telescope_hist}. This figure also demonstrates that the lowest limits ($\sim 10^{36}\rm ~erg~s^{-1}$) are provided by Chandra and XMM.

We find a SN detection frequency of approximately $\sim10\%$ in the combined sample of Chandra and XMM observations (see Table~\ref{tab:detected_ratio}). No detections are found with Swift or NuSTAR. Although the statistics are uncertain due to the small number of detections, Chandra appears to have a slightly higher detection rate than XMM. This is likely attributable to the lower background in Chandra, which makes it possible to detect fainter sources, since the distance distributions for the observed and detected SNe are similar for the two telescopes. Counting all SNe with multiple observations, there are 43 detections in total, 33 with Chandra and 10 with XMM. For Chandra, we detect 29 sources on the ACIS-S3 chip, two on ACIS-S2 and two on ACIS-I3 (see Table~\ref{tab:detected_SN} for further details). Among the 12 detected SNe, 9 are located within 20~Mpc, while the most distant detection is SN~2003gk at $\sim 49$~Mpc.  Table~\ref{tab:detected_ratio} also lists the detections for type II and stripped SNe separately. Half of the detected sample is type II SNe, but only stripped types are detected within 20~yrs post-explosion due to the epoch cuts discussed in Section~\ref{sec:sample_selection}.

\begin{deluxetable*}{l c c c  c c c}[ht] 
\tablenum{4}
\tablecaption{SN Detection Rates
 \label{tab:detected_ratio}}
\tablewidth{0pt}
\tablehead{ 
  & \multicolumn{3}{c}{\textless{}20~yrs} & \multicolumn{3}{c}{\textgreater{}20~yrs} \\
  & Chandra \& XMM   & Chandra  & XMM  & Chandra \& XMM   & Chandra   & XMM   \\
}
\startdata
    Stripped & 4(10\%) & 3(8\%) & 2(25\%) & 2(13\%) & 2(14\%) & 0(0\%) \\
    Type II  &   \dots      &  \dots      &   \dots      & 6(11\%) & 6(12\%) & 1(6\%) \\ 
    Total    & 4(10\%) & 3(8\%) & 2(25\%) & 8(11\%) & 8(12\%) & 1(5\%)\\ 
\enddata
\tablecomments{ \footnotesize{The detection rates of SNe divided into categories based on the following attributes: epochs—before and after 20~yrs; SN type—stripped, type II and combined; and telescopes—Chandra, XMM or both. The number of detected SNe in each category is presented alongside the corresponding detection ratio (\%) calculated specifically for that category.
}}
\end{deluxetable*}

The epoch cuts are also apparent in Figure~\ref{fig:obsflux}, which shows that the full sample ranges from $\sim2$~yrs all the way up to $\sim100$~yrs, with detected SNe between $\sim$ 5 and 60~yrs. The oldest detected SN in our sample is SN~1941C.
Figure~\ref{fig:obsflux} also shows that there is a wider range of luminosity limits at earlier epochs ($\leqslant 40$~yrs). This is due to the larger range of distances and SN types at those times, as clearly illustrated in Figure~\ref{fig:hist_1}. Additionally, most Swift observations are within this epoch range and further increase the upper luminosity bound of the limits (Figure~\ref{fig:telescope_hist}). Conversely, more constrained- and deeper luminosity limits are seen at later epochs due to the close proximity of these SNe.

\begin{figure}[ht]
\includegraphics[width=\columnwidth]{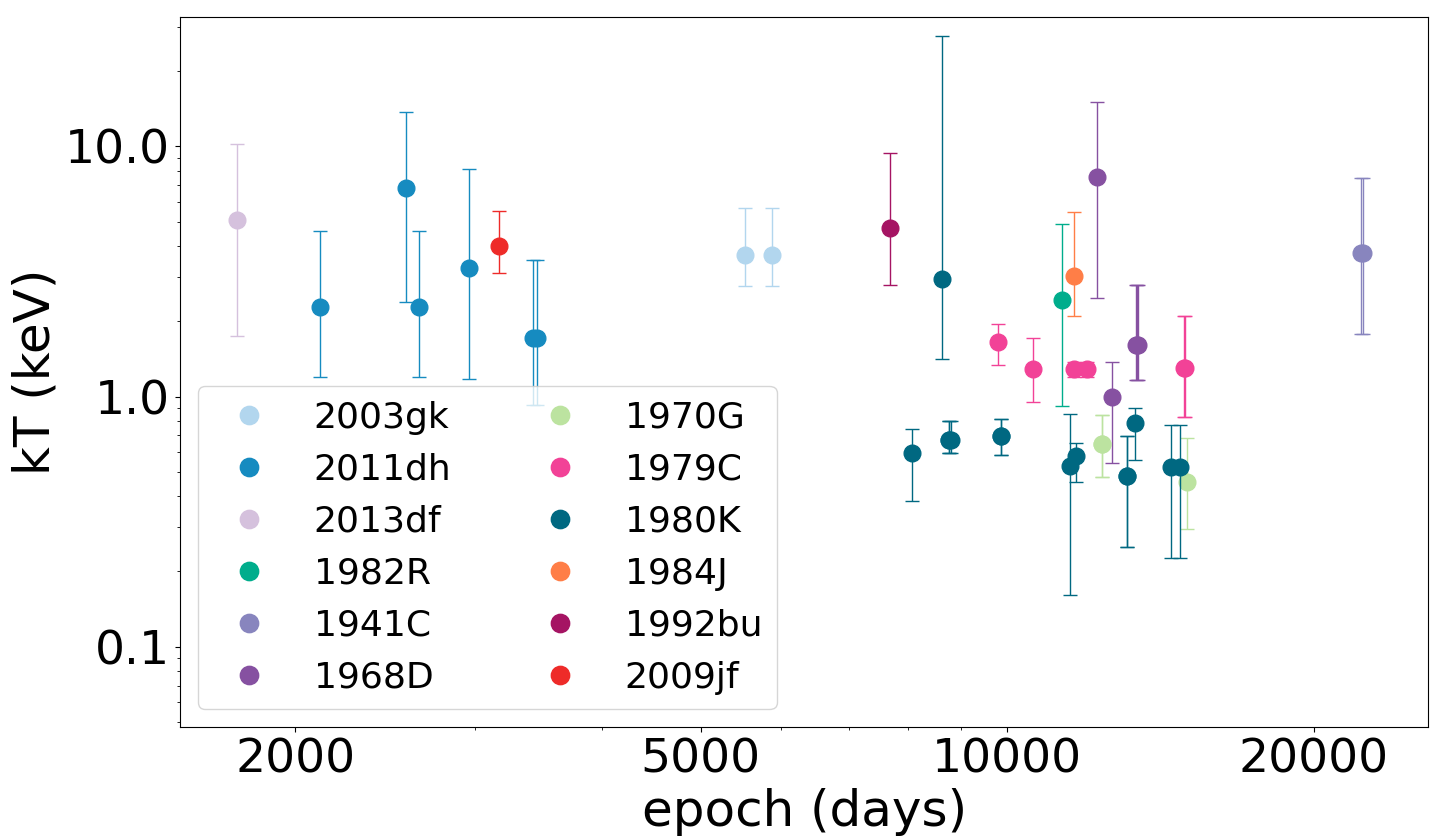} 
\caption{\footnotesize{Plasma temperature evolution for all SNe detected with Chandra and XMM, where unique SNe are indicated by color. The error bars show 90\% confidence intervals. }\label{fig:kT_evol} }
\end{figure}

All 43 observations of the 12 detected SNe are shown separately in the bottom panel of Figure~\ref{fig:obsflux}. Details of these results are presented in Table~\ref{tab:detected_SN}, while examples of spectral fits for each SN are shown in Appendix~\ref{sec:appendix_SN_fits}. The overall pattern in the bottom panel of Figure~\ref{fig:obsflux} is a decreasing SN luminosity with time, as expected. In line with the decreasing luminosities, the plasma temperatures also generally decrease with time, as seen in Figure~\ref{fig:kT_evol}. 

To our knowledge, four (SN~1982R, SN~1984J, SN~1992bu, and SN~2003gk) out of the 12 detected SNe in our sample have not previously been detected in X-rays, while four other (SN~1980K, SN~2009jf, SN~2011dh, and SN~2013df) have only previously been detected closer to the time of explosion and not at these late epochs \citep{2008Fridriksson,2008Soria,2011Pooley,2012Campana,2012Soderberg,2014Maeda,2016Heikkila,2016Kamble}. In particular, studies of SN~1980K \citep{2008Fridriksson,2008Soria} found soft X-ray emission in Chandra observations ranging from 8063~days to 8802~days with average luminosities of $3\times10^{37}~\rm erg~s^{-1}$ in the 0.3--7~keV energy range  \citep{2008Fridriksson} and $4\times10^{37}~\rm erg~s^{-1}$ over 0.3-8~keV \citep{2008Soria}. This is consistent with the 0.5--8~keV luminosities obtained from our analysis of the same Chandra data (see Table~\ref{tab:detected_SN} for 0.5--10~keV luminosities). 
We further detect this SN in seven XMM observations ranging from 8629~days to 14768~days, where the source shows a mildly decreasing luminosity, although the uncertainties are large (see Figure~\ref{fig:obsflux} and Table~\ref{tab:detected_SN} for further details).

\subsection{Compact Object Emission} \label{sec:result_CO}
\begin{figure*}[ht]
\includegraphics[width=\textwidth]{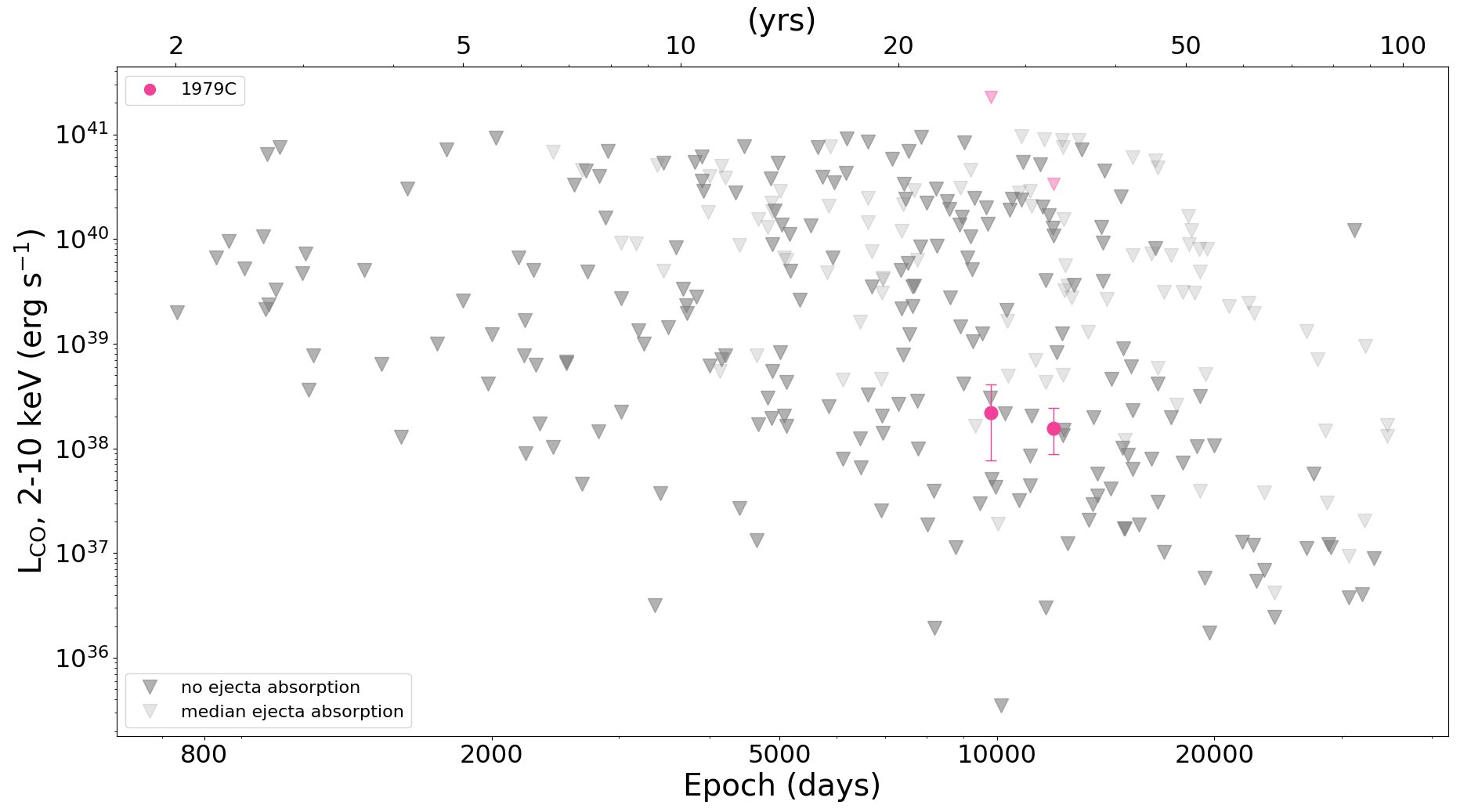} 
\caption{\footnotesize{CO luminosities and limits assuming a favorable viewing angle of negligible ejecta absorption (solid markers) and median absorption (semi-transparent markers) in the energy interval 2--10~keV. The colored markers show detected power-law components of SN~1979C assuming negligible ejecta absorption, while the corresponding 3$\sigma$ upper limits for median absorption are plotted with semi-transparent markers.}\label{fig:COlum_noabs_med} }
\end{figure*}

\begin{figure*}[ht]
\includegraphics[width=\textwidth]{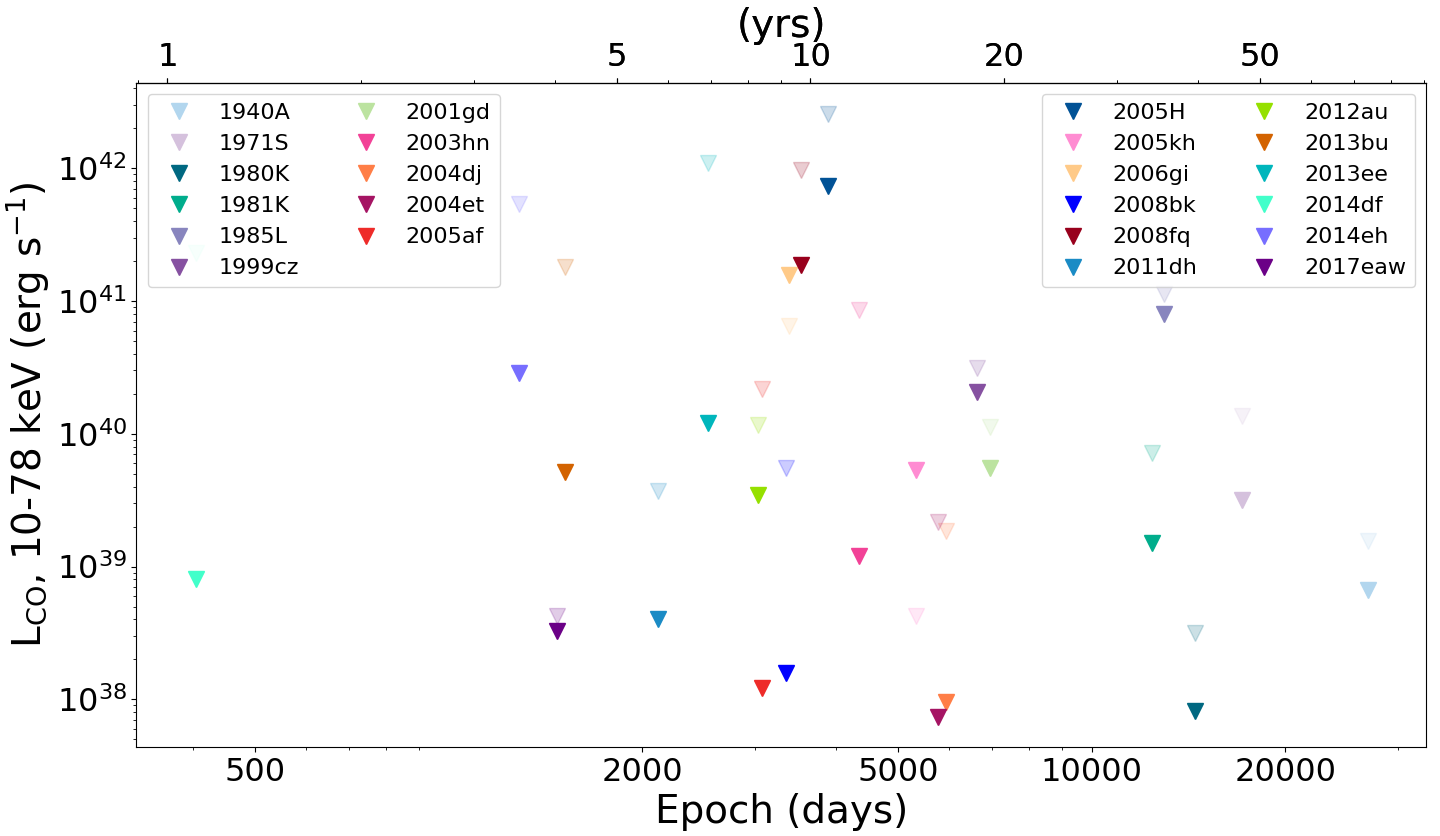} 
\caption{\footnotesize{CO luminosity limits from NuSTAR observations in the energy interval 10--78~keV, assuming a favorable viewing angle of negligible ejecta absorption (solid markers) and median absorption (semi-transparent markers). }\label{fig:COlum_nustar} }
\end{figure*}

The limits on the CO luminosities are derived assuming four different degrees of ejecta absorption. In Figure~\ref{fig:COlum_noabs_med} we present two cases: $L_{\rm CO, 0,abs}$ (darker grey/solid color) and $L_{\rm CO,50,abs}$ (light grey/semi-transparent) for each SN. As for the SN luminosities in Figure~\ref{fig:obsflux}, the deepest CO luminosity limit is shown for each SN, where observations by the same telescope of SNe close in time have been simultaneously fitted as described in Section~\ref{sec:Analysis_of_detected_sources}. Chandra and XMM observations commonly give $L_{\rm CO,0,abs}\leqq 10^{39}~\rm erg~s^{-1}$ and $L_{\rm CO,50,abs}\leqq 10^{43}~\rm erg~s^{-1}$, while typical limits from Swift and NuSTAR are less constraining (following the same trends as in Figure~\ref{fig:telescope_hist}). Since CO luminosity limits above $\sim 10^{41}~\rm erg~s^{-1}$ do not yield meaningful constraints for the pulsar population discussed in Section~\ref{sec:discussion}, we choose not to include them in Figure~\ref{fig:COlum_noabs_med}. We further present the most constraining CO-luminosity limit (for negligible and median ejecta absorption) for each SNe in our NuSTAR sample in the higher energy band 10--78~keV in Figure~\ref{fig:COlum_nustar}.

\begin{figure}[ht]
\hspace{-0.1cm}\includegraphics[width=0.48\textwidth]{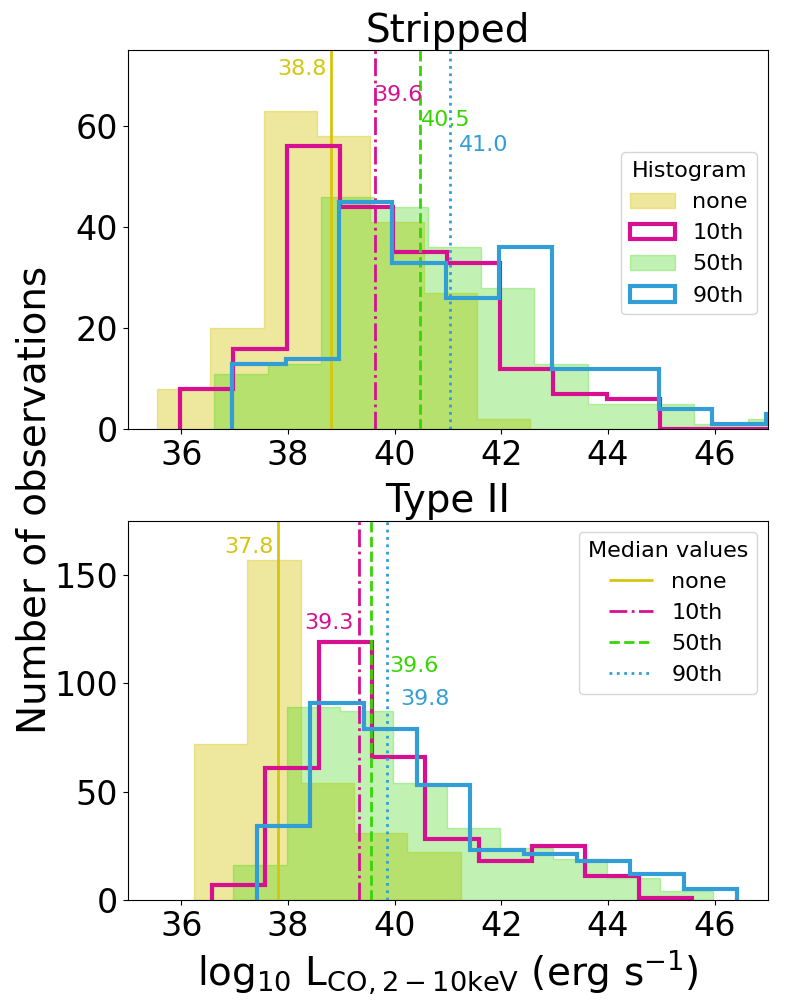} 
\caption{\footnotesize{Histograms of the 3$\sigma$ CO luminosity limits for different degrees of ejecta absorption, represented by different colors. Median luminosity limits for each distribution are shown by vertical lines of corresponding colors. The top panel shows the distribution for stripped type SN and the bottom panel shows type II. Note that upper panel is cut at luminosity limits of $10^{47}~\rm erg~s^{-1}$ as such high luminosities are not physically meaningful.}  \label{fig:abs_scale_hist} }
\end{figure}

As shown in Figure~\ref{fig:COlum_noabs_med}, the CO luminosity limits decrease over time under median absorption conditions. This trend is expected across all levels of ejecta absorption as the expansion of ejecta results in a reduced optical depth over time as $\tau \propto t^{-2}$. In Figure~\ref{fig:abs_scale_hist}, we show histograms of the CO luminosity limits for stripped and type II SNe assuming different levels of ejecta absorption, as well as the corresponding median luminosity limits. 

The median luminosity limits increase with increasing ejecta absorption for all SN types as expected.
Furthermore, the difference $L_{\rm CO,10, abs}-L_{\rm CO,0, abs}$ is greater for type II than for stripped type SNe owing to their higher ejecta masses, while the difference between luminosity limits assuming higher degrees of ejecta absorption are smaller for type II. The latter is expected since the ejecta models are more asymmetric for the IIb model than all the type II models (see cumulative distributions of $\tau$ in Figure~5 in \citealt{2018Alp}).

\begin{figure*}[ht]
\gridline{\fig{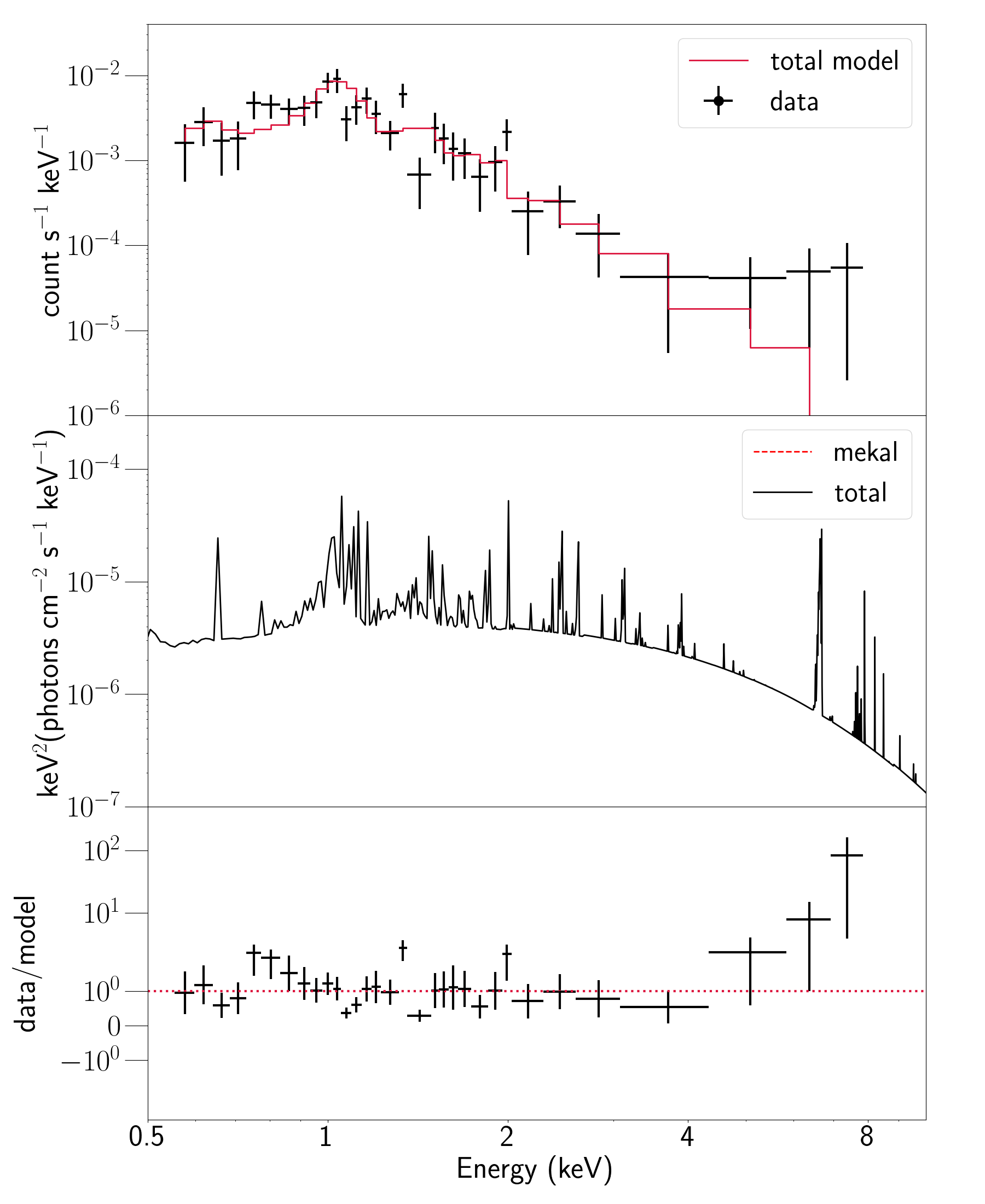}{0.42\textwidth}{(a) obsid--6727 }
          \fig{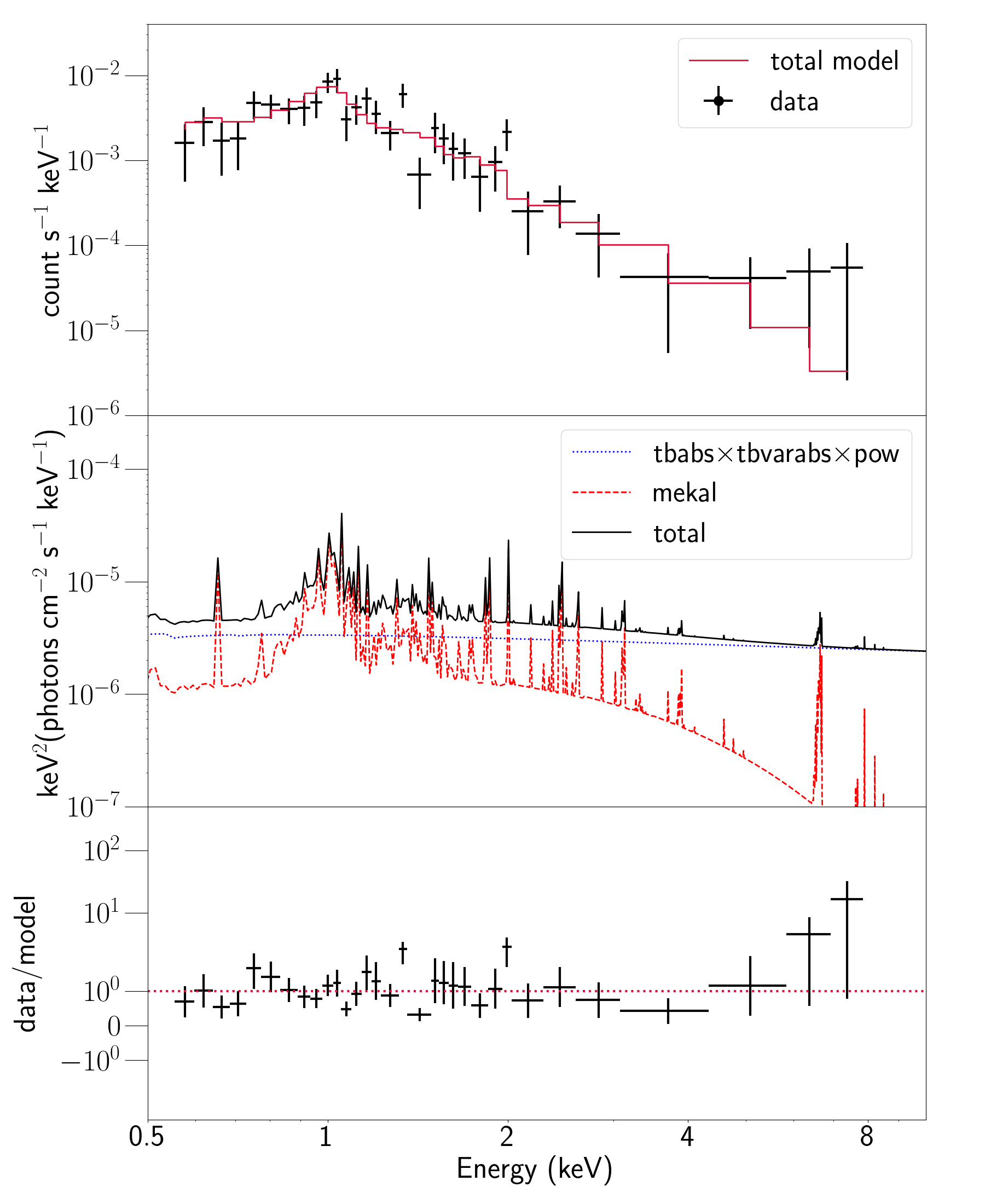}{0.42\textwidth}{(b) obsid--6727}~
          }
\gridline{\fig{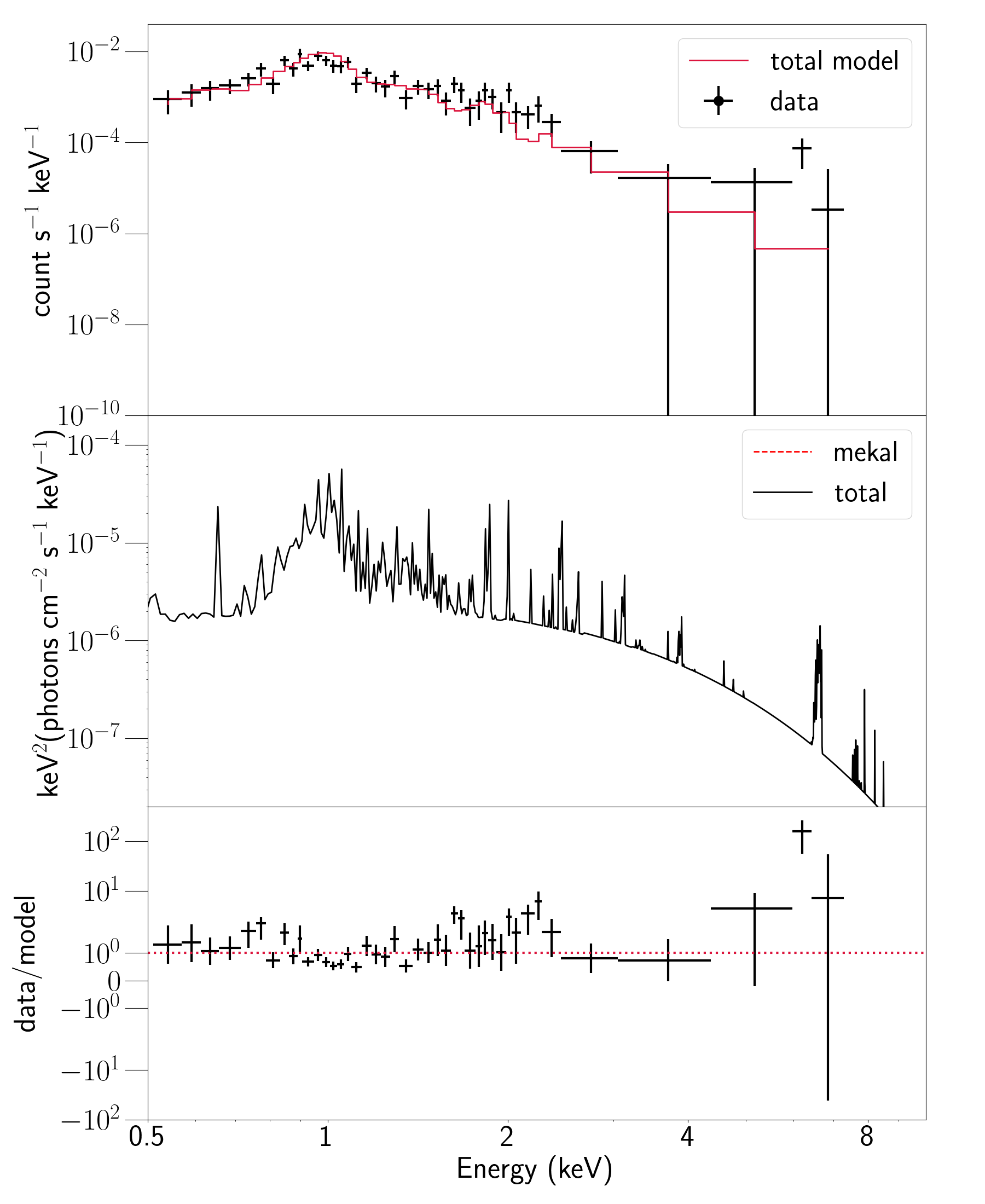}{0.42\textwidth}{(c) obsid--14230}
          \fig{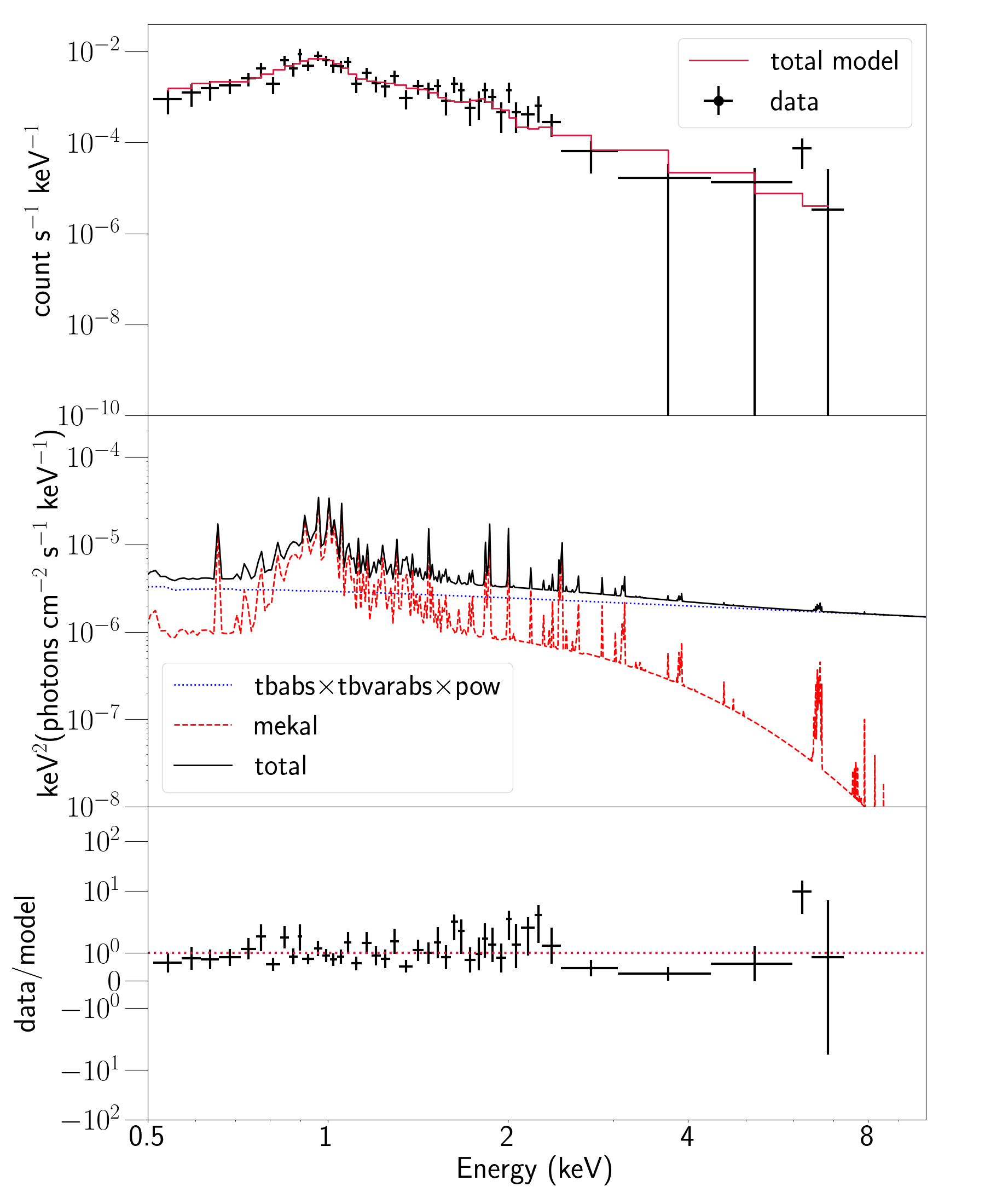}{0.42\textwidth}{(d) obsid--14230}
          }
\caption{\footnotesize{Fits for SN~1979C. The first column (a and c) show the spectra fitted with a \texttt{tbabs*mekal} model and the second column (b and d) show the fits after adding a power-law component to the model: model--\texttt{tbabs*(mekal+pow)}. The upper most panels show the spectra and fitted model, the middle panels show the total model and different model components, and the bottom panels show the log ratio of data to model. All spectra have been rebinned with three counts per bin for visual clarity. Both observations were taken with Chandra ACIS S3.}}\label{fig:fit_1979C}
\end{figure*}

Of the 12 detected SNe, an additional power-law component associated with a CO could be constrained in 10, but only in the scenario of no ejecta absorption. Among these 10 SNe, the power law was significant at 3$\sigma$ in only one case: SN~1979C (in two observations by Chandra). Fits of SN~1979C with a \texttt{mekal} model with and without a power-law component are shown in Figure~\ref{fig:fit_1979C}, while the best-fit parameters are presented in Table~\ref{tab:fit_1979C}. This table also presents fits with two \texttt{mekal} components for comparison, discussed further in Section~\ref{sec:detected_CO_discussion}.

\begin{deluxetable*}{l l l l l l l l l l l}[ht] 
\tablenum{5}
\tablecaption{Fit Results for SN~1979C
\label{tab:fit_1979C}}
\tablewidth{0pt}
\tablehead{ 
 \colhead{Obsid} & \colhead{Exp.} & \colhead{Obs. Date} & \colhead{Epoch}  & \colhead{cstat} & \colhead{dof} &  \colhead{$kT$}  &\colhead{} & \colhead{} &  \colhead{} &  \colhead{L$_{\rm SN}$} \\ 
 \colhead{} & \colhead{(ks)} & \colhead{} & \colhead{(Days)}  & \colhead{} & \colhead{} & \colhead{(keV)}& \colhead{}   & \colhead{} &  \colhead{} &  \colhead{($10^{38}\ \rm erg~s^{-1}$)}
}
\startdata
        6727 &37.87 & 2006-02-18 & 9802 & 89.42 & 93 &  $1.65_{-0.30}^{+0.31}$  &  &  &  & $1.95_{-0.79}^{+0.76}$ \\
        9121 & 14.92& 2008-04-20 & 10594 & 43.0 & 50 &  $1.29_{-0.34}^{+0.42}$  & &  &   & $1.19_{-0.73}^{+1.20}$ \\
        12696$^*$ & 14.86& 2011-02-24 & 11634 & 155.52 & 155 &  $1.29_{-0.09}^{+0.08}$  &  & &  & $0.86_{-0.24}^{+0.30}$ \\
        14230$^*$ & 79.03& 2012-02-16 &11991 & 155.52 & 155&  $1.29_{-0.09}^{+0.08}$ &   &  &  & $1.05_{-0.19}^{+0.22}$ \\
        23140$^\dagger$ &9.96& 2020-02-15 & 14912 & 17.56 & 21 & $1.30_{-0.47}^{+0.81}$  &  &  & & $0.77_{-0.51}^{+1.21}$ \\
        23141$^\dagger$ &9.96 & 2020-03-15 & 14939  & 17.56 & 21 &  $1.30_{-0.47}^{+0.81}$  & &  & &  $0.78_{-0.50}^{+1.34}$ \\
        \hline
        & &  &  &   &  & $kT$ & $\Gamma$   &  \colhead{L$_{\rm th,0,abs}$} & \colhead{L$_{\rm CO,0,abs}$} &  \colhead{L$_{\rm SN}$}\\
         & &  & &    &  & (keV) & &  \colhead{($10^{38}\ \rm erg~s^{-1}$)} &  \colhead{($10^{38}\ \rm erg~s^{-1}$)} &  \colhead{($10^{38}\ \rm erg~s^{-1}$)}\\[2mm]
        \hline
        6727 &37.87 & 2006-02-18 & 9802 & 79.33 & 91  & $1.24_{-0.24}^{+0.27}$ & $2.17_{-0.53}^{+0.56}$  & $0.52_{-0.31}^{+0.50}$& $2.2_{-1.4}^{+1.9}$  & $2.7_{-1.1}^{+1.7}$ \\
        9121 & 14.92& 2008-04-20 & 10594 & 34.82 & 48  & $1.00_{-0.28}^{+0.30}$ & $1.78_{-1.30}^{+0.86}$  &  $0.36_{-0.26}^{+0.43}$ & $<25.29$  & $3.6_{-2.1}^{+4.8}$ \\
        12696$^*$ & 14.86& 2011-02-24 & 11634 & 121.62 & 152  & $1.01_{-0.13}^{+0.10}$ & $2.29_{-0.38}^{+0.35}$ & $<0.44$ & $2.6_{-1.3}^{+1.8}$    & $2.7_{-1.2}^{+1.8}$ \\
        14230$^*$ & 79.03& 2012-02-16 &11991 & 121.62 & 152  & $1.01_{-0.13}^{+0.10}$& $2.29_{-0.38}^{+0.35}$  & $0.32_{-0.14}^{+0.15}$ & $1.55_{-0.68}^{+0.88}$  & $1.86_{-0.60}^{+0.84}$ \\
        23140$^\dagger$ &9.96& 2020-02-15 & 14912 & 16.84 & 19  & $0.90^{+0.96}$  & [2.0] & $<4.64$ & $<6.42$   & $2.0_{-1.7}^{+2.5}$ \\
        23141$^\dagger$ &9.96 & 2020-03-15 & 14939 & 16.84 & 19  & $0.90^{+0.96}$ &  [2.0] & $<4.45$ & $<1.01$  &   $1.3_{-1.2}^{+5.3}$ \\
        \hline
         & &  &  & &  & $kT1$ & $kT2$ &   \colhead{L$_{\rm kT1,0,abs}$} &  \colhead{L$_{\rm kT2,0,abs}$} &  \colhead{L$_{\rm SN}$}\\
         & &  &  & &    & (keV) & (keV)&  \colhead{($10^{38}\ \rm erg~s^{-1}$)} &  \colhead{($10^{38}\ \rm erg~s^{-1}$)} &  \colhead{($10^{38}\ \rm erg~s^{-1}$)}\\[2mm]
        \hline
        6727 &37.87 & 2006-02-18 & 9802 & 79.69 & 91  & $1.02_{-0.46}^{+0.31}$ & $2.8_{-1.2}^{+6.5}$ & $2.4_{-1.0}^{+1.9}$ & $0.24_{-0.22}^{+0.56}$  & $2.6_{-1.2}^{+1.9}$ \\
        9121 & 14.92& 2008-04-20 & 10594 & 34.24 & 48   & $0.83_{-0.27}^{+0.29}$& $3.1_{-1.7}$  & $<8.71$ & $0.18_{-0.14}^{+0.49}$  & $2.8_{-1.8}^{+4.3}$ \\
        12696$^*$ & 14.86& 2011-02-24 & 11634  & 119.33 & 152 & $0.83_{-0.12}^{+0.10}$ & $2.25_{-0.52}^{+1.04}$ &  $2.17_{-0.95}^{+1.45}$ & $<0.27$  & $2.21_{-0.93}^{+1.42}$ \\
        14230$^*$ & 79.03& 2012-02-16 &11991 & 119.33 & 152 & $0.83_{-0.12}^{+0.10}$ & $2.25_{-0.52}^{+1.04}$   & $1.48_{-0.49}^{+0.69}$ & $0.16_{-0.085}^{+0.096}$  & $1.65_{-0.50}^{+0.70}$ \\
        23140$^\dagger$ &9.96& 2020-02-15 & 14912 & 15.6 & 18 & 0.72 & $1.57_{-0.55}$  & $<7.10$  & $<2.70$ & $1.10_{-0.69}^{+2.09}$ \\
        23141$^\dagger$ &9.96 & 2020-03-15 & 14939 & 15.6 & 18 & 0.72 & $1.57_{-0.55}$  & $<6.56$ & $<12.71$  & $0.57_{-0.46}^{+4.08}$ 
\enddata
\tablecomments{\footnotesize{The spectra were fitted with \texttt{tbabs*mekal} in the upper most section, \texttt{tbabs*(mekal+pow)} in the middle section and \texttt{tbabs*(mekal+mekal)} in the lower section. The outermost column $L_{\rm SN}$ includes all model components, however all luminosities in this table have been corrected for Galactic absorption. Uncertainties are 90\% and upper limits $3\sigma$. Parameters frozen during fitting are given in square brackets and values without upper and-or lower error bars indicate that the confidence intervals reach the hard boundaries of the parameters. The observations marked with $^*$ and $^\dagger$ represent the cases where $kT$ and/or $\Gamma$ were tied in the fits (separately for $^*$ and $^\dagger$) and the reported cstat and dof are correspondingly for the joint fit. All spectral fitting was done in the energy interval 0.5--8~keV and the luminosities are derived for the 2--10~keV interval by extrapolation of the model. All observations are taken by Chandra and the distance adopted for the luminosity calculations was 16~Mpc as given by the OSNC.\footnote{We note that a slightly different distance estimate of 15.2~Mpc is provided by \cite{2001Freedman}. While the results presented here are based on the somewhat larger value from the OSNC, adopting 15.2~Mpc would lower our luminosities by $\sim$10~\%.} The two observations with a significant power-law component are 6727 and 14230. }}
\end{deluxetable*}

\section{Population Synthesis and Discussion}\label{sec:discussion}

The luminosity limits of the COs estimated in this work can be used to impose constraints on the physical parameters of the COs themselves. The power-law component is in principle applicable to PWN/pulsar emission as well as BH accretion, though the observed limits are most constraining for the former, which are expected to be brighter. In Section~\ref{sec:dipole} below, we present a population synthesis of pulsars, where we extensively explore the parameter space of pulsar properties as well as the fraction of SNe that do not produce pulsars. We then discuss the evidence for a compact object in SN~1979C in Section~\ref{sec:detected_CO_discussion}, and finally comment on the late-time CSM interaction for the 12 detected SNe in Section~\ref{sec:mass_loss_rate}.

\subsection{Pulsar Population Synthesis} \label{sec:dipole}

\subsubsection{Model Setup}\label{sec:pulsar_model}
Pulsars are commonly modeled as rotating magnetic dipoles in vacuum (\citealt{1969Ostriker};\citealt{1983Shapiro}), where the loss of rotational energy, also referred to as spin-down power, is the dominant source of the pulsar radiation. The time derivative of the rotational energy of a rotating rigid body is described by 
\begin{equation}\label{eqn:Erot_dot}
    \dot{E}_{\rm rot}=-4\pi^2I\frac{\dot{P}}{P^3} \hspace{5mm},
\end{equation}
where $I$ is the moment of inertia, set to $I=10^{45}~\rm g~cm^2$, $P$ is the spin period and $\dot{P}$ its time derivative. The energy emitted from the surrounding PWN corresponds to the spin-down power of the pulsars. We adopt this vacuum-dipole model with a magnetic field ($B$) as in \cite{2016Shibata}
\begin{equation}\label{eqn:b_field}
    B^2= \frac{3}{2}\frac{Ic^3}{(2\pi)^2R^6}P\dot{P},
\end{equation}

where $R$ is the radius of the NS, assumed to be 12~km, and $c$ is the speed of light. This expression assumes an equatorial surface magnetic field, but can be generalized by including a factor $\sin^2(\alpha)$ in the denominator, where $\alpha$ is the angle between the magnetic and rotational axes. 

From equations \ref{eqn:Erot_dot} and \ref{eqn:b_field}, we see that the spin-down power scales as $\dot{E}_{\rm rot} \propto B^2/P^4$. This will decrease over time as pulsars spin down, with $P$ following the relation
\begin{equation}\label{eqn:p}
    P=\left[ P_0^2+\left(\frac{16\pi^2R^6B^2}{3Ic^3}\right)t\right]^{1/2},
\end{equation}
where $P_0$ is the birth spin period, $t$ is the time, and B is assumed to stay constant.

Part of the spin-down power is converted into X-rays, defined by the efficiency $\eta=L_{\rm x}/\dot{E}_{\rm rot}$. We assume the $L_{\rm x}$ to be a real-time indicator of the spin-down power of the pulsar, as the synchrotron timescale is short in comparison to the ages of the sources in our sample (see \citealt{2008Perna} for further discussion). Studies of nearby pulsars and PWNe have shown that $\eta$ increases with $\dot{E}_{\rm rot}$. We adopt the $\dot{E}_{\rm rot}-L_{\rm X}$ relation for PWNe as derived by \cite{2008Li}
\begin{equation}
\label{eqn:final_Lx}
     L_{\rm x,2-10~keV}=10^{-19.6\pm3.0} \dot{E}_{\rm rot}^{1.45\pm0.08},
\end{equation}
where the uncertainties include both the intrinsic spread in the sample and measurement uncertainties. The relation for pulsars is somewhat flatter ($L_{\rm x} \propto \dot{E}_{\rm rot}^{\alpha}$, $\alpha \sim $0.8--1.3 \citealt{2008Kargaltsev,2008Li, 2016Shibata, 2025Xu}), though their contribution to the total X-ray emission is typically lower than that of the PWNe, as discussed in Section~\ref{sec:pow_comp} \citep{2008Li,2008Kargaltsev,2021Hsiang}. In line with eq.~\ref{eqn:final_Lx}, \cite{2002Possenti} found $L_{\rm x} \propto \dot{E}_{\rm rot}^{1.34\pm0.03}$ for the combined PWN+pulsar emission, and \cite{2004Cheng} found $L_{\rm x} \propto \dot{E}_{\rm rot}^{1.4\pm0.1}$ for PWNe.  

An important consideration is that that we apply this relation to systems that are younger and brighter than the Galactic pulsars/PWN used in the above studies. There is no obvious reason why this relation cannot be extrapolated to more energetic systems, though values of $\eta \gtrsim 1$ are clearly unphysical. To be conservative, we impose a maximum value of $\eta=0.1$ in our simulations, which is compatible with the most efficient systems found by \cite{2004Cheng} and \cite{2008Li}. In addition, we assume eq.~\ref{eqn:final_Lx} to hold also in the hard 10--78~keV energy range observed by NuSTAR, with the constant adjusted assuming a power-law spectrum with $\Gamma = 2.1$ (motivated by the value of $2.106\pm0.006$ measured for the Crab in the energy interval 3-50~keV, \citealt{2017Madsen}).

For comparison, we also perform simulations where we use the $\eta$ measured for the Crab PWN+pulsar, irrespective of the value of $\dot{E}_{\rm rot}$. The value of $\eta$ is calculated by dividing the Crab luminosity in the relevant energy range by the spin-down power of the Crab, $\dot{E}_{\rm rot}=4.6\times 10^{38}~\rm erg~s^{-1}$ \citep{2016Ansoldi}. The luminosity in each energy range is determined by extrapolating the power-law model ($\Gamma=2.106$, \citealt{2017Madsen}), giving $\eta=0.022$ and $\eta=0.019$ in the 2--10 and 10--78~keV energy intervals, respectively.

We also need to account for the fact that not all SNe in our sample will produce pulsars. Therefore, we explore different non-pulsar fractions ($f_{\rm np}$), defined as the fraction of SNe that produce a BH or a NS without any non-thermal emission from a pulsar/PWN. The latter would be newly created analogs of the class central compact objects (CCO) in our Galaxy (see \citealt{2017DeLuca} for a review). For both BHs and CCOs, we assume that the hard X-ray emission would be too faint to be detected.

The fraction of massive stars that create BHs is uncertain, with model predictions varying in the range 1--40\% \citep{2003Heger,2011OConnor,2012Ugliano,2015Kochanek}. The vast majority of these are expected to form in failed/dim explosions, implying that we expect a smaller BH fraction in our sample of regular, bright SNe. The BHs in our sample would have been created either promptly in the explosions of very massive stars \citep{2023Burrows,2024Eggenberger} or as a result of ejecta fallback (e.g., \citealt{2020Chan}). The fraction of CCO created in SNe is also uncertain. About a dozen such objects are known in our Galaxy (\citealt{2017DeLuca}), but this is hard to translate to a birth rate given observational selection effects and the poor understanding of the lack of pulsar activity. These objects may simply have slow rotation and weak magnetic fields, but another scenario is that they have strong magnetic fields that are buried by fallback accretion and only emerge on long time scales of $>1000$~yrs (e.g., \citealt{2016TorresForne}). Considering the uncertainties in the fraction of BHs and CCOs created in SNe, we explore a wide range of $f_{\rm np}$ from 1--40\% in our simulations.

For the SNe that do produce pulsars, $P_{\rm 0}$ and $B_{\rm 0}$ are sampled from log-normal and normal distributions determined from population synthesis of pulsars, mostly Galactic radio pulsars (see later discussion in Section~\ref{sec:pop_synt_disc}), where we further constrain $P_{\rm 0}$ to physical values of $P_{\rm 0}\ge 1~$ms. Seven such pulsar birth distributions are explored and listed in the first column of Table~\ref{tab:pulsar_results}.

\subsubsection{Simulation Procedure}\label{sec:simulation_procedure}
To explore the pulsar parameter space, we perform a series of Monte Carlo simulations based on the pulsar models outlined in the previous section. These simulations compare theoretical pulsar luminosities with the observational luminosity limits derived in this study. For each observational limit ($L_{\rm CO}$), we compare to a predicted luminosity ($L_{\rm model}$), which is derived as follows:

First, we determine whether $L_{\rm model}$ is based on a pulsar model or non-pulsar model (i.e., a BH or CCO). This is done by comparing a randomly sampled value with the parameter $f_{\rm np}$. If the object is classified as a non-pulsar, $L_{\rm model}=10^{35}~\rm erg~s^{-1}$, which is below our deepest limits and hence never detected. If the object is classified as a pulsar, $P_{\rm 0}$ and $B_{\rm 0}$ are sampled from the distributions in Table~\ref{tab:pulsar_results}, and the corresponding spin-down power at the epoch of the observation is calculated as outlined in Section~\ref{sec:pulsar_model}. Finally, the $L_{\rm model}$ in the relevant energy range is determined using either eq.~\ref{eqn:final_Lx} (which is sampled including the uncertainties) or the $\eta$ of the Crab.

Once $L_{\rm model}$ has been determined, it is compared to the corresponding $L_{\rm CO}$  for different levels of ejecta absorption. For $L_{\rm CO}=L_{\rm CO, 0, abs}$, the modeled and observational values are directly compared. In the case where ejecta absorption is included, the observational luminosity limit is sampled from one of the three estimated CO luminosites with varying degrees of ejecta absorption. This sampling is weighted in favor of the $L_{\rm CO,50, abs}$ luminosity limit (60\%), and lesser for $L_{\rm CO,10, abs}$ and $L_{\rm CO,90, abs}$ (20\% each). This procedure provides a rough approximation of the varying degrees of absorption due to the different lines of sight through the asymmetric ejecta. If $L_{\rm model}>L_{\rm CO}$, the model predicts that the CO should be detected.

In a single realization of the simulations, this process is repeated for the full SN sample for a given model and observational limits, either with or without absorption. For model comparisons to observational limits, we exclude extreme outliers with luminosity limits $L_{\rm CO,2-10~keV}>10^{41}~\rm erg~s^{-1}$, as these do not contribute to meaningful constraints. This results in a sample size of 350 observations and 80 SN. The 2--10~keV luminosity limits without ejecta absorption naturally lower and the full sample comprises 645 observations and 233 SN. All observational limits from NuSTAR in the 10--78~keV energy band are included both with and without ejecta absorption, which adds 50 observations of 23 SN to the samples. For SNe with multiple observations at different epochs, we assume the same line of sight or degree of ejecta absorption for observed luminosity limits, as well as the same initial $P_{\rm 0}$ and $B_{\rm 0}$ for model predictions. This implies that the $L_{\rm model}$ for different observations of the same SN in a given realization only differs due to the pulsar spin-down and $\eta$.

\begin{figure}[ht]
    \hspace{-0.3cm}\fig{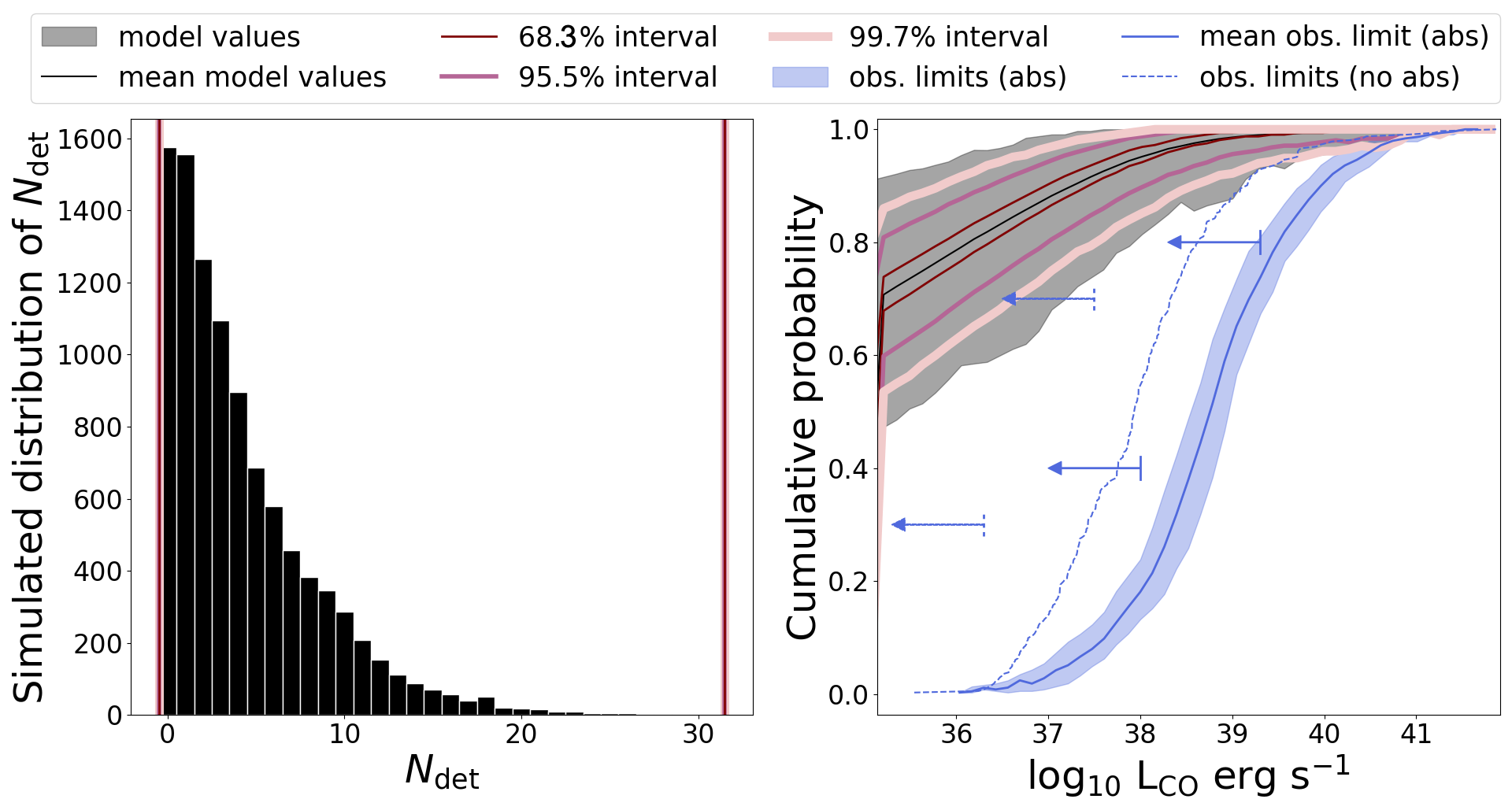}{0.5\textwidth}{\footnotesize{(a) Example of a compatible model: $B_{\rm 0, mean}$ and $P_{\rm 0, mean}$ from \cite{2024Pardo_Araujo}}}
    
    \hspace{-0.3cm}\fig{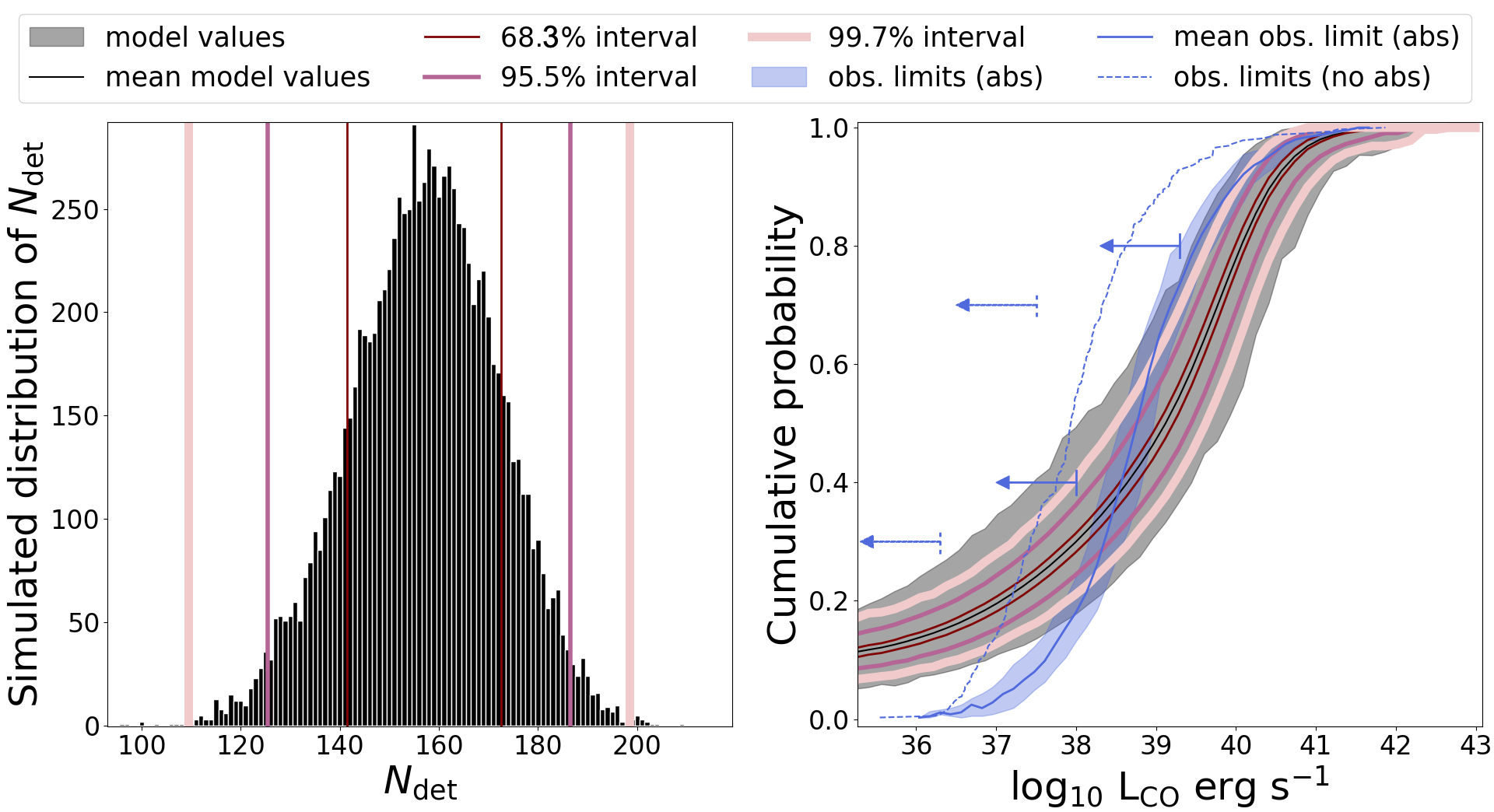}{0.5\textwidth}{\footnotesize{(b)Example of an excluded model: $B_{\rm 0, mean}$ and $P_{\rm 0, mean}$ from \cite{2002Arzoumanian}}}
    
    \caption{\footnotesize{Two examples of the pulsar population synthesis. Both models assume spin-down efficiency according to eq.~\ref{eqn:final_Lx} and $f_{\rm np}=10$\%. \textit{Left panels}: final histogram of number of predicted detections. \textit{Right panels:} corresponding cumulative distributions of observed luminosity limits assuming no absorption (blue dashed line) and a distribution of ejecta absorption (blue solid line), together with the range of modeled values (grey).} \label{fig:pulsar_cumu_dist_hist}}
\end{figure}

For each realization, we record the number of predicted detections. A full simulation of 10,000 realizations\footnote{The number of realizations is increased to 100,000 to improve accuracy in cases where the agreement between the model and simulations are near the boundaries of the relevant confidence intervals.} then provides the distribution of expected detections for a given synthesis model, defined by combinations of $B_{\rm 0}$-$P_{\rm 0}$ distributions, $\eta$, degree of ejecta absorption and $f_{\rm np}$. In Figure ~\ref{fig:pulsar_cumu_dist_hist}, we show two examples of such distributions in the left panels. 
If the actual number of detections (zero assuming ejecta absorption and two assuming no absorption) is within the 68.3\%, 95.5\%, 99.7\% confidence intervals of the model predictions ($N_{\rm det}$), the synthesis model is considered compatible at the corresponding level of significance. We treat the two observations of SN~1979C with significant power-law components as detections of a pulsar/PWN for the purposes of this comparison, though we note that there are significant uncertainties, see Section~\ref{sec:detected_CO_discussion}. This does not affect the main conclusions of the population synthesis. The confidence intervals are determined by starting at the mode of the histogram. We then compare the $N_{\rm det}$ of adjacent bins, successively expanding the interval to the side with the highest predicted number of detections per sample until the sum of the histogram within the interval is $\approx$68.3\%, 95.5\%, and 99.7\% of the total sample. In panel (a) of Figure ~\ref{fig:pulsar_cumu_dist_hist}, we present an example of a compatible model (0 detections is within the 68.3\% interval), while panel (b) shows an example of a non-compatible model, where the predicted number of detections is between $\sim$110 and $\sim$200 (99.7\%) significantly higher than what the observational results suggest. The right panels of Figure~\ref{fig:pulsar_cumu_dist_hist} show the cumulative distributions of $L_{\rm model}$ and $L_{\rm CO}$ for the two examples, as well as $L_{\rm CO}$ for the case of no absorption, which is incompatible with both models.

This procedure allows us to explore the parameter space of $f_{\rm np}$, $B_{\rm 0, mean}$ and $P_{\rm 0, mean}$ distributions, $\eta$ and the impact of ejecta absorption. By doing so, we can assess whether a given synthesis model is compatible with our observational results, and to what degree. Alternatively, we can determine if the model should be rejected.

\subsubsection{Population Synthesis Results and Interpretation}\label{sec:pop_synt_disc}

In Table~\ref{tab:pulsar_results}, we present a summary of the results from the pulsar population synthesis, indicating which models are consistent with our observational limits. Column 4--6 details the compatibility of each synthesis model with the observational results, while column 7 lists the predicted number of detected pulsars based on the peak values of the simulations for $f_{\rm np}=10\%$. This shows large variations in model compatibility, illustrating how X-ray limits help constrain pulsar birth distributions, $f_{\rm np}$ and $\eta$. In one extreme end, we find the birth parameter distributions from \cite{2002Arzoumanian} which are completely excluded, as all synthesis models based on these distributions predict too many detected pulsars, regardless of the assumptions about $\eta$ and $f_{\rm np}$ (see also Figure~\ref{fig:pulsar_cumu_dist_hist}b). These models have a large fraction of pulsars with very short spin periods, leading to predicted luminosities well above our upper limits. Given $f_{\rm np}=10~\%$ and ejecta absorption, this model predicts that pulsars should be detected in $>40~\%$ of our X-ray observations, which is clearly excluded. 

\begin{deluxetable*}{lllllll}[ht]
    \tablenum{6}
    \tablecaption{Pulsar Population Synthesis Results
    \label{tab:pulsar_results}}
    \tablewidth{0pt}
    \tablehead{ (1) & (2) & (3) & (4) & (5) & (6) & (7)\\
        $B_0$, $P_0$ Distribution & Ejecta Abs. & $\eta$  & \multicolumn{3}{l}{Model Compatibility} & $N_{\rm det}(f_{\rm det})$\\
        $\log B_{\rm0,mean}(\sigma\log B_{\rm0,mean})$ & & & 99.7\% & 95.5\% & 68.3\%  &  \footnotesize{assuming $f_{\rm np}=10\%$} \\  
        $\log P_{\rm0,mean}$($\sigma \log P_{\rm0,mean}$)/$P_{\rm0,mean}(\sigma P_{\rm0,mean})$ & & & & & & 
    }
    \startdata
    \cite{2002Arzoumanian} & no & Crab $\eta$  & \ding{56} & \ding{56} & \ding{56}  & 447~(69\%)\\
    $\bm{\log B_{\rm0,mean}:}$~12.35(0.4)~G       &  & eq.~\ref{eqn:final_Lx}   & \ding{56} & \ding{56} & \ding{56} & 397~(62\%)\\
    $\bm{\log P_{\rm0,mean}:}$~-2.3(0.2)~s        & yes & Crab $\eta$  & \ding{56} & \ding{56} & \ding{56} &164~(47\%)\\
                     & & eq.~\ref{eqn:final_Lx} & \ding{56} & \ding{56} & \ding{56}  &150~(43\%)\\
    \hline
    \cite{2006Kaspi} & no & Crab $\eta$  & \ding{51} & \ding{51} & \ding{51} & 0~(0\%)\\
    $\bm{\log B_{\rm0,mean}:}$~12.65(0.55)~G      &  & eq.~\ref{eqn:final_Lx}   & \ding{51} & \ding{51} & \ding{51} & 0~(0\%)\\
    $\bm{P_{\rm0,mean}:}$~0.3(0.15)~s      & yes & Crab $\eta$   & \ding{51} & \ding{51} & \ding{51}& 0~(0\%)\\
                     & & eq.~\ref{eqn:final_Lx}  & \ding{51} & \ding{51} & \ding{51}& 0~(0\%)\\
    \hline
    \cite{2015Gullon} & no & Crab $\eta$   & \ding{51} & \ding{51} & \ding{51}& 2~(0\%)\\
    $\bm{\log B_{\rm0,mean}:}$~13.23(0.72)~G  &  &  eq.~\ref{eqn:final_Lx}  & \ding{51} & \ding{51} & \ding{51} &   5~(1\%)\\
    $\bm{P_{\rm0,mean}:}$~0.32(0.19)~s     & yes & Crab $\eta$   & \ding{51} & \ding{51} & \ding{51}& 0~(0\%)\\
                     & & eq.~\ref{eqn:final_Lx}  & \ding{51} & \ding{51} & \ding{51} &0~(0\%)\\
    \hline
    \cite{2020Cieslar} & no & Crab $\eta$  & \ding{56} & \ding{56} & \ding{56}& 72~(11\%)\\
    $\bm{\log B_{\rm0,mean}:}$~12.67(0.34)~G      &  & eq.~\ref{eqn:final_Lx}   & \ding{56} & \ding{56} & \ding{56}& 69~(11\%)\\
    $\bm{P_{\rm0,mean}:}$~0.05(0.07)~s     & yes & Crab $\eta$   & \ding{51}  & \ding{51} & \ding{51}$\ge$25\%& 5~(1\%)\\
                     & & eq.~\ref{eqn:final_Lx}  & \ding{51}  & \ding{51}$\ge$15\%  & \ding{56}& 12~(3\%) \\
    \hline
    \cite{2022Igoshev} & no & Crab $\eta$    & \ding{56}  & \ding{56}  & \ding{56}& 48~(7\%)\\
    $\bm{\log B_{\rm0,mean}:}$~12.44(0.44)~G      &  & eq.~\ref{eqn:final_Lx}    & \ding{56} & \ding{56} & \ding{56}& 49~(8\%)\\
    $\bm{\log P_{\rm0,mean}:}$~-1.04(0.53)~s      & yes & Crab $\eta$  & \ding{51} & \ding{51} & \ding{51} & 4~(1\%)\\
                     & & eq.~\ref{eqn:final_Lx}  & \ding{51}  & \ding{51}  & \ding{51}$\ge$15\% & 5~(1\%)\\
    \hline
    \cite{2024Graber} & no & Crab $\eta$   & \ding{56} & \ding{56} & \ding{56}& 46~(7\%) \\
    $\bm{\log B_{\rm0,mean}:}$~13.10(0.45)~G      &  & eq.~\ref{eqn:final_Lx}  & \ding{56} & \ding{56} & \ding{56}& 49~(8\%)\\
    $\bm{\log P_{\rm0,mean}:}$~-1.00(0.38)~s      & yes & Crab $\eta$ & \ding{51}  & \ding{51} & \ding{51} & 0~(0\%)\\
                     & & eq.~\ref{eqn:final_Lx} & \ding{51} & \ding{51} & \ding{51} & 3~(1\%)\\
    \hline
    \cite{2024Pardo_Araujo} & no & Crab $\eta$  & \ding{51}$\ge$10\% & \ding{56} & \ding{56} & 25~(4\%)\\
    $\bm{\log B_{\rm0,mean}:}$~13.09(0.5)~G       &  & eq.~\ref{eqn:final_Lx}   & \ding{51}$\ge$40\% & \ding{56} & \ding{56}& 26~(4\%)\\
    $\bm{\log P_{\rm0,mean}:}$~-0.67(0.55)~s      & yes & Crab $\eta$ & \ding{51} & \ding{51} & \ding{51} & 0~(0\%)\\
                     & & eq.~\ref{eqn:final_Lx} & \ding{51} & \ding{51} & \ding{51}& 0~(0\%)\\
    \enddata
    
    \tablecomments{\footnotesize{The models of the population synthesis are defined by (1) the birth distributions, (2) ejecta absorption (yes/no), (3) efficiency scaling. Each model has been run for $f_{\rm np}$ between $1-40\%$ in 5\% steps. Columns 4--6 show model compatibility with observed detections (0 and 2, with/without absorption) for given confidence intervals. Compatible models are denoted by \ding{51} and non-compatible ones by  \ding{56}. The lowest value of $f_{\rm np}$ for which a given model is compatible is also provided when relevant. If no number is given, the result is the same for all $f_{\rm np}$ in the range 1--40\%. Column 7 lists the most probable $N_{\rm det}$, i.e.\ the predicted number of detections, for $f_{\rm np}=10\%$, as well as the corresponding predicted detection fraction $f_{\rm det}$. $f_{\rm det}$ is based on 645 SNe (no absorption) and 350 (with absorption), including NuSTAR data in 10--78~keV energy range.}}
\end{deluxetable*}

At the other extreme, the $B_0,P_0$--distributions from \cite{2006Kaspi} and \cite{2015Gullon} are always consistent with our limits within the 68.3~\% confidence interval, even in the extreme case when no ejecta absorption is considered. Individual objects with a negligible amount of absorption may be present in our sample due to favorable viewing angles in highly asymmetric explosions, or due to ionization of the ejecta by the pulsar/PWN \citep{2014Metzger}, but this is extremely unlikely to hold for the sample as a whole. The birth distributions from \cite{2006Kaspi} and \cite{2015Gullon} therefore imply a very low probability for future X-ray detections of pulsars in SNe. Compared to the \cite{2002Arzoumanian} models, the \cite{2006Kaspi} and \cite{2015Gullon} models have much larger $P_{\rm0,mean}$ and wider distributions of $P_{\rm0}$, whereas the   $B_{\rm0}$ distributions are more similar.

For the four remaining birth parameter distributions  \citep{2020Cieslar,2022Igoshev,2024Graber,2024Pardo_Araujo}, we find broadly similar results, with the models being consistent with the observational limits only when ejecta absorption is considered, albeit with some variations depending on  $\eta$ and $f_{\rm np}$ (see Table~\ref{tab:pulsar_results} for details). The most important difference between these models is that the distribution from \cite{2020Cieslar} predicts a higher number of detections due to the lower  $P_{\rm0,mean}$  of 50~ms, which reduces the compatibility with the observations. The other three models all have $P_{\rm0,mean} \gtrsim 100$~ms, which gives predictions that are consistent with the observations even at the 68.3~\%  level.

The birth distributions we have explored have been derived using different observational samples, different models for pulsar emission and evolution, as well as different methods for inferring the optimal birth distributions. Most of the studies are based on large samples ($\sim$500--2500) of isolated Galactic radio pulsars \citep{2002Arzoumanian,2006Kaspi,2015Gullon,2020Cieslar,2024Graber,2024Pardo_Araujo}, where \cite{2015Gullon} also accounts for magnetars and pulsars with thermal X-ray emission, while the study by \cite{2022Igoshev} is based on 56 objects associated with SNRs. It is notable that the birth distributions derived from these studies have similar distributions of $B_{\rm0}$, while $P_{\rm0}$ exhibit considerably greater variability, both in terms of peak values and the shape of the distribution. The latter is harder to determine as the information about $P_{\rm 0}$ is lost during the pulsar evolution. As illustrated by our results, the limits on X-ray emission from young pulsars help to constrain the $P_0$ distribution, with the observations favoring the models with $P_{\rm 0, mean} \gtrsim 100$~ms.
There are a number of known pulsars in core-collapse SNRs with short spin periods, including the Crab pulsar and PSR~0540, which have $P=33$ and $51$~ms, respectively, with corresponding inferred birth periods of $P_0 \sim 17$ and $32$~ms \citep{Lyne2015,Marshall2016,Tenhu2024}. The existence of pulsars with such low $P_0$ is consistent with the population synthesis results, but indicate that these pulsars are at the low end of the $P_0$ distributions and do not represent the most common pulsars created in SNe.

In terms of predicting X-ray luminosities from these pulsar populations, it is clear that accounting for ejecta absorption has a major impact (Figure~\ref{fig:pulsar_cumu_dist_hist} and Table~\ref{tab:pulsar_results}).
Our absorption model (Section~\ref{sec:ejecta_models}) is based on 3D simulations of neutrino-driven explosions, which have been shown to be able to reproduce the main asymmetries observed in SN~1987A and Cas~A \citep{2019Alp,2020Jerkstrand}. However, our approach for accounting for the absorption is clearly simplified in that we only consider three different progenitors (which are scaled to typical ejecta masses and explosion energies for different SN classes) and three lines of sight through the asymmetric ejecta. These assumptions imply that we most likely underestimate the diversity of absorption properties within our sample. The uncertain classification for the oldest objects in our sample, which are all classified as type IIs, also imply that we have likely overestimated the level of absorption in some cases.

In order to explore the effects of the absorption, we perform additional simulations for the last four models in Table~\ref{tab:pulsar_results}, where we leave the level of absorption fixed at the 10th, 50th, and 90th percentiles, respectively, instead of sampling them with 20-60-20$\%$ probability. 
The results in terms of model compatibility are as follows: The 50th percentile results are very close to the models with varying absorption in Table~\ref{tab:pulsar_results}. 
The 90th percentile results typically improve the agreement by one step (e.g., going from 99.7 to 95.5~\%) for models that were not already fully consistent with the limits. Assuming absorption at the 10th percentile level has the opposite effect on the distributions from \cite{2020Cieslar} and \cite{2022Igoshev}, while the other distributions remain fully consistent with the observations. We thus conclude that the main results and overall trends regarding the birth-parameter distributions in Table~\ref{tab:pulsar_results} hold even when allowing for relatively large variations in ejecta absorption. However, as noted above, the greater variations expected in practice, including cases with extremely low absorption, means that future X-ray detections of pulsars/PWN are plausible for many of the birth distributions considered. 

Other uncertainties in our populations synthesis include the conversion efficiency of spin-down power to X-ray emission, which has only been measured for older systems, as discussed above, as well as the simplified vacuum-dipole model adopted for the pulsars. We also note that it is plausible that the pulsar samples used for these population syntheses are not fully representative of the distribution of pulsars in our sample, which were all accompanied by regular SNe, without any exotic, extremely luminous or underluminous events. 

Additional X-ray observations of $\sim$decades old nearby SNe are needed to assess the various model assumptions and better constrain the pulsar properties. In this context, we note that NuSTAR observations are especially interesting since the level of ejecta absorption in the hard X-ray band is lower, which should provide more tightly constrained pulsar luminosity limits at earlier times, and in turn, more stringent constraints on the population synthesis results. However, after incorporating 50 NuSTAR observations into our analysis, we find only small differences in the outcome compared to the 2--10~keV results from XMM, Chandra, and Swift. Specifically, we find improved constraints corresponding to few steps in the confidence levels for a few combinations of model parameters, but in the majority of models the $f_{\rm det}$ changes by no more than $\sim1~\%$. We attribute this primarily to the limited number of NuSTAR observations at early epochs (see Figure~\ref{fig:COlum_nustar}), which are insufficient to produce a significant impact on the results.

\subsection{Compact Object in SN~1979C}\label{sec:detected_CO_discussion}
As presented in Section~\ref{sec:result_CO}, we find that SN~1979C exhibits signs of an additional power-law component when assuming negligible ejecta absorption. It is detected in two out of six observations by Chandra at epochs 9802 \& 11991~days post-explosion.
Previous studies have suggested that SN~1979C may host either a PWN or a BH \citep{2005Immler,2011Patnaude}. \cite{2005Immler1979C} presented luminosities and limits in the 0.3--2~keV energy interval from an early observation at 0.7~yrs and then between $\sim16$--23~yrs post-explosion, where the upper limit at 0.7~yrs is $<4.47\times10^{39}~\rm erg~s^{-1}$ and the later observations show a constant luminosity at $\sim 8\times 10^{38}~\rm erg~s^{-1}$. \cite{2005Immler1979C} also presented a more detailed analysis of the XMM observation at $\sim$ 23~yrs (7731~days). They found that a two-temperature thermal plasma emission model ($kT$=$0.78_{-0.17}^{+0.25}$~keV and $kT$=$4.1_{-2.3}^{+76}$~keV) provided the best fit to the spectrum, which they interpreted as CSM interaction. However, they could not rule out alternative models including a single thermal bremsstrahlung spectrum with $kT=0.53_{-0.33}^{+1.40}$~keV or a single power-law model with $\Gamma=4.48_{-0.48}^{+0.5}$. 
We have not considered this XMM observation in our study as the SN position is on a chip gap in the EPIC-pn image and we are not using MOS data.

The earliest observation included in our study is a Chandra observation (obsid 6727, exposure time 38 ks and epoch 9802~days), which was also analyzed in \cite{2005Immler1979C}. We find that a two-component model, consisting of a thermal plasma and a power-law component, provides a significantly better fit compared to a single thermal plasma model (see Figure~\ref{fig:fit_1979C} and Table~\ref{tab:fit_1979C}). This is also true for the only other long exposure (obsid 14230, exposure time 79 ks and epoch 11991~days). For both observations, we find a plasma temperature around $\sim1~$keV and a power-law index of $\sim$2.2 (precise values are found in Table~\ref{tab:fit_1979C}). These results suggest the presence of an additional emission component, potentially associated with a CO. All other observations of this SN in our sample have much shorter exposure times ($<15$~ks), which limits the possibility to significantly detect such a component, although a single CSM component is always detected (see Tables~\ref{tab:fit_1979C} \&\ref{tab:detected_SN}). 

A more recent study by \cite{2011Patnaude} proposed that the CO in SN~1979C is an accreting BH. They argue that the steady light curve and presence of a hard spectral component support this interpretation. Our analysis covers later epochs than \cite{2011Patnaude}, sharing only the first two Chandra observations of our sample at 9802 and 10594~days. We find that the luminosity of the power-law component remains approximately constant throughout the later epochs (20--40~yrs after explosion), while we find a decreasing trend for the thermal plasma component. 
For the Chandra observations included in both studies, we find good agreement in the best-fit parameters. In particular, \cite{2011Patnaude} obtained $kT=1.1_{-0.12}^{+0.14}$~keV and $\Gamma = 2.2_{-0.4}^{+0.3}$ for the observation at 9802~days, which is consistent with our results (see second section of Table~\ref{tab:fit_1979C}). Although \cite{2011Patnaude} argue that the additional component indicates an accreting BH, they also note that a PWN cannot be statistically ruled out based on the spectral fit. 

\begin{figure}[t]
\includegraphics[width=\columnwidth]{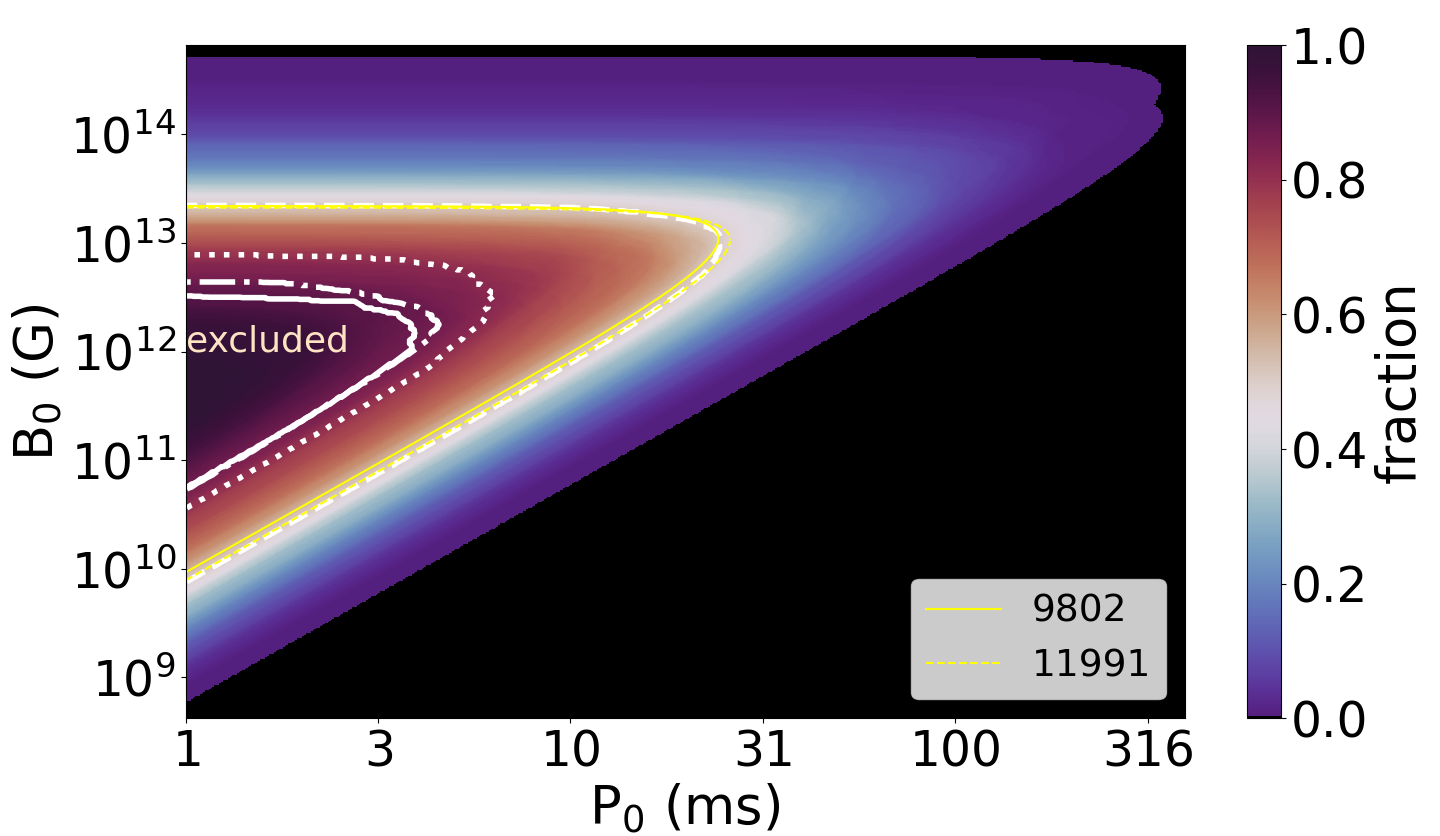}
\caption{\footnotesize{Pulsar birth parameters for SN~1979C compared to limits from the full sample. The yellow lines mark the values derived from the observations of SN~1979C, where the label indicates the epoch. The color gradient is based on $L_{\rm CO,0,abs}$ and demonstrates the fraction of the sample excluded at each point of the parameter space. Each white line (dashed, dotted, dotted-dashed \& solid) represents the contour at 0.5 for different lines of sight with increasing levels of ejecta absorption (0, 10th, 50th and 90th percentiles). The parameter constraints were derived using the model from Section~\ref{sec:pulsar_model} with the Crab $\eta$. The most constraining luminosity limit for each SN was used to generate both the color gradient and the contour lines.} \label{fig:B_P}}
\end{figure}

We further explore the possibility that the hard spectral component in SN~1979C originates from a pulsar/PWN. The two yellow lines in Figure~\ref{fig:B_P} show the pulsar birth parameter combinations ($P_0$, $B_0$) obtained from $L_{\rm CO,0,abs}$ of the two aforementioned observations, at 9802~days (yellow solid line) and 11991~days (yellow dashed line). The parameters were derived assuming the pulsar model outlined in Section~\ref{sec:pulsar_model} with $\eta$ from the Crab.  The constraints for SN~1979C are plotted on top of a birth-parameter exclusion map derived from the 3$\sigma$ upper luminosity limits for the full sample. The color gradient is based on limits in the scenario without ejecta absorption, illustrating the fraction of the parameter space excluded at each point, while the white contour lines illustrate how the results change when considering absorption. 

The figure shows that the lines for SN~1979C are at $\sim$50\% of the excluded region of the $B_0$--$P_0$ space without ejecta absorption (exclusion fraction is $\sim0.5$). This indicates that the properties of the possible pulsar in SN~1979C are rather unusual. Alternatively, in the more realistic scenario where ejecta absorption is accounted for, we find that pulsars with similar properties as inferred for SN~1979C may be present in the vast majority of the sample. In this case, the putative pulsar in SN~1979C is instead exceptional because it is not affected by ejecta absorption. 

In order to investigate the time evolution in the pulsar scenario, we take  the values $\log B_0\approx13.1~$G and $P_0\approx21.7~$ms, which correspond to the intersection of the  lines derived from the two observations in Figure~\ref{fig:B_P}. We stress that this should just be considered an illustrative example, as a wide range of $B_0$ and $P_0$ are possible when considering the statistical uncertainties. We find that the expected luminosity of this pulsar is consistent with the upper limit at 0.7~yr if we add a small amount of absorption at early times, corresponding to $\rm nH=3.6(2.8)\times10^{20}~\rm cm^{-2}$. Furthermore, since $\tau\propto t^{-2}$, we find that the remaining ejecta absorption at the time of the last observation (11991~days) is negligible ($\rm nH=1.62\times10^{17}~\rm cm^{-2}$). This pulsar model is thus consistent with all the observational constraints, including the low absorption and slow time evolution at late times. For comparison, \cite{2011Patnaude} exclude a magnetar model in SN~1979C based on the upper limit at 0.7~yrs and the subsequent steady evolution, but do no consider the effects of ejecta absorption.

Although our results suggest that the CO in SN~1979C is consistent with being a pulsar, we also note that the spectra are equally well fit with a double plasma model (\texttt{mekal+mekal} instead of \texttt{mekal+pow}), as the cstat/dof values are similar for both models (Table~\ref{tab:fit_1979C}). For the observation at 11991~days, the double plasma model gives $kT=0.83^{+0.11}_{-0.13}$~keV and $kT=2.20^{+1.21}_{-0.57}$~keV, which is consistent with the results of \cite{2005Immler}. We therefore cannot exclude that the hard component is due to CSM interaction. The scenario of an accreting black hole suggested by \cite{2011Patnaude} also remains a possibility.  

\subsection{Detected SNe and Late Time CSM Interaction}\label{sec:mass_loss_rate}

Since this study focuses on detecting or constraining CO emission, we have excluded SN types and epochs known to be dominated by interaction. Nevertheless, 12 SN were detected (seven at multiple epochs) with X-ray spectra well described by a thermal plasma model, likely due to late-time interaction between ejecta and CSM or ISM (we include SN~1979C among these 12 since it also has a clear thermal component). Six of these 12 SNe are of type II, detected between 8063~days ($\sim22~$yrs) to 22333~days ($\sim61~$yrs) after explosion, and the remaining six are different kinds of stripped types, detected between 1753~days ($\sim5~$yrs) to 11313~days ($\sim31~$yrs) after explosion. Four of the detected SNe have, to our knowledge, not previously been detected in X-rays (SN~1982R, SN~1984J, SN~1992bu and SN~2003gk). 

In the bottom panel of Figure~\ref{fig:obsflux}, we show the time evolution of the luminosity of the detected SNe, and similarly the temperature evolution in Figure~\ref{fig:kT_evol}. For CSM interaction with a steady wind medium, we expect a luminosity and temperature decrease over time following $L_{\rm X}\propto t^{-1}$, or somewhat flatter for the more narrow energy band that Chandra and XMM observes in \citep{1996Fransson, 2012Dwarkadas}. We confirm this overall decreasing trend for both the sample as a whole and for individual SNe with multiple observations (the 0.5--10~keV luminosities show the same time evolution as the 2--10~keV light curves plotted in Figure~\ref{fig:obsflux}, just shifted to higher values). However, we cannot draw firm conclusions regarding details of the time evolution due to low statistics in the spectra. 

To explore the nature of the late time interaction, we estimate the mass-loss rates of the progenitor stars based on the X-ray luminosities in the 0.5--10~keV energy range, and compare the results with expected values for different types of SNe and their corresponding progenitors. The mass-loss rate is derived using the following relation given by \cite{2017Chevalier}:

 \begin{equation}
    Lx\approx 3\times 10^{39}g_{ff}C_n\left(\frac{\dot{M_{-5}}}{v_{w1}}\right)^2t_{10}^{-1},
\end{equation}
where $C_n=(n-3)(n-4)^2/[4(n-2)]$ is for the reverse shock, where we assume $\rm n=10$, $g_{ff}$ is the gaunt factor of order unity, and $\dot{M_{-5}}$ is the mass-loss rate in units of $10^{-5}~M_\Sun~\rm yr^{-1}$. The wind velocity, $v_{1w}$, is in units of 10~km~s$^{-1}$, which is typical for type II SN progenitors, whereas $v_{1w} \sim$1000~km~s$^{-1}$ is expected for stripped type progenitors \citep{2014Smith,2017Chevalier}. Finally, $t_{10}$ is time in units of 10 days. 

With this simplified calculation we find mass-loss rates for the progenitors of the type II SNe in the range $7\times10^{-6}\lesssim\dot{M}\lesssim5\times10^{-5}~M_\Sun~\rm yr^{-1}$, which are typical values for RSG progenitors \citep{2014Smith,2017Chevalier}. The luminosities for these SNe are similar to those presented by \cite{2020Ramakrishnan}, who studied late time X-ray emission from CSM interaction in a  sample of type II SNe. For the SNe that overlap between our samples (SN~1941C, SN~1968D and SN~1970G), we find good agreement between our 0.5--8.0~keV luminosities and the 0.3--8~keV luminosities in \cite{2020Ramakrishnan}. Similarly, we discussed in Section~\ref{sec:result_obsflux} that our luminosity estimates for SN~1980K are consistent with previous studies \citep{2008Fridriksson,2008Soria}. Furthermore, the mass-loss estimate for SN~1980K by \cite{2008Soria} ($\dot{m}\approx 3\times10^{-5}~M_\Sun~\rm yr^{-1}$) is close to our estimates: $0.7-1.8\times10^{-5}~M_\Sun~\rm yr^{-1}$.

For the stripped type SNe in our sample, we find mass-loss rates of $4\times10^{-4}\lesssim\dot{M}\lesssim2\times10^{-2}~M_\Sun~\rm yr^{-1}$. These values are higher than expected for Wolf Rayet progenitors \citep{2014Smith,2017Chevalier}, and for some SNe (e.g SN~1992bu, SN~2003gk and SN~2009jf) the estimated $\dot{M_{-5}}$ are even in the range of interacting SNe ($\dot{M}_{-5}>10^{-2}$ ). These results could be affected by incorrectly classified SNe. In particular, we note that SN~1992bu is not spectroscopically classified \citep{2011RomeroCanizales}. SN~2009jf and SN~2003gk (listed as type Ib in Table~\ref{tab:detected_SN}) have also been discussed as type Ib/c and Ic respectively \citep{2017GalYam,2014Bietenholz}, though this won't affect our inferred mass-loss rates. 

Another possible scenario is that these SNe are evolving into interacting or SN~2014C-like events \citep{2014Corsi,2017Anderson,2022Brethauer}, where instead of a steady wind, the progenitor may have experienced eruptive mass loss, creating a denser shell of material that the ejecta now interact with. This phenomenon has been seen for other stripped type SNe such as SN~1993J, SN~2001em and SN~2004dk \citep{2009Chandra,2009Schinzel,2014Dwarkadas,2019Pooley,2020Chandra}. In line with this, such a scenario has previously been discussed based on radio observations of SN~2003gk \citep{2014Bietenholz}, which is in our sample, while a study of early radio and X-ray emission for the type IIb SN~2013df, which we also detect at late times, has revealed  properties similar to SN~1993J \citep{2016Kamble}. Such extreme mass loss is common for luminous blue variables, which are known to be the progenitors of type IIn SNe \citep{2014Smith}. Alternatively, this denser CSM shell structure could have been created via binary interaction which is also viewed as a potential progenitor system for stripped SNe. 

It is not surprising that many of the interacting SNe we detect show signs of unusually high mass loss, given that we detected them at such late times. We stress, however, that the range of $\dot{M_{-5}}$  we derive should only be seen as indicative, given the many simplifying assumptions, including a steady wind, only a reverse shock contribution, and a fixed cooling function.

\section{Summary \& Conclusion}\label{sec:summary_conclusion}
We have presented an extensive search for late-time X-ray emission from 242 SNe within 60~Mpc. The full sample is based on 607 observations from the Chandra, XMM, Swift and NuSTAR telescopes. 
Our primary goal was to constrain any emission from COs, which is expected to emerge as hard X-ray components years to decades after the explosions when the absorption by ejecta has decreased. We therefore only included observations obtained at epochs $>$ 2 and 20~yrs post-explosion for stripped (IIb, Ib, Ic) and type II SNe, respectively, though earlier observations were included for NuSTAR due to the lower absorption in the hard X-ray band. 
By using the results of asymmetric ejecta absorption models from \cite{2018Alp}, which are based on 3D simulations of neutrino-driven explosions \citep{2015Wongwathanarat}, we placed conservative constraints on the X-ray luminosities of COs within these SNe. 
 We also performed a population synthesis based on the results to place constraints on the birth-parameter distribution of pulsars. Finally, we obtained simple estimates of the mass-loss rates of the progenitor stars of the detected SNe that showed signs of CSM interaction. Our main findings are as follows:

\begin{enumerate}

\item We detected 12 SNe, half of which are classified as different stripped types (SN~1982R, SN~1992bu, SN~2003gk, SN~2009jf, SN~2011dh, and SN~2013df) and the other half as type II (SN~1941C, SN~1968D, SN~1970G, SN~1979C, SN~1980K, and SN~1984J). Of these 12, four have not previously been detected in X-rays (SN 1982R, SN 1984J, SN 1992bu and SN 2003gk), while we detect several of the others at later epochs than in previous studies. The oldest detected SN in our sample is SN~1941C at approximately 60~yrs post-explosion. The X-ray emission in all these SNe, with the possible exception of SN~1979C, is consistent with being dominated by CSM interaction. 

\item We detected a hard X-ray component in SN~1979C, modeled by a power law with $\Gamma \sim 2.2$, which we interpret as potential emission from a CO. This component is detected in the two longest Chandra observations (at $\sim$27~yrs and $\sim$32~yrs post-explosion), but only when assuming negligible ejecta absorption. We find that a model where the emission originates from a pulsar/PWN is consistent with the time evolution at late times, as well as with previous measurements at earlier epochs \citep{2005Immler1979C, 2011Patnaude}. However, we note that the spectra are equally well fitted with a double plasma model, implying that the observed emission could alternatively arise from ejecta interaction with the CSM or ISM. The hard component may also be due to an accreting BH, as previously suggested by \cite{2011Patnaude}.
  
\item 
A comparison of population synthesis results based on seven different pulsar birth distributions derived from studies of Galactic pulsars illustrates that X-ray limits add important constraints on the population of pulsars created in SNe. We found that models with mean initial spin periods longer than $\sim100$~ms are all compatible with the X-ray limits, given expected levels of ejecta absorption, while the compatibility of models with somewhat shorter spin periods ($50~{\rm ms} \lesssim P_{\rm 0, mean}\lesssim100~\rm ms$) depend on the fraction of SNe that produce pulsars and assumptions about the conversion efficiency of spin-down power to X-rays. Finally, birth distributions with even shorter spin periods, including the one from \cite{2002Arzoumanian}, are inconsistent with the X-ray limits. This holds even when allowing for up to 40\% of SNe not producing pulsars, but instead BHs or thermally emitting NSs that would not be detected in our study.  

\item For the 12 detected SNe, we estimated the mass-loss rates of their progenitor stars, assuming that the X-ray emission originates from interaction with a wind medium. We found mass-loss rates that are at the high end of typical values for the winds of the assumed progenitors, or much higher for some of the stripped SNe (especially  SN 1992bu, SN 2003gk and SN 2009jf). This indicates that the progenitors have had episodic or eruptive mass losses, forming denser CSM shells which enhance the X-ray emission. Alternatively, binary interaction could have contributed to denser CSM environments. However, the details of the CSM interaction in these SNe are uncertain given the limited statistics in the spectra and simplified models.

\end{enumerate}

\begin{acknowledgments}
This research has made use of data from the Chandra Data Archive and Chandra Source Catalog, both provided by the Chandra X-ray Center (CXC) and employs a list of Chandra datasets, obtained by the Chandra X-ray Observatory, contained in the Chandra Data Collection (CDC) ~\dataset[doi:10.25574/cdc.435]{https://doi.org/10.25574/cdc.435}.This research has made use of software provided by the Chandra X-ray Center (CXC) in the application package CIAO. Based on observations obtained with XMM-Newton, an ESA science mission with instruments and contributions directly funded by ESA Member States and NASA. This research has made use of software provided by the High Energy Astrophysics Science Archive Research Center (HEASARC), which is a service of the Astrophysics Science Division at NASA/GSFC. This work made use of data supplied by the UK Swift Science Data Centre at the University of Leicester. This research has made use of data from the NuSTAR mission, a project led by the California Institute of Technology, managed by the Jet Propulsion Laboratory, and funded by the National Aeronautics and Space Administration. Data analysis was performed using the NuSTAR Data Analysis Software (NuSTARDAS), jointly developed by the ASI Science Data Center (SSDC, Italy) and the California Institute of Technology (Caltech, USA).

\facilities{CXO(ACIS), XMM-Newton(EPIC-PN), Swift(XRT), NuSTAR}

\software{CIAO/CALDB \citep{2006SCIAO},
          HEAsoft \citep{2014Nasa},
          XRT Products \citep{2009Evans,2020Evansswift},
          SAOImage DS9 \citep{2003SAOImage},
          \texttt{astropy} \citep{2013astropy,2018astropy,2022astropy},
          \texttt{numpy} \citep{2020numpy},
          \texttt{scipy} \citep{2020SciPy},
          XSPEC \citep{1996Arnaud}
          }
\end{acknowledgments}

\bibliography{ref}{}
\bibliographystyle{aasjournal}

\appendix
\renewcommand{\thetable}{A.\arabic{table}}
\renewcommand{\thefigure}{A.\arabic{figure}} 
\setcounter{table}{0}
\setcounter{figure}{0}

\section{Detected supernovae}\label{sec:appendix_A}
In Table~\ref{tab:detected_SN} we present the results of all observations resulting in detections of SNe emission.

\movetabledown=5mm
\begin{longrotatetable} 
\tabletypesize{\footnotesize}
\startlongtable
\begin{deluxetable}{l l l l l l l l l l l l}
\tablecaption{Observations and Best-fit Parameters of Detected SNe \label{tab:detected_SN}}
\tablewidth{1pt}
\tablehead{ 
\colhead{Name} & \colhead{Type} & \colhead{Dist}  & \colhead{Obsid} &Exp. & \colhead{Obs. Date}& \colhead{Epoch}  & \colhead{$kT$} & \colhead{$\rm F_{\rm SN, 0.5-10~keV}$} & \colhead{$\rm L_{\rm SN, 0.5-10~keV}$}& \colhead{$\rm F_{\rm SN, 2-10~keV}$} & \colhead{$\rm L_{\rm SN, 2-10~keV}$} \\ 
\colhead{} & \colhead{} & \colhead{(Mpc)} & \colhead{} & \colhead{(ks)}& \colhead{} & \colhead{(Days)}  & \colhead{(keV)} & \colhead{($10^{-15}\rm erg~s^{-1}~cm^{-2}$)} & \colhead{($10^{37}\rm erg~s^{-1}$)} & \colhead{($10^{-15}\rm erg~s^{-1}~cm^{-2}$)} & \colhead{($10^{37}\rm erg~s^{-1}$)}
 }
\startdata
1941C &
  II &
  9.0 &
  $2920^{\rm C,S3}$ &
  18.52 &
  2002-03-07 &
  22240 &
  $3.7_{-2.0}$ &
  $5.0_{-2.5}^{+4.6}$ &
  $4.9_{-2.4}^{+4.5}$ &
  $2.8_{-1.9}^{+3.9}$ &
  $2.7_{-1.8}^{+3.8}$ \\
 &
   &
   &
  $2921^{\rm C,S3}$ &
  19.71 &
  2002-06-08 &
  22333 &
  $3.7_{-2.0}$ &
  $4.8_{-2.4}^{+4.8}$ &
  $4.7_{-2.3}^{+4.7}$ &
  $2.7_{-1.8}^{+4.1}$ &
  $2.6_{-1.8}^{+4.0}$ \\
  1968D &
  II &
  4.7 &
  $1043^{\rm C,S3}$ &
  58.29 &
  2001-09-07 &
  12244 &
  $7.5_{-5.1}$ &
  $8.7_{-3.6}^{+3.6}$ &
  $2.30_{-0.96}^{+0.95}$ &
  $6.0_{-3.5}^{+2.8}$ &
  $1.58_{-0.93}^{+0.73}$ \\
 &
   &
   &
  $4404^{\rm C,S3}$ &
  29.76 &
  2002-11-25 &
  12688 &
  $1.00_{-0.45}^{+0.38}$ &
  $6.2_{-2.4}^{+2.7}$ &
  $1.63_{-0.62}^{+0.70}$ &
  $0.59_{-0.48}^{+0.77}$ &
  $0.15_{-0.13}^{+0.20}$ \\
 &
   &
   &
  $4631^{\rm C,S3}$ &
  29.66 &
  2004-10-22 &
  13385 &
  $1.6_{-0.4}^{+1.2}$ &
  $3.9_{-1.8}^{+2.5}$ &
  $1.03_{-0.47}^{+0.66}$ &
  $1.08_{-0.67}^{+1.79}$ &
  $0.29_{-0.18}^{+0.47}$ \\
 &
   &
   &
  $4632^{\rm C,S3}$ &
  27.29 &
  2004-11-06 &
  13400 &
  $1.6_{-0.4}^{+1.2}$ &
  $5.7_{-2.3}^{+3.0}$ &
  $1.50_{-0.61}^{+0.81}$ &
  $1.57_{-0.91}^{+2.35}$ &
  $0.41_{-0.24}^{+0.62}$ \\
 &
   &
   &
  $4633^{\rm C,S3}$ &
  26.62 &
  2004-12-03 &
  13427 &
  $1.6_{-0.4}^{+1.2}$ &
  $6.0_{-2.5}^{+3.2}$ &
  $1.61_{-0.65}^{+0.83}$ &
  $1.7_{-1.0}^{+2.4}$ &
  $0.45_{-0.27}^{+0.64}$ \\
1970G &
  II &
  7.0 &
  $5340^{\rm C,S2}$ &
  54.42 &
  2004-07-08 &
  12397 &
  $0.65_{-0.17}^{+0.20}$ &
  $3.2_{-1.0}^{+1.3}$ &
  $1.87_{-0.64}^{+0.76}$ &
  $1.7_{-1.4}^{+19.9}\times10^{-2}$ &
  $1.0_{-0.8}^{+11.7}\times10^{-2}$ \\
 &
   &
   &
  $4734^{\rm C,S2}$ &
  35.48 &
  2004-07-11 &
  12400 &
  $0.65_{-0.17}^{+0.20}$ &
  $3.8_{-1.3}^{+1.6}$ &
  $2.21_{-0.74}^{+0.94}$ &
  $0.14_{-0.10}^{+0.12}$ &
  $8.0_{-6.0}^{+7.3}\times 10^{-2}$ \\
 &
   &
   &
  $14341^{\rm C,S3}$ &
  49.08 &
  2011-08-27 &
  15003 &
  $0.46_{-0.16}^{+0.23}$ &
  $2.79_{-0.86}^{+1.21}$ &
  $1.63_{-0.51}^{+0.71}$ &
  $2.5_{-1.9}^{+6.3} \times 10^{-2}$ &
  $1.5_{-1.1}^{+3.7}\times 10^{-2}$ \\
1979C &
  IIL &
  16.0 &
  $6727^{\rm C,S3}$ &
  37.87 &
  2006-02-18 &
  9802 &
  $1.65_{-0.31}^{+0.30}$ &
  $22.2_{-3.1}^{+3.5}$ &
  $67.9_{-9.6}^{+10.7}$ &
  $6.4_{-2.6}^{+2.5}$ &
  $19.5_{-7.9}^{+7.6}$ \\
 &
   &
   &
  $9121^{\rm C,S3}$ &
  14.92 &
  2008-04-20 &
  10594 &
  $1.29_{-0.34}^{+0.42}$ &
  $21.3_{-4.4}^{+5.0}$ &
  $65.3_{-13.5}^{+15.4}$ &
  $3.9_{-2.4}^{+3.9}$ &
  $11.9_{-7.3}^{+12.0}$ \\
 &
   &
   &
  $12696^{\rm C,S3}$ &
  14.86 &
  2011-02-24 &
  11634 &
  $1.29_{-0.09}^{+0.08}$ &
  $15.3_{-3.6}^{+4.3}$ &
  $46.9_{-11.1}^{+13.1}$ &
  $2.79_{-0.78}^{+0.97}$ &
  $8.5_{-2.4}^{+2.9}$ \\
 &
   &
   &
  $14230^{\rm C,S3}$ &
  79.03 &
  2012-02-16 &
  11991 &
  $1.29_{-0.09}^{+0.08}$ &
  $18.7_{-1.9}^{+2.0}$ &
  $57.4_{-5.7}^{+6.1}$ &
  $3.42_{-0.62}^{+0.70}$ &
  $10.5_{-1.9}^{+2.2}$ \\
 &
   &
   &
  $23140^{\rm C,S3}$ &
  9.96 &
  2020-02-15 &
  14912 &
  $1.30_{-0.47}^{+0.81}$ &
  $13.6_{-5.6}^{+7.7}$ &
  $41.7_{-17.3}^{+23.6}$ &
  $2.5_{-1.7}^{+3.9}$ &
  $7.7_{-5.1}^{+12.1}$ \\
 &
   &
   &
  $23141^{\rm C,S3}$ &
  9.96 &
  2020-03-13 &
  14939 &
  $1.30_{-0.47}^{+0.81}$ &
  $13.7_{-6.0}^{+8.2}$ &
  $42.0_{-18.3}^{+25.0}$ &
  $2.5_{-1.6}^{+4.4}$ &
  $7.8_{-5.0}^{+13.4}$ \\
1980K &
  II &
  4.7 &
  $4404^{\rm C,S3}$ &
  29.76 &
  2002-11-25 &
  8063 &
  $0.59_{-0.21}^{+0.15}$ &
  $21.1_{-6.0}^{+8.8}$ &
  $5.6_{-1.6}^{+2.3}$ &
  $0.43_{-0.29}^{+0.33}$ &
  $0.11_{-0.08}^{+0.09}$ \\
 &
   &
   &
  $0200670301^{\rm X}$ &
  13.83 &
  2004-06-13 &
  8629 &
  $2.9_{-1.5}$ &
  $19.8_{-9.5}^{+15.4}$ &
  $5.2_{-2.5}^{+4.1}$ &
  $9.8_{-7.1}^{+15.0}$ &
  $2.6_{-1.9}^{+4.0}$ \\
 &
   &
   &
  $4631^{\rm C,S3}$ &
  29.66 &
  2004-10-22 &
  8760 &
  $0.67_{-0.08}^{+0.12}$ &
  $15.8_{-3.7}^{+4.4}$ &
  $4.18_{-0.98}^{+1.16}$ &
  $0.48_{-0.16}^{+0.32}$ &
  $0.13_{-0.04}^{+0.09}$ \\
 &
   &
   &
  $4632^{\rm C,S3}$ &
  27.29 &
  2004-11-06 &
  8775 &
  $0.67_{-0.08}^{+0.12}$ &
  $15.4_{-3.7}^{+4.5}$ &
  $4.07_{-0.99}^{+1.18}$ &
  $0.47_{-0.15}^{+0.32}$ &
  $0.12_{-0.04}^{+0.08}$ \\
 &
   &
   &
  $4633^{\rm C,S3}$ &
  26.62 &
  2004-12-03 &
  8802 &
  $0.67_{-0.08}^{+0.12}$ &
  $14.4_{-3.7}^{+4.4}$ &
  $3.81_{-0.97}^{+1.16}$ &
  $0.44_{-0.15}^{+0.31}$ &
  $0.12_{-0.04}^{+0.08}$ \\
 &
   &
   &
  $0500730201^{\rm X}$ &
  33.32 &
  2007-11-02 &
  9866 &
  $0.69_{-0.11}^{+0.12}$ &
  $16.9_{-4.0}^{+4.5}$ &
  $4.5_{-1.1}^{+1.2}$ &
  $0.55_{-0.22}^{+0.35}$ &
  $0.13_{-0.05}^{+0.09}$ \\
 &
   &
   &
  $0500730101^{\rm X}$ &
  28.41 &
  2007-11-08 &
  9872 &
  $0.69_{-0.11}^{+0.12}$ &
  $11.6_{-3.7}^{+4.3}$ &
  $3.1_{-1.0}^{+1.1}$ &
  $0.38_{-0.16}^{+0.27}$ &
  $9.2_{-3.8}^{+6.5}\times10^{-2}$ \\
 &
   &
   &
  $13435^{\rm C,S3}$ &
  20.44 &
  2012-05-21 &
  11528 &
  $0.53_{-0.36}^{+0.33}$ &
  $17.4_{-8.8}^{+119.6}$ &
  $4.6_{-2.3}^{+31.6}$ &
  $0.25_{-0.25}^{+0.61}$ &
  $6.6_{-6.5}^{+1.6}\times10^{-2}$ \\
 &
   &
   &
  $0691570101^{\rm X}$ &
  114.34 &
  2012-10-21 &
  11681 &
  $0.58_{-0.12}^{+0.07}$ &
  $4.40_{-0.66}^{+0.80}$ &
  $1.19_{-0.18}^{+0.21}$ &
  $0.19_{-0.08}^{+0.13}$ &
  $5.6_{-2.4}^{+3.7}\times10^{-2}$ \\
 &
   &
   &
  $17878^{\rm C,S3}$ &
  40.04 &
  2016-09-28 &
  13119 &
  $0.48_{-0.23}^{+0.22}$ &
  $21.7_{-7.8}^{+27.0}$ &
  $5.7_{-2.1}^{+7.1}$ &
  $0.24_{-0.20}^{+0.34}$ &
  $6.3_{-5.3}^{+8.9}\times10^{-2}$ \\
 &
   &
   &
  $19887^{\rm C,S3}$ &
  18.55 &
  2016-09-28 &
  13119 &
  $0.48_{-0.23}^{+0.22}$ &
  $21.0_{-10.1}^{+34.9}$ &
  $5.6_{-2.7}^{+9.2}$ &
  $0.15_{-0.12}^{+0.43}$ &
  $4.0_{-3.1}^{+11.5}\times10^{-2}$ \\
 &
   &
   &
  $0794581201^{\rm X}$ &
  47.04 &
  2017-06-01 &
  13365 &
  $0.78_{-0.23}^{+0.11}$ &
  $16.2_{-3.4}^{+3.8}$ &
  $4.28_{-0.91}^{+1.02}$ &
  $0.76_{-0.47}^{+0.32}$ &
  $0.20_{-0.13}^{+0.09}$ \\
 &
   &
   &
  $0870830101^{\rm X}$ &
  14.70 &
  2020-07-08 &
  14498 &
  $0.52_{-0.30}^{+0.24}$ &
  $9.9_{-5.0}^{+6.3}$ &
  $2.6_{-1.3}^{+1.7}$ &
  $0.14_{-0.13}^{+0.31}$ &
  $3.7_{-3.6}^{+8.1}\times10^{-2}$ \\
 &
   &
   &
  $0870830301^{\rm X}$ &
  12.45 &
  2021-04-04 &
  14768 &
  $0.52_{-0.30}^{+0.24}$ &
  $11.2_{-4.8}^{+6.0}$ &
  $3.0_{-1.3}^{+1.6}$ &
  $0.16_{-0.15}^{+0.32}$ &
  $4.2_{-4.0}^{+8.5}\times10^{-2}$ \\
1982R &
  Ib &
  18.0 &
  $16487^{\rm C,I3}$ &
  40.51 &
  2013-10-14 &
  11313 &
  $2.4_{-1.5}$ &
  $2.8_{-1.7}^{+3.5}$ &
  $11.0_{-6.7}^{+13.6}$ &
  $1.1_{-1.0}^{+3.2}$ &
  $4.8_{-4.0}^{+12.5}$ \\
1984J &
  II &
  14.0 &
  $16745^{\rm C,G,S3}$ &
  45.40 &
  2016-06-09 &
  11640 &
  $3.0_{-1.0}^{+2.4}$ &
  $13.2_{-3.5}^{+4.7}$ &
  $31.0_{-8.2}^{+11.1}$ &
  $6.7_{-2.7}^{+4.2}$ &
  $15.8_{-6.3}^{+9.9}$ \\
1992bu &
  Ib/c? &
  45.0 &
  $15077^{\rm C,G, S3}$ &
  51.90 &
  2013-03-13 &
  7674 &
  $4.7_{-1.9}$ &
  $18.1_{-7.2}^{+6.2}$ &
  $4.4_{-1.8}^{+1.5} \times 10^{+2}$ &
  $11.2_{-4.7}^{+5.3}$ &
  $2.7_{-1.1}^{+1.3} \times 10^{+2}$ \\
2003gk &
  Ib &
  49.1 &
  $21004^{\rm C,S3}$ &
  14.89 &
  2018-08-21 &
  5530 &
  $3.7_{-0.9}^{+2.0}$ &
  $71.9_{-13.9}^{+17.5}$ &
  $2.07_{-0.40}^{+0.51} \times10^{+3}$ &
  $40.3_{-11.0}^{+15.4}$ &
  $1.16_{-0.32}^{+0.44}\times10^{+3}$ \\
 &
   &
   &
  $22271^{\rm C,S3}$ &
  14.89 &
  2019-08-05 &
  5879 &
  $3.7_{-0.9}^{+2.0}$ &
  $57.2_{-12.5}^{+15.8}$ &
  $1.65_{-0.36}^{+0.46}\times 10^{+3}$ &
  $32.0_{-9.2}^{+13.0}$ &
  $9.2_{-2.7}^{+3.7}\times 10^{+2}$ \\
2009jf &
  Ib &
  31.0 &
  $0824450601^{\rm X,G}$ &
  79.74 &
  2018-05-30 &
  3167 &
  $4.0_{-0.9}^{+1.5}$ &
  $28.0_{-3.8}^{+4.4}$ &
  $3.22_{-0.43}^{+0.51}\times 10^{+2}$ &
  $16.1_{-3.4}^{+4.2}$ &
  $1.85_{-0.40}^{+0.48}\times 10^{+2}$ \\
2011dh &
  IIb &
  7.2583 &
  $19522^{\rm C,I3}$ &
  37.76 &
  2017-03-17 &
  2116 &
  $2.3_{-1.1}$ &
  $3.5_{-2.1}^{+4.1}$ &
  $2.2_{-1.3}^{+2.6}$ &
  $1.4_{-1.1}^{+4.0}$ &
  $0.88_{-0.67}^{2.5}$ \\
 &
   &
   &
  $0830191601^{\rm X,G}$ &
  59.76 &
  2018-06-15 &
  2571 &
  $6.8_{-4.4}$ &
  $8.7_{-3.5}^{+4.7}$ &
  $5.5_{-2.2}^{+3.0}$ &
  $5.8_{-3.4}^{+4.7}$ &
  $3.7_{-2.2}^{+3.0}$ \\
 &
   &
   &
  $20998^{\rm C,S3}$ &
  19.82 &
  2018-08-31 &
  2648 &
  $2.3_{-1.1}$ &
  $5.6_{-3.4}^{+6.0}$ &
  $3.6_{-2.1}^{+3.8}$ &
  $2.3_{-1.6}^{+5.9}$ &
  $1.4_{-1.0}^{+3.7}$ \\
 &
   &
   &
  $0852030101^{\rm X, G}$ &
  70.01 &
  2019-07-11 &
  2962 &
  $3.3_{-2.1}^{+4.9}$ &
  $7.8_{-4.2}^{+3.2}$ &
  $4.9_{-2.6}^{+2.0}$ &
  $4.1_{-3.8}^{+3.2}$ &
  $2.6_{-2.4}^{+2.0}$ \\
 &
   &
   &
  $23472^{\rm C,S3}$ &
  33.62 &
  2020-10-13 &
  3422 &
  $1.7_{-0.8}^{+1.8}$ &
  $3.2_{-1.8}^{+2.7}$ &
  $2.0_{-1.1}^{+1.7}$ &
  $0.98_{-0.76}^{+1.75}$ &
  $0.62_{-0.48}^{+1.10}$ \\
 &
   &
   &
  $23473^{\rm C,S3}$ &
  34.51 &
  2020-11-18 &
  3458 &
  $1.7_{-0.8}^{+1.8}$ &
  $5.8_{-2.7}^{+3.2}$ &
  $3.7_{-1.7}^{+2.0}$ &
  $1.8_{-1.3}^{+2.5}$ &
  $1.10_{-0.82}^{+1.56}$ \\
2013df &
  IIb &
  10.588 &
  $21005^{\rm C,S3}$ &
  21.67 &
  2018-03-26 &
  1753 &
  $5.1_{-3.3}$ &
  $8.5_{-4.9}^{+7.1}$ &
  $11.3_{-6.5}^{+9.6}$ &
  $5.3_{-4.0}^{+5.8}$ &
  $7.1_{-5.3}^{+7.7}$
\enddata
\tablecomments{\footnotesize{C=Chandra and X=XMM, for Chandra observations we further specify the chip on which the source is located. Uncertainties are $90\%$.  *=Previously not detected. G=Prominent galaxy background. $kT$ values given without upper error margins reached the upper limit of 10~keV in the error calculation. All fluxes and luminosities have been corrected for Galactic absorption.}}
\end{deluxetable}
\end{longrotatetable}

\clearpage
\onecolumngrid

\renewcommand{\thetable}{B.\arabic{table}}
\renewcommand{\thefigure}{B.\arabic{figure}}
\setcounter{table}{0} 
\setcounter{figure}{0}

\section{Spectral Fits of Detected SNe}\label{sec:appendix_SN_fits}

\begin{figure*}[ht!]
\gridline{\fig{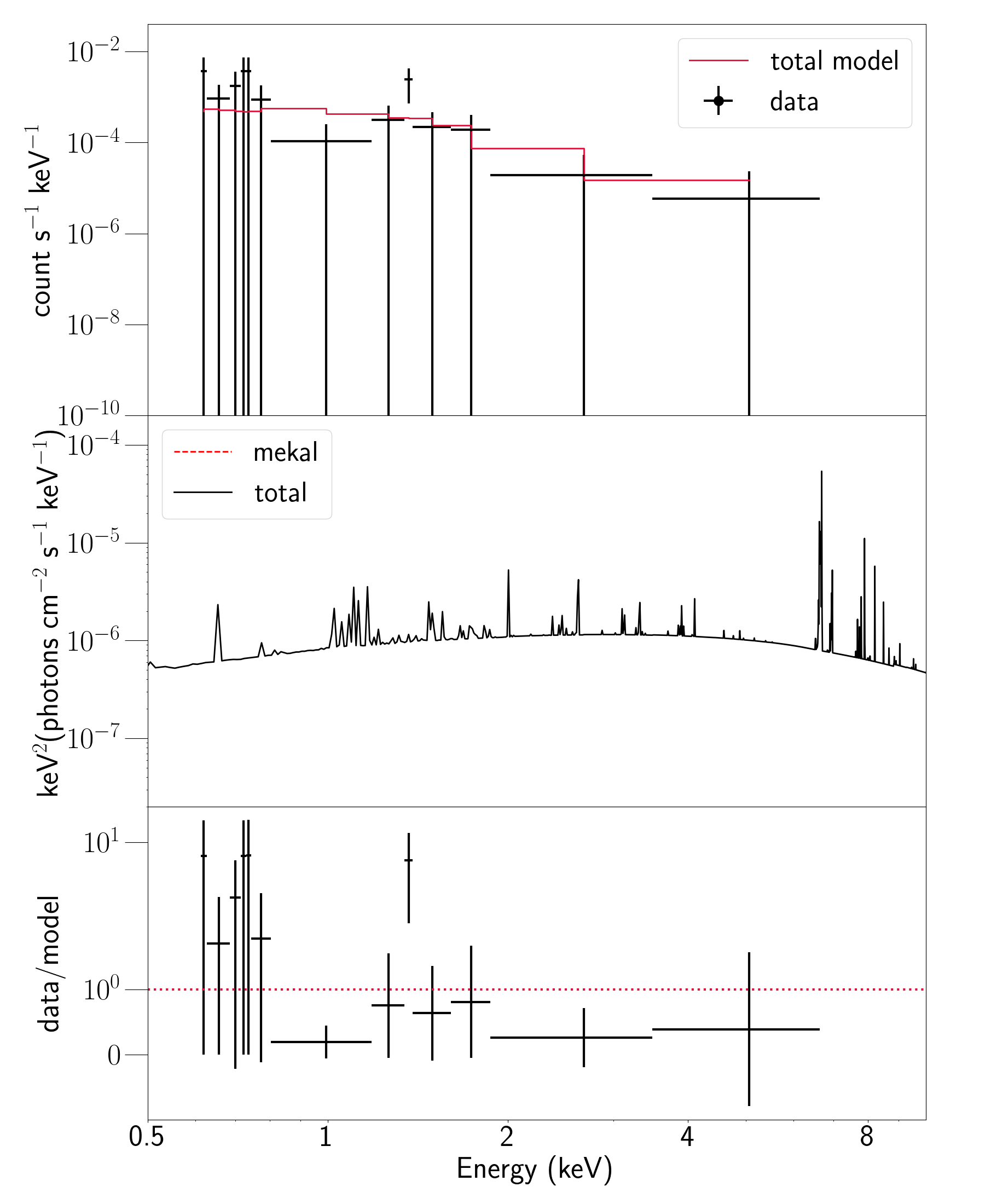}{0.375\textwidth}{(a) SN~1941C, Chandra (S3), epoch--22240d, obsid--2920, binning=1}
          \fig{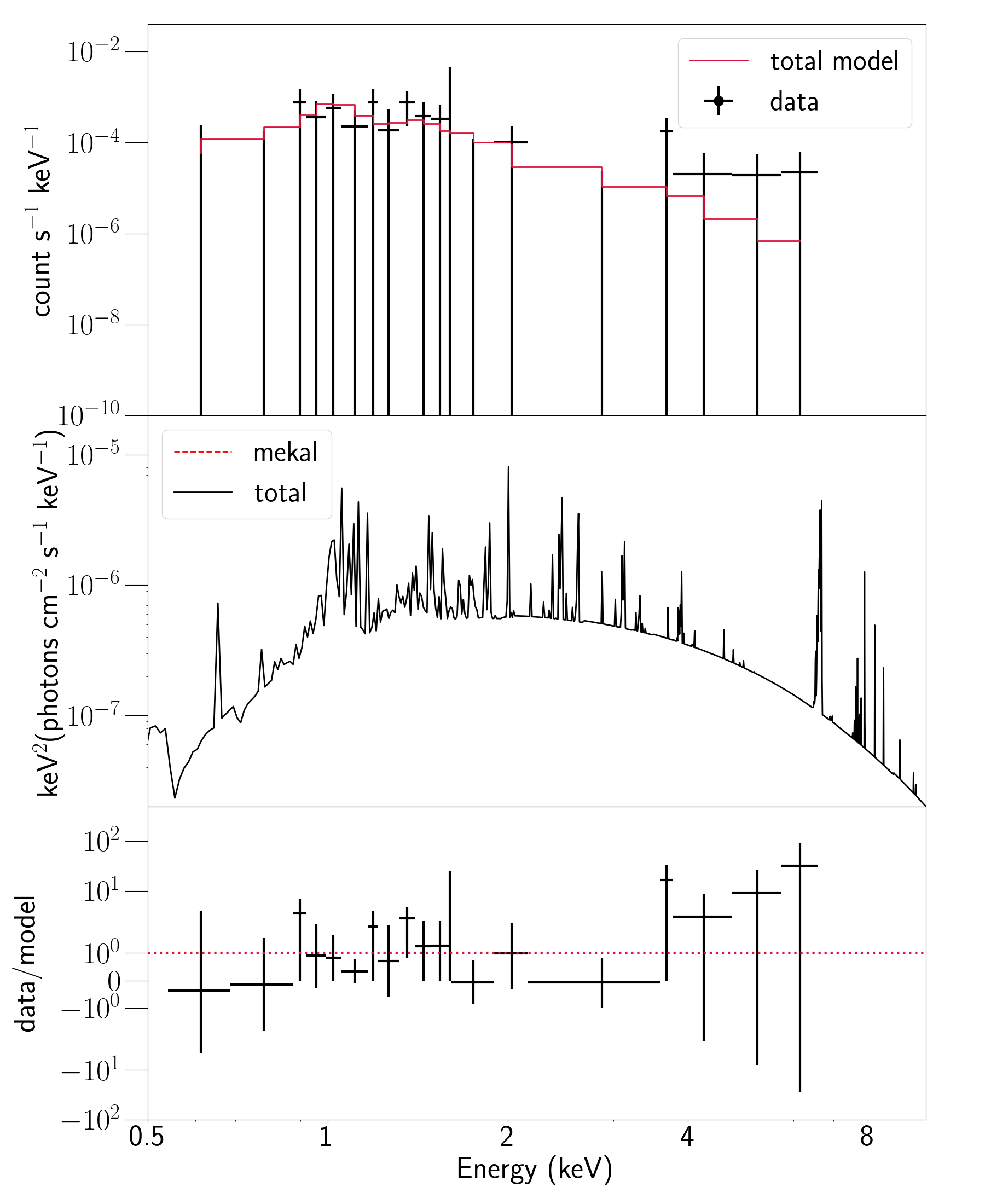}{0.375\textwidth}{(b) SN~1968D, Chandra (S3), epoch--13385d, obsid--4631, binning=1}~
          }
\gridline{\fig{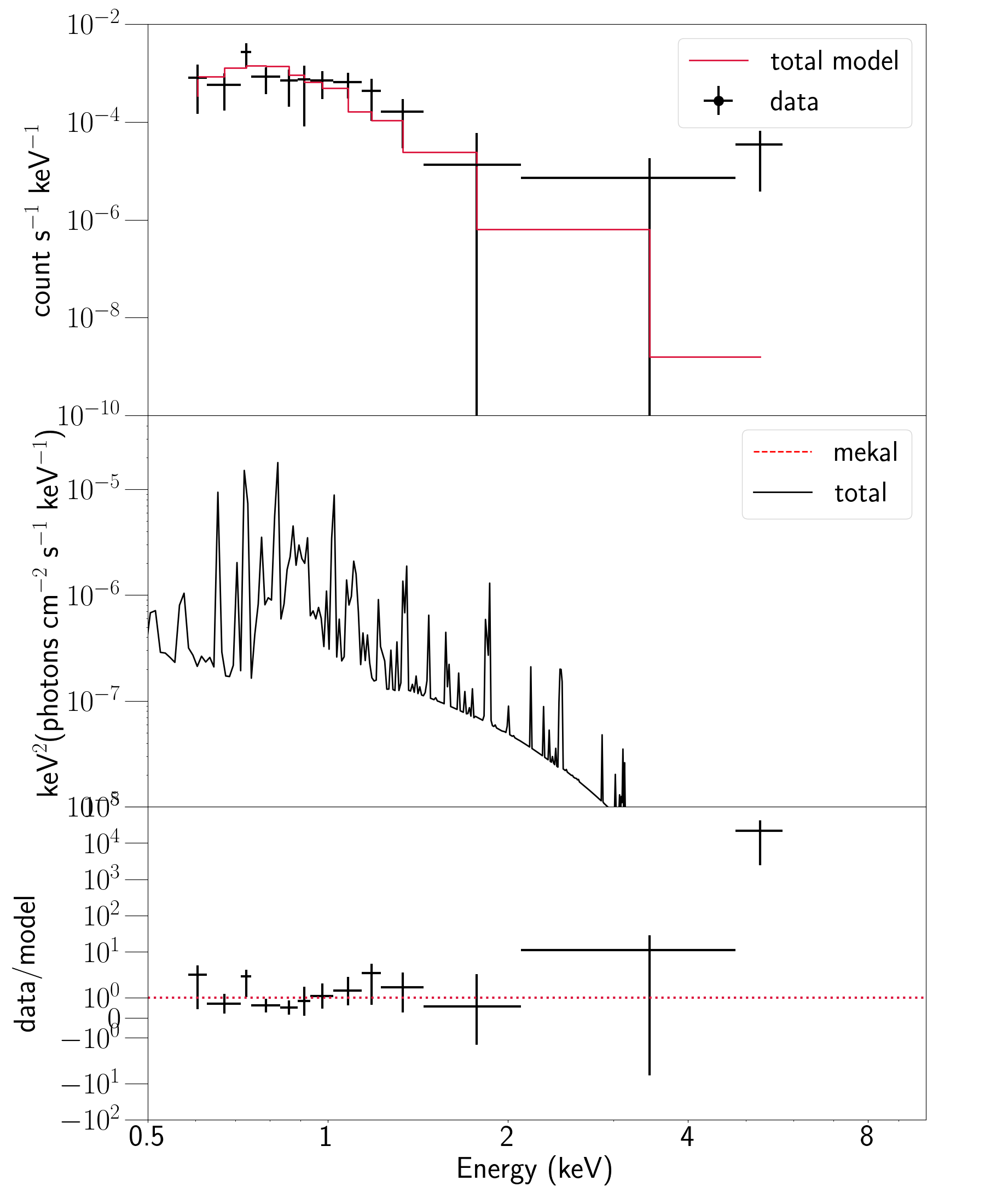}{0.375\textwidth}{(c) SN~1970G, Chandra (S3), epoch--15003d, obsid--14341, binning=2}
          \fig{1979C_14230_none_obsflux.png}{0.375\textwidth}{(d) SN~1979C, Chandra (S3), epoch--11991d, obsid--14230, binning=3}
          }
\caption{\footnotesize{Example fits of the detected SNe. } \label{fig:det_SN_1}}
\end{figure*}

\begin{figure*}[ht!]
\gridline{\fig{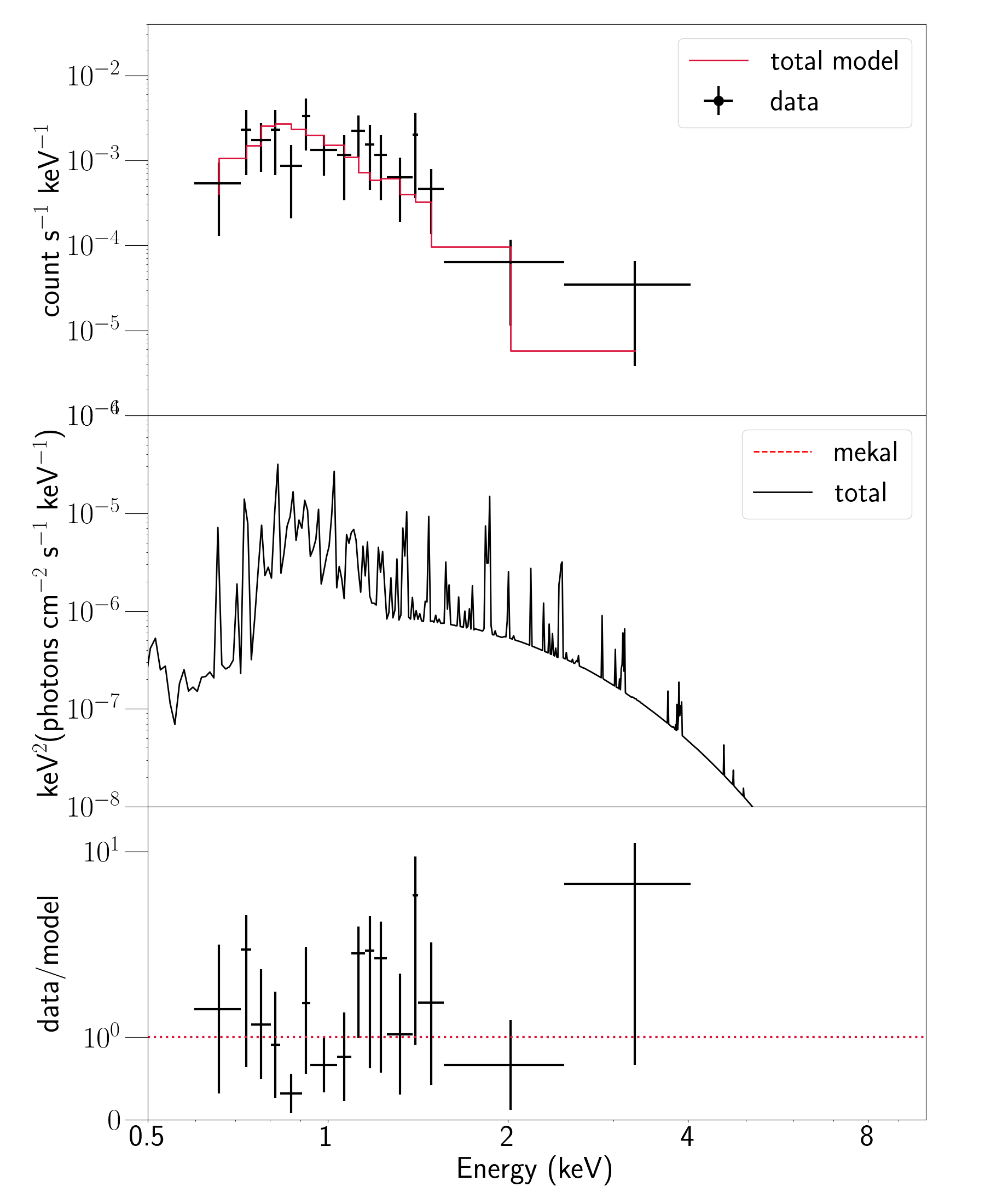}{0.375\textwidth}{(a) SN~1980K, Chandra, epoch--8063d, obsid--4404, binning=2}
          \fig{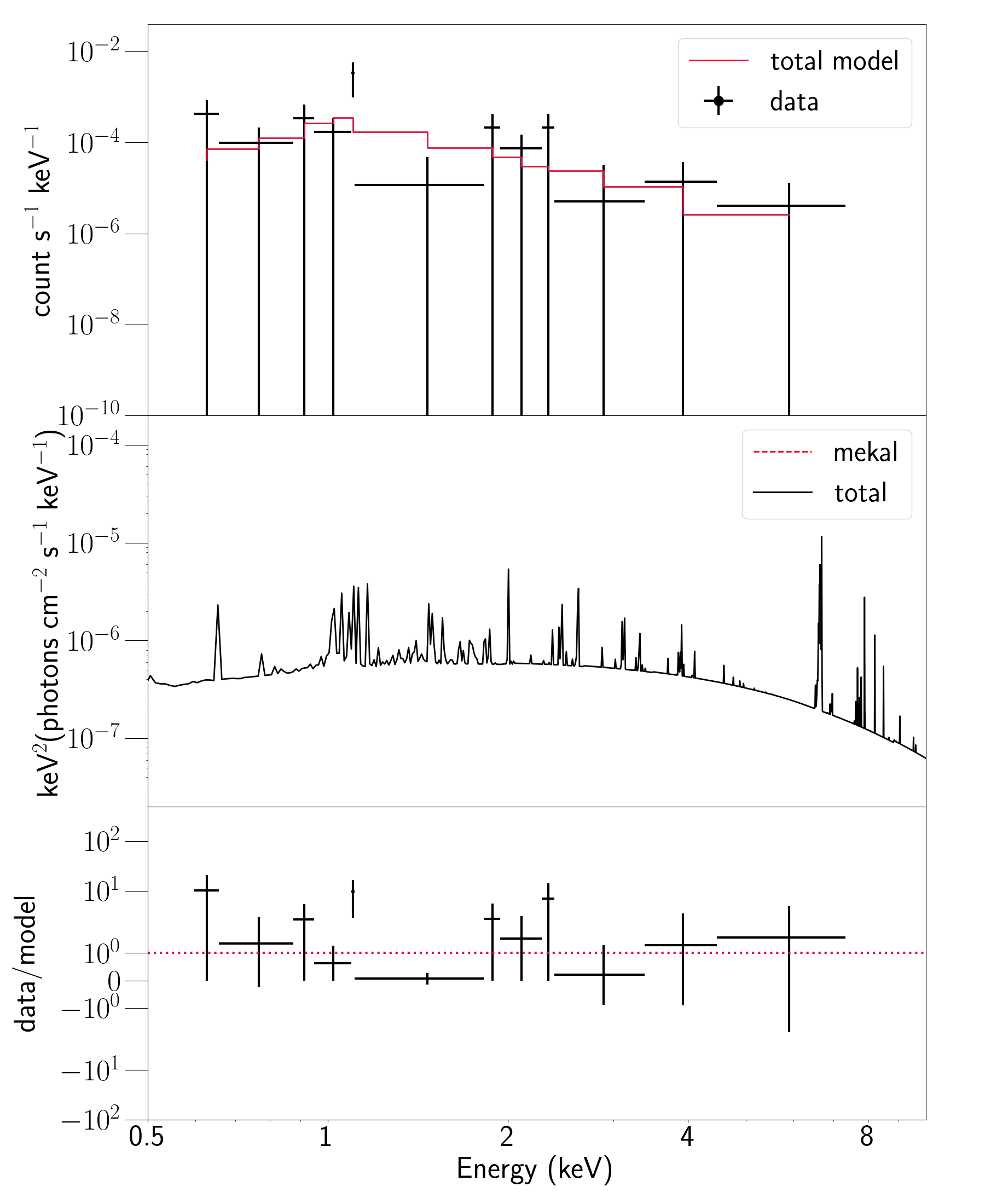}{0.375\textwidth}{(b) SN~1982R, Chandra (I3), epoch--11313d, obsid--16487, binning=1}~
          }
\gridline{\fig{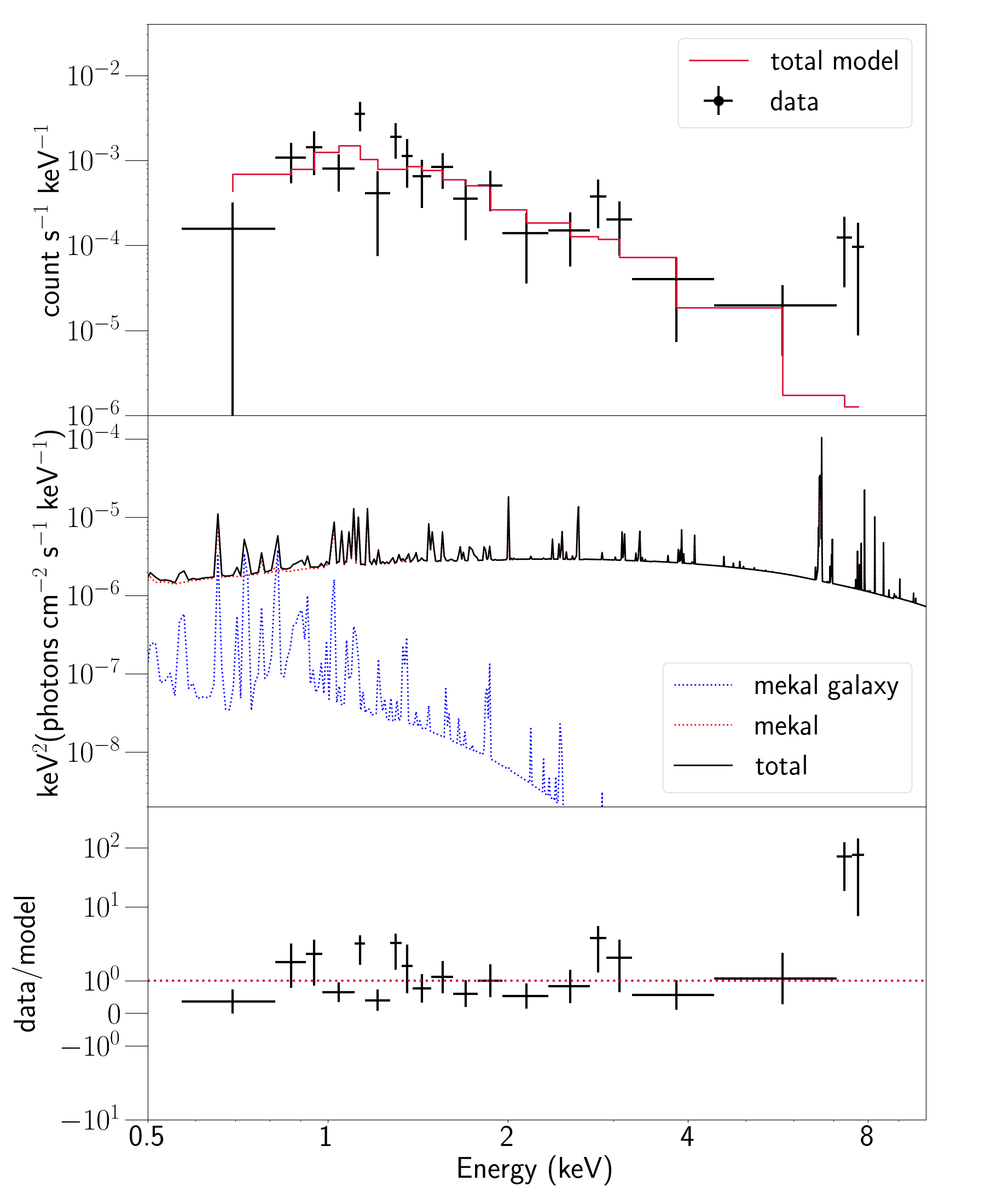}{0.375\textwidth}{(c) SN~1984J, Chandra (S3), epoch--11640d, obsid--16745, binning=3}
         \fig{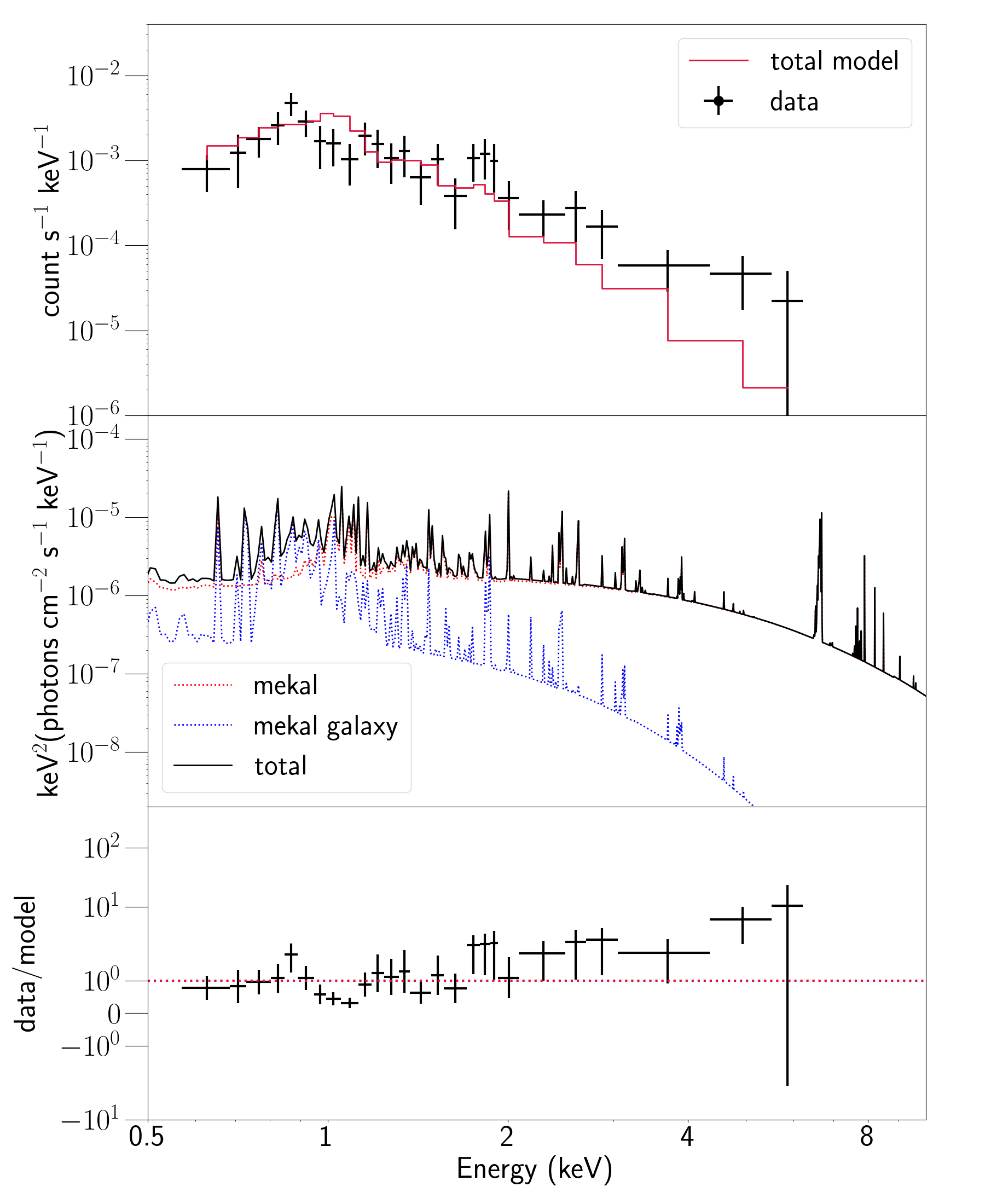}{0.375\textwidth}{(d) SN~1992bu, Chandra (S3), epoch--7674d, obsid--15077, binning=3}
          }
\caption{\footnotesize{Example fits of the detected SNe.} \label{fig:det_SN_2}}
\end{figure*}

\begin{figure*}[ht!]
\gridline{\fig{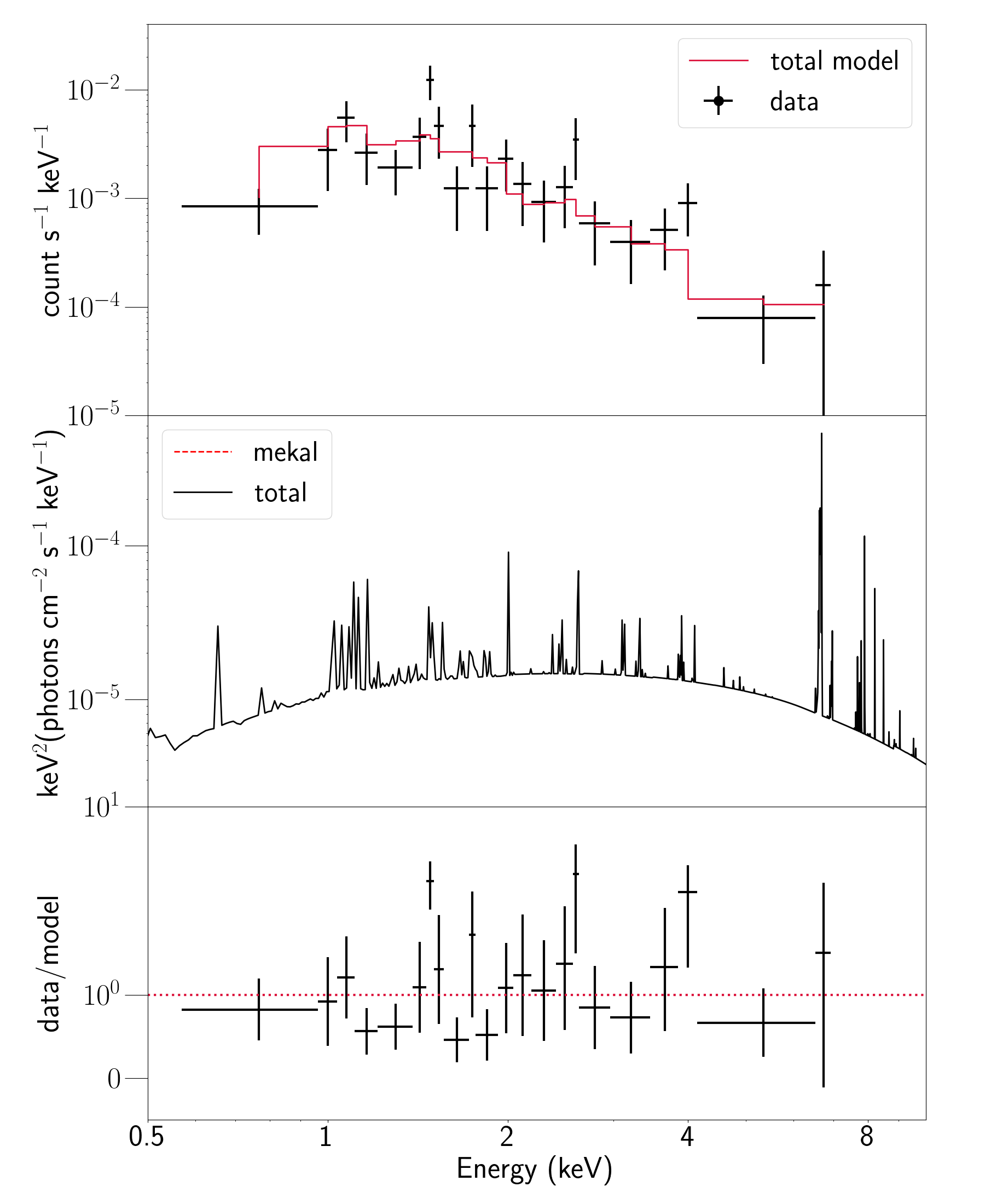}{0.375\textwidth}{(a) SN~2003gk, Chandra (S3), epoch--5530d, obsid--21004, binning=3}~
\fig{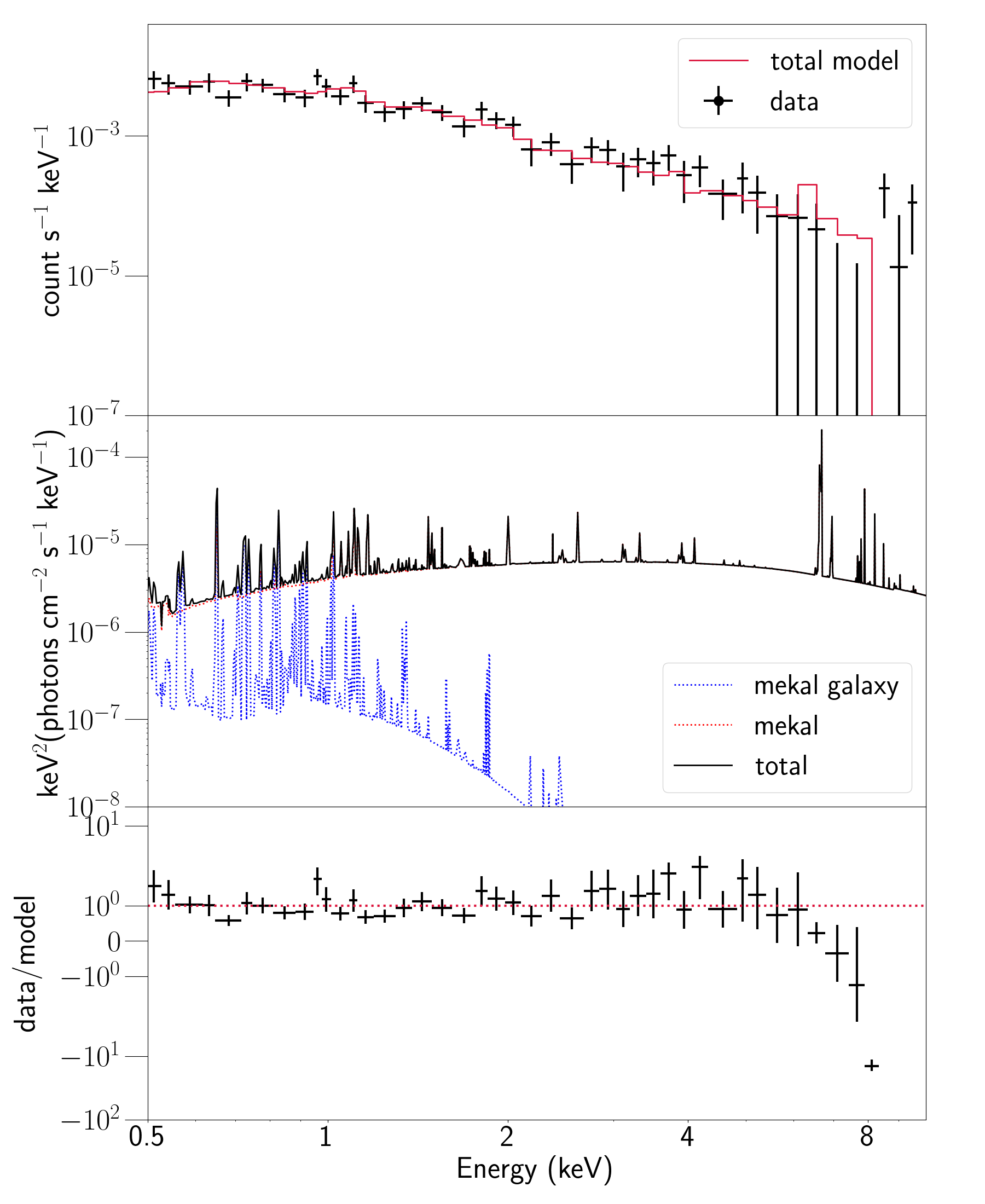}{0.375\textwidth}{(b) SN~2009jf, XMM, epoch--3167d, obsid--0824450601, binning=3}
          }
\gridline{\fig{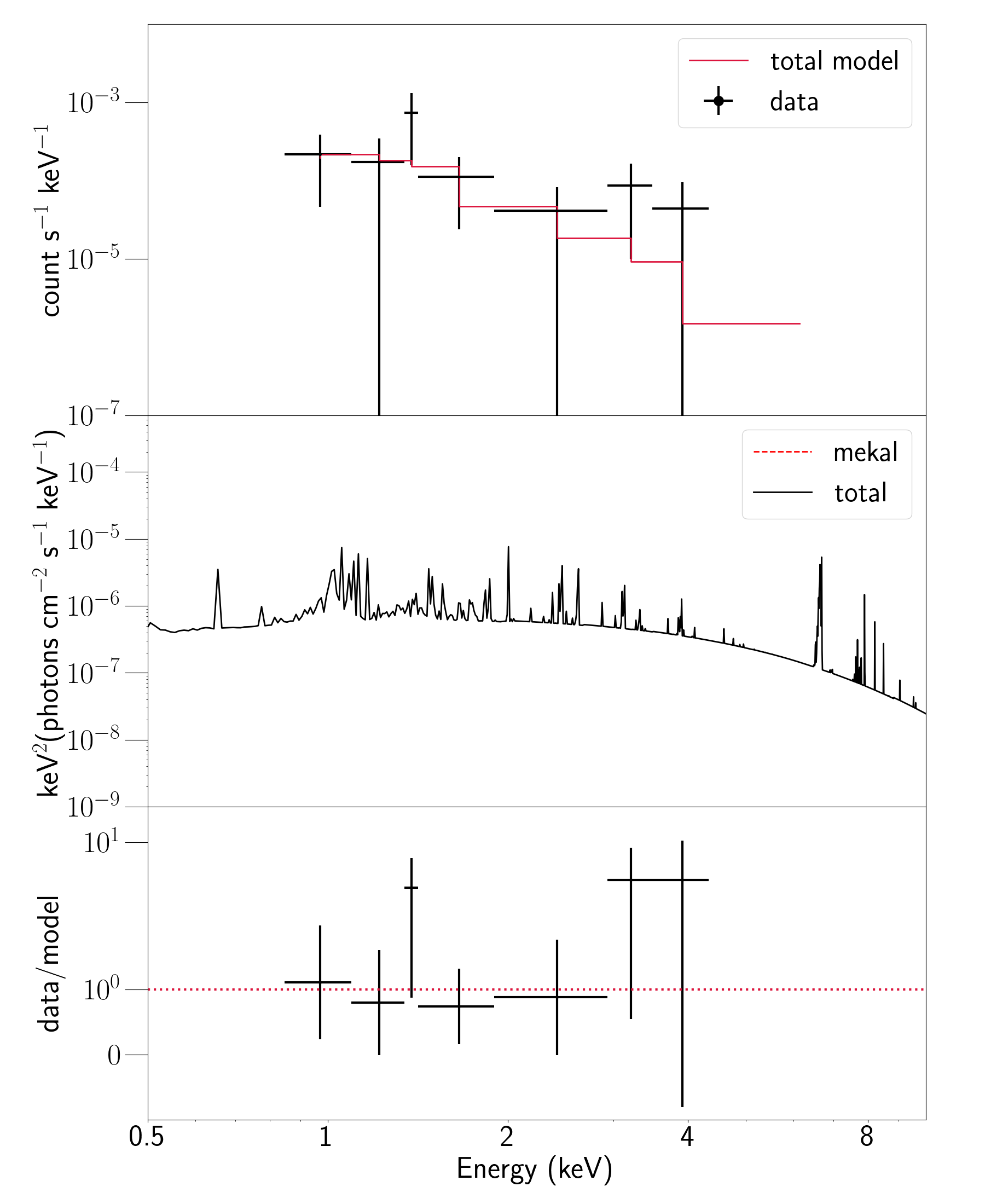}{0.375\textwidth}{(c) SN~2011dh, Chandra (S3), epoch--3422d, obsid--23472, binning=2}~
\fig{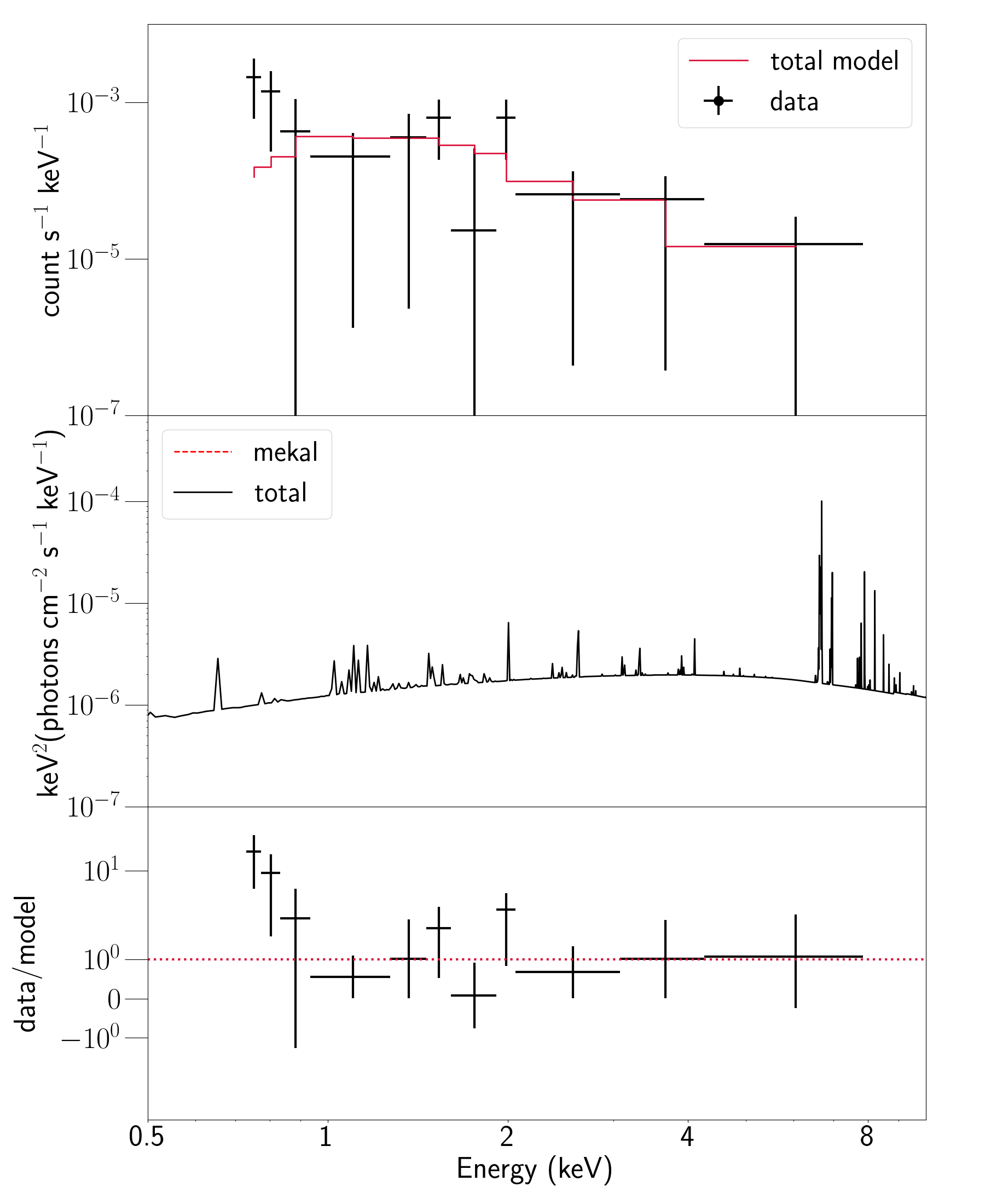}{0.375\textwidth}{(d) SN~2013df, Chandra (S3), epoch--1753d, obsid--21005, binning=2}
}
\caption{\footnotesize{Example fits of the detected SNe.} \label{fig:det_SN_3}}
\end{figure*}

In this section we include the spectral fits (see Figures~\ref{fig:det_SN_1}, \ref{fig:det_SN_2}, \ref{fig:det_SN_3}) of the detected SNe. For each SN, we show the observation with the best parameter constraints from Table~\ref{tab:detected_SN}. Each figure shows in the top panel the spectral data and fitted model, the middle panel shows the total model and different model components, and the bottom panel shows the residuals as log ratio of data to model. Some spectra have been rebinned for visual clarity. The number of counts per bin is specified in each panel as \textit{binning}.

\end{document}